%% file: 2HDMSpot.tex
\documentclass[a4paper,11pt]{article}
\usepackage{amsmath}
\usepackage{a4wide}
\usepackage{pstricks,pst-node,pst-text,pst-3d}
\usepackage{amsfonts}
\usepackage{amssymb}
\usepackage{graphicx,graphics,epsfig}
\usepackage{geometry}
\usepackage[english]{babel}
\usepackage[utf8]{inputenc}
\usepackage[T1]{fontenc}
 \usepackage{float}
\usepackage{hyperref}
\usepackage{cite}
\usepackage{ulem}

\numberwithin{equation}{section} 
\def\noi{\noindent}
\def\non{\nonumber}

\newcommand{\tb}{t_\beta}
\newcommand{\cosb}{c_\beta}
\newcommand{\sinb}{s_\beta}
\newcommand{\sbt}{s_{2\beta}}

\newcommand{\drbar}{\overline{\rm DR}}

\def\l{\lambda}
\def\lt{{\tilde \lambda}}

\def\b{\beta}
\def\g{\gamma}
\newcommand{\Lag}{\mathcal{L}}

\def\k{\kappa}

\def\sbb{\sin\!2\b}
\def\cb{\cos\!\b}
\def\sb{\sin\!\b}
\def\ba{\begin{array}}
\def\ea{\end{array}}
\def\bea{\begin{eqnarray}}
\def\eea{\end{eqnarray}}
\newcommand{\Omegah}{\Omega_{\chi} h^{2}}

\newcommand{\neuto}{{\tilde{\chi}}_1^0}

\newcommand{\ra}{\rightarrow}
\def\lsim{\;\raise0.3ex\hbox{$<$\kern-0.75em\raise-1.1ex\hbox{$\sim$}}\;}
\def\gsim{\;\raise0.3ex\hbox{$>$\kern-0.75em\raise-1.1ex\hbox{$\sim$}}\;}

\begin{document}
\begin{titlepage}
 \vspace*{0.1cm}
 \rightline{TTP12-032}

\begin{center}
 {\bf Analysis of the Higgs potentials for two doublets and a singlet.}

\vspace{0.5cm}
{G.~Chalons, F.~Domingo}\\

\vspace{4mm}
{\it Institut f\"ur Theoretische Teilchenphysik, \\
Karlsruhe Institute of Technology, Universit\"at Karlsruhe \\
Engesserstraße 7, 76128 Karlsruhe, Germany}
\vspace{10mm}

\abstract{
We consider the most general CP-conserving renormalizable effective scalar
potential involving two doublets 
plus one singlet Higgs and satisfying the electroweak gauge symmetry. After
deriving the electroweak-symmetry 
breaking conditions, we focus on special cases, characterized by
specific symmetry properties
and/or relations to supersymmetry-inspired extensions of the Standard Model
({e.g.}\ n/NMSSM, UMSSM). We then
investigate the question of the reconstruction of the potential parameters from
the Higgs masses and mixing 
angles and show that in some specific cases, such as the one of an underlying
NMSSM, an accuracy at the 
order of leading-logarithms is achievable with minimal effort. We finally study
a few phenomenological consequences for this latter model. More specifically,
we consider how our parameter reconstruction modifies the outcome of two
publicly available codes : \texttt{micrOMEGAs} and \texttt{NMSSMTools}. We
observed noteworthy effects in regions of parameter space where Higgs-to-Higgs
decays are relevant, impacting the collider searches for light Higgs states and
the prediction of the Dark-Matter relic density.
}
\end{center}
\normalsize
\end{titlepage}

\section*{Introduction}

\par\noi The origin of ElectroWeak Symmetry Breaking (EWSB) stands as one of
the 
critical questions in high-energy physics and a central goal of the 
Large Hadron Collider (LHC) is to reveal its nature. The recent
discovery 
of a new massive boson around $125$~GeV \cite{higgsday12}, reported by both the
ATLAS 
and CMS collaborations \cite{XLHC}, and
supported by the broad excess seen at TeVatron \cite{Xtev0712},
represents a first step towards the identification of the Higgs boson and the
measurement of the underlying Higgs potential, a task which however only the
next generation of colliders will probably complete. Although essentially
compatible with the Higgs boson of the Standard Model (SM), this new state may
already be hinting towards some new physics, in that the peaks of the diphoton
and $ZZ\ra4l$ decays differ from what one would expect in the SM. 
The stronger signal in the $H \ra \g\g$ channel, in particular, seems of
importance because this loop-induced process is particularly sensitive to
physics beyond the SM. One should also consider the non-observation
of events at CMS -- although supported by very little statistics -- in the
$H\ra\tau\tau$ channel. Testing the SM-nature of this would-be Higgs state,
inspecting possible deviations in its coupling to SM particles shall represent a
major undertaking of modern particle physics and a probable probe into the
mechanism of EWSB. 

\par\noindent The `Higgs mechanism'\cite{higgsmech},
involving scalar elementary fields, is the most efficient way to generate masses
for the fermions and gauge-bosons. 
Its implementation within the SM is the minimal one: only one scalar field,
transforming 
as a doublet under $SU(2)_L$, is introduced to break the electroweak (EW)
symmetry through
its vacuum expectation value (v.e.v.). Nevertheless the Higgs sector is still
essentially 
undetermined and there is no reason to stick to minimality if some benefits
should emerge 
from a more elaborate scalar sector. For instance, introducing a second Higgs
doublet allows 
for an implementation of CP violation through this sector \cite{lee73}: CP 
violation appears in this context because some of the parameters in the
potential of the Two Higgs Doublet Model (2HDM) 
can be chosen complex (non-real). Yet the requirements relative to neutral
flavor conservation constrain this possibility significantly. Large
flavour-changing couplings of neutral Higgs bosons can be avoided
in the so-called `2HDM of type II', where the Higgs doublets $H_u$ and $H_d$, of
opposite hypercharges $Y=\pm 1$, enter separately, and respectively, up- and
down-type Yukawa terms (at tree level). Another (more exotic) possibility
consists in requiring the alignement of the Yukawa coupling matrices in flavor
space: see \cite{pich2HDM}. Although such 2HDM's may hold as autonomous
extensions of the SM, they can also be embedded within more elaborate models: 
Left-Right gauge models and their Grand-Unification Theory (GUT) ramifications
-- Pati-Salam, $SO(10)$, etc.-- offer a first framework for this
operation, in which the question of CP-violation was originally central
\cite{LRmod}.

\par\noindent From another angle, the well-documented `Hierarchy Problem'
\cite{HierProb}
underlines the theoretical difficulties for understanding the stability of a
Higgs mass at the electroweak scale, with respect to new-physics at very-high
energies (GUT, Planck scales). Regarding the SM as the low-energy effective
theory of some more-fundamental model, the quadratic sensitivity 
of scalar squared masses to new-physics masses would lead to a technically
unnatural fine-tuning of the Higgs-mass parameter in the more-fundamental theory
with the radiative corrections resulting from the integrated-out new-physics
states\ldots\ Unless new-physics appears sufficiently close to the electroweak
scale: typically at the TeV scale. Among the proposed solutions, Supersymmetry
(SUSY) allows to stabilize a scalar Higgs mass at the electroweak scale, due to
the renormalization properties of supersymmetric theories.
However, SUSY being obviously not realized in low-energy particle physics,
viable SUSY-inspired models need to include SUSY-breaking effects, which are
parametrized within the Lagrangian through the so-called `soft terms', generate
e.g.\ mass terms for all non-SM particles and trigger the Higgs mechanism. This
ad-hoc setup could yet remain an acceptable solution to the Hierarchy Problem
only if the supersymmetry-breaking scale is near the
electroweak scale.
Other attractive properties of SUSY-inspired models lie in the 
possibility of one-step unification, due to the more-convergent running of
SM-gauge couplings in the presence of the enlarged SUSY field-content
\cite{gaugerun}, or in the
dark-matter (DM) sector, the lightest supersymmetric particle being a stable (or
long-lived) and viable candidate in the presence of (approximate) R-parity
\cite{SUSYDM}.

\par\noindent Holomorphicity of the superpotential (cancellation of
gauge-anomalies) dictates the
requirement for at least two $SU(2)_L$ Higgs doublets in a SUSY-inspired model,
intervening in a 
Type II 2HDM fashion, so that both up-type and down-type masses be generated.
The simplest 
implementation of a SUSY-inspired SM, known as the Minimal Supersymmetric
Standard Model (MSSM) \cite{MSSMrev} confines to this
minimal 2HDM requirement. There, the quartic Higgs couplings are determined by 
the EW gauge couplings, which results in tight constraints on the tree-level
mass of the lightest Higgs boson: the latter is indeed bounded from above by the
$Z^0$-boson mass $M_Z$. Radiative corrections improve this feature and can
arrange for fairly heavy Higgs masses provided the SUSY-scale is large enough,
see for example \cite{Djouadi_Susy_Higgs}. Yet this last necessity tends to
conflict with the naturalness-dictated $\lsim1$~TeV SUSY-breaking scale.
Accommodating for a Higgs state at $125$ GeV in the MSSM hence severely
constrains the parameter space of this model \cite{MSSMscan}. Another 
criticism to this minimal setup, the so-called `$\mu$-problem' \cite{kim83},
points out the necessity of tuning a supersymmetric mass-term, the
conventionally-baptized $\mu$ parameter, at the electroweak/TeV scale 
in order to ensure EWSB: being of supersymmetric origins, this parameter is in
principle unrelated to the SUSY-breaking scale and would thus coincide with it
out of sheer coincidence.

\par\noindent The introduction of an additional gauge-singlet superfield $S$
addresses both 
shortcomings of the MSSM. The $\mu$-term can indeed be generated effectively
through a $\lambda SH_u\cdot H_d$
term when the singlet takes a v.e.v.\ $s$: $\mu_{\mbox{\tiny eff.}}\equiv\lambda
s$ \cite{fayet}. Concerning the lightest Higgs mass, the
presence of a new superfield coupling to the Higgs doublets induces additional
contributions to the Higgs mass matrix, so that the MSSM limit can be exceeded,
already at tree-level \cite{barbieri06,nmssmhiggsbound}. It is also
worth to mention that the lightest CP-even Higgs state in this context 
might well be dominantly of a singlet nature, hence, the singlet decoupling from
SM-fermions and gauge bosons, essentially invisible at colliders: the SM-like
Higgs state would then be the second lightest and a small mixing effect with the
singlet would thus shift its mass towards slightly higher values. In short,
radiative corrections are no longer the only mechanism able to generate a
SM-like Higgs-state heavier than $M_Z$ in such a singlet-extension.

\par\noindent The simplest version of such a model with singlet-enlarged
superfield content is known as the 
Next-to-Minimal Supersymmetric Standard Model (NMSSM)
\cite{nmssmrev-ellwanger,nmssmrev-maniatis}. 
It relies on a $\mathbb{Z}_3$ discrete symmetry in order to forbid all
dimensional parameters (including 
$\mu$) in the superpotential, so that the soft-terms provide the only relevant
scale in the scalar 
potential, triggering the EWSB. Several other SUSY-models engaging a singlet in
addition to the two Higgs 
doublets are to be found in the literature, including the nearly Minimal
Supersymmetric Standard Model 
(nMSSM, sometimes MNSSM) \cite{nMSSM,nMSSM-tad}, $U(1)'$-extended MSSM's, with
their simplest version known 
as the UMSSM \cite{UMSSM}, models based on the $E_6$ exceptional group
\cite{E6SSM}, SUSY/compositeness
hybrids, such as `fat Higgs models' \cite{fathiggs} or models
using the Seiberg Duality
\cite{csaki11}, etc.

\par\noi In the present paper, we aim at studying the effective Higgs potential
involving 
$2$-doublet$+1$-singlet Higgs fields. The relations between physical input,
represented by
the mass matrices and mixing angles, and the parameters of the potential, as
well as with the 
trilinear Higgs couplings, shall be at the center of this discussion, in view of
a possible
reconstruction of the potential from such input, at, and beyond, leading order
(LO). Similar 
analyses for the $1$-doublet setup \cite{1doubHP}, or the $2$-doublet
setup,
for instance in \cite{me-3h-2hdm}, with the MSSM as a background-model, have
already been 
proposed in the literature. Given that the singlet-extensions of the MSSM offer
a natural 
origin to our $2$-doublet$+1$-singlet setup, we shall refer and return
explicitly to such 
models in the course of our discussion: specific attention will be dedicated in
particular 
to the n/NMSSM or the UMSSM. Most of our discussion should however be
generalizable to 
other models resulting in a $2$-doublet$+1$-singlet Higgs potential\footnote{We
have already
referred to Left-Right
models and their GUT extensions as an alternative approach to the 2HDM
framework. Note that the addition
of a SM-gauge singlet is essentially an undemanding requirement and may be
arranged within such 
models as well.}, as long as matching
conditions or/and symmetry properties are satisfied. The first part of the
present paper
shall be dedicated to the presentation of the general framework, including
notations, the
discussion of residual symmetries and the pattern of EWSB leading to the Higgs
spectrum.
In the second part, we shall focus on the question of the reconstruction of the
potential
from a measurement of the Higgs masses and mixing angles: beyond the general
case where a
large number of undetermined parameters remain, the possibility of a
reconstruction in 
constrained models will be discussed at leading order. The analysis of the large
logarithms 
appearing in the Coleman-Weinberg \cite{colemanweinberg,rge}
approach
shall convince us, in particular, that a full reconstruction at the order of
leading logarithms
should be achievable in the $\mathbb{Z}_3$-symmetric case represented by an
underlying NMSSM.
Concentrating on the NMSSM in the last part, we shall analyse the
phenomenological consequences
for this model, both in terms of constraints from Higgs-to-Higgs decays and
computation of
the Dark-Matter relic density. The decay $h_i^0 \ra \g\g$ \cite{ellwanger-gg}
will also be 
revisited, although little impact is expected there. This phenomenological
analysis will rely 
on the numerical output of several public codes, including
\texttt{NMSSMTools\_3.2.0} 
\cite{nmssmtools,nmhdecay}, \texttt{micrOMEGAs\_2.4.1}
\cite{micromegas,belanger-nmssm} and a version 
of \texttt{SloopS} \cite{boudjema05,baro08} adapted to the NMSSM
\cite{chalons11}.

\section{Two Higgs doublet plus one singlet potential}
\subsection{General parametrization}
New-Physics (NP) effects are most conveniently encoded in terms of effective
Lagrangians. Under the guidelines of Lorentz and gauge invariance, as well as
possible additional symmetries, one can write a list of all the operators,
classified according to their mass-dimension. For the two $SU(2)_L$ 
doublets and the singlet, we shall use the notations (with $v_d$, $v_u$ and $s$
representing the v.e.v.'s of these fields):
\begin{equation}
\label{conventions}
  H_d = \begin{pmatrix}
         v_d + (h_d^0 + i a_d^0)/\sqrt{2} \\ H_d^-
        \end{pmatrix},\hspace{2mm}
 H_u = \begin{pmatrix}
         H_u^+ \\ v_u + (h_u^0 + i a_u^0 )/\sqrt{2}
        \end{pmatrix},\hspace{2mm}
 S =  s + (h_s^0 + i a_s^0 )/\sqrt{2}
\end{equation}\noi  
The most general Higgs potential involving these fields
and compatible with
the electroweak gauge 
symmetry then reads, when one restricts to renormalizable terms: 
\begin{eqnarray}
\label{genradpot}
{\cal V}^S_{\rm
eff.}&=&m_{H_u}^2|H_u|^2+m_{H_d}^2|H_d|^2-\left(m_{12}^2H_u\cdot
H_d+h.c.\right)+\frac{\lambda_1}{2}|H_d|^4+\frac{\lambda_2}{2}
|H_u|^4+\lambda_3|H_u|^2|H_d|^2 \non \\
&+&\lambda_4|H_u\cdot H_d|^2+\left[\frac{\lambda_5}{2}(H_u\cdot
H_d)^2+(\lambda_6|H_u|^2+\lambda_7|H_d|^2)H_u\cdot H_d+h.c.\right] \non \\
&+&m_S^2|S|^2+\kappa^2|S|^4+\left[\lambda_TS+\frac{\mu^2_S}{2}
S^2+\frac{A_S}{3}S^3+\frac{\tilde{A}_S}{3}S|S|^2+\frac{\kappa_S^2}{4}S^4+\frac{
\tilde{\kappa}_S^2}{4}S^2|S|^2+h.c.\right] \non \\
&+&\left[A_{ud}SH_u\cdot H_d+\tilde{A}_{ud}S^*H_u\cdot
H_d+\lambda_M|S|^2H_u\cdot H_d+\lambda_P^MS^{*\,2}H_u\cdot
H_d+\tilde{\lambda}_P^MS^{2}H_u\cdot H_d+h.c.\right] \non \\
&+&\lambda_P^u|S|^2|H_u|^2+\lambda_P^d|S|^2|H_d|^2+\left[(A_{us}S+\tilde{\lambda
}
_P^uS^2)|H_u|^2+(A_{ds}S+\tilde{\lambda}_P^dS^2)|H_d|^2+h.c.\right]
\end{eqnarray}\noi 
The first two lines comprise the usual 2HDM potential, the third one,
the pure-singlet terms and the latter two, the singlet-doublet mixing-terms. 
$m_{H_u}^2$, $m_{H_d}^2$, $\lambda_1$, $\lambda_2$, $\lambda_3$,
$\lambda_4$, $m_{S}^2$, $\kappa^2$, 
$\lambda_P^u$ and $\lambda_P^d$ are (10) real parameters, while $m_{12}^2$,
$\lambda_5$, $\lambda_6$, $\lambda_7$, $\lambda_T$,
$\mu^2_S$, $A_S$, $\tilde{A}_S$, $\kappa^2_S$, $\tilde{\kappa}^2_S$, $A_{ud}$,
$\tilde{A}_{ud}$, $\lambda_M$, $\lambda_P^M$, 
$\tilde{\lambda}_P^M$, $A_{us}$, $A_{ds}$, $\tilde{\lambda}_P^u$ and
$\tilde{\lambda}_P^d$ are (19) in-principle-complex parameters. One parameter 
({e.g.}\ $\lambda_T$) is superfluous and may be absorbed in a translation of
the singlet; three others ($m^2_S$, $m_{H_u}^2$ 
and $m^2_{H_d}$) can be traded for the field vacuum expectation values through
the minimization conditions. From now on, we will consider, for simplicity,
that 
all the parameters are real, hence barring the possibility of CP-violation.
(We will however continue to refer to the 19 potentially non-real parameters
as `complex' parameters.)

\subsection{Symmetry classification}
By imposing additional symmetries, the form of the potential in
Eq.(\ref{genradpot}) can be further constrained at the classical level and the 
remaining parameters\footnote{We shall use the notation `$\l_i$' in order to 
concisely refer to \textit{any} parameter entering Eq.(\ref{genradpot}).} 
$\l_i^{\rm cl.}$ will be called `classical' parameters. At the quantum level,
all the
eliminated terms $\l_j^{\rm qm}$ may reappear, in principle, if the symmetry is
broken, either directly by the quantum fluctuations, or spontaneously, when the
fields 
acquire v.e.v.'s. In the later case, symmetry-violation is a relic 
from higher-dimensional operators at the non-symmetric vacuum, due to the truncation
of the potential to dimension $\leq4$ terms. To be definite, if at high energy,
beyond a certain scale
$\Lambda$,
the symmetry holds, the potential $\mathcal{V}$ is then well approximated by
its 
classical form (the symmetry-violating effects being negligible) and the
$\l_i^{\rm cl.}$ 
at the scale $\Lambda$ may be chosen as boundary conditions for the general
parameters of
Eq.(\ref{genradpot}), 
\begin{equation}
 \l_i^{\rm cl.}=\l(\Lambda)\,;\qquad \l_j^{\rm qm}=0
\end{equation}\noi 
such that ${\cal V}\equiv {\cal V}(\lambda_i^{\rm cl.}(\Lambda))$. At scales
$\mu \ll \Lambda$, however, symmetry-violating effects are no longer negligible
so that
non-trivial values of $\l^{\rm qm}$ are generated by the renormalization group 
equations. 

\par\noindent We shall now enumerate possible symmetries one can impose to 
the potential of Eq.(\ref{genradpot}): 
\begin{itemize}
 \item Discrete $\mathbb{Z}_n$-symmetries: they are characterized by the
transformations
$\Phi\mapsto e^{\frac{2\imath\pi }{n}Q_{\Phi}}\Phi$, where $\Phi=S,H_u,H_d$ and
$Q_{S,H_u,H_d}$ are the charges 
under the discrete symmetry group. They allow for significant selectivity
among the complex terms of the general potential, while avoiding the problem of
an axion 
(unless the potential they induce is also accidentally $U(1)$-invariant).
Spontaneous 
breakdown of these symmetries (through Higgs v.e.v.'s) however generically leads
to 
cosmological difficulties, in the form of a domain-wall problem
\cite{vilenkin85}, which should then be addressed separately.
\begin{enumerate}
\item The complex doublet-terms are controlled by $Q_{H_u}+Q_{H_d}$:
$Q_{H_u}+Q_{H_d}\equiv0\left[n\right]$ causes no constraint; for even $n$, 
$Q_{H_u}+Q_{H_d}\equiv\frac{n}{2}\left[n\right]$ allows only for $\lambda_5$; 
other choices forbid all the corresponding terms.
\item Complex mixing-terms are governed by both $Q_{H_u}+Q_{H_d}$ and $Q_S$.
Only in the case $\{Q_{H_u}+Q_{H_d}\equiv0\left[n\right]$,$Q_{S}\equiv0[n]\}$ are
they all
allowed by the $\mathbb{Z}_n$-symmetry. Otherwise, the relative choice of $Q_S$
and 
$Q_{H_u}+Q_{H_d}$ constrains them, with the specific values 
$Q_S\equiv\pm(Q_{H_u}+Q_{H_d})\left[n\right]$,
$2Q_S\equiv\pm(Q_{H_u}+Q_{H_d})\left[n\right]$ 
and up to the exclusion of all these terms.
\item The complex singlet-terms are governed by $Q_S$, ranging from conservation
of all ($Q_S\equiv0\left[n\right]$) to exclusion of all, with the special cases 
$2Q_S\equiv0\left[n\right]$, $3Q_S\equiv0\left[n\right]$ and
$4Q_S\equiv0\left[n\right]$.
\end{enumerate}
A typical example for such a discrete symmetry and deserving particular
attention
is that of the $\mathbb{Z}_3$-symmetry with charges $Q_{S,H_u,H_d}=1$: this
corresponds 
to the case of an underlying NMSSM.Invariance under $\Phi\mapsto
e^{\frac{2\imath\pi}{3}}\Phi$ reduces the potential to the form:
\begin{eqnarray}
 \label{Z3pot}
{\cal
V}^S_{\mathbb{Z}_3}&=&m_{H_u}^2|H_u|^2+m_{H_d}^2|H_d|^2+\frac{\lambda_1}{2}
|H_d|^4+\frac{
\lambda_2}{2} |H_u|^4+\lambda_3|H_u|^2|H_d|^2+ \lambda_4|H_u\cdot H_d|^2 \non \\
&+&m_S^2|S|^2+\kappa^2|S|^4+\left[\frac{A_S}{3}S^3+h.c.\right] \non \\
&+&\lambda_P^u|S|^2|H_u|^2+\lambda_P^d|S|^2|H_d|^2+ \left[A_{ud}SH_u\cdot
H_d+\lambda_P^MS^{*\,2}H_u\cdot H_d+h.c.\right] 
\end{eqnarray}\noi
The tree-level conditions resulting from the NMSSM read:
\begin{equation}
\label{NMSSMmatch}
 \begin{array}{llll}
\displaystyle
\l_1=\frac{g^2+g'^2}{4}=\l_2
 &;\,\displaystyle\lambda_3=\frac{g^2-g'^2}{4}
&;\,\displaystyle\lambda_4=\lambda^2-\frac{g^2}{2}
&;\,\displaystyle\lambda_P^u=\lambda^2=\lambda_P^d\ ;\\
\displaystyle\lambda_P^M=\lambda\kappa
&;\,\displaystyle A_S=\kappa A_{\kappa}
&;\,\displaystyle A_{ud}=\lambda A_{\lambda}
&;\,\displaystyle\kappa^2=\kappa^2
\end{array}
\end{equation}\noi 
Our notations for the SUSY parameters follow those of \cite{nmssmrev-ellwanger},
except for the 
electroweak gauge couplings which we denote as $g'$ and $g$ for, respectively,
the hypercharge 
$U(1)_Y$ and the $SU(2)_L$ symmetry.
\item Continuous global symmetries: here we mean essentially global phase
transformations $\Phi\mapsto e^{\imath Q_{\Phi}\alpha}\Phi$, that is
$U(1)$-Peccei-Quinn (P.Q.)
symmetries \cite{PQsym}. Such symmetries are spontaneously broken by the
v.e.v.'s of the Higgs fields so that they produce massless axions. They are
also chiral in nature, so that anomalies will be generated at the quantum level
(unless the field-content is enlarged so as to cancel them).
Such symmetries are thus likely to stand only as approximate limiting cases.
\begin{enumerate}
 \item $\{Q_{H_u}+Q_{H_d}=0, Q_{S}=0\}$ is automatically satisfied: this is the
hypercharge.
 \item $\{Q_{H_u}+Q_{H_d}=0, Q_{S}\neq0\}$ preserves the doublet potential while
constraining drastically the singlet couplings:
\begin{eqnarray}
{\cal V}^{S-S}_{PQ}
&=&m_{H_u}^2|H_u|^2+m_{H_d}^2|H_d|^2-\left(m_{12}^2H_u\cdot
H_d+h.c.\right)+\frac{\lambda_1}{2}|H_d|^4+\frac{\lambda_2}{2}
|H_u|^4+\lambda_3|H_u|^2|H_d|^2 \non \\
&+&\lambda_4|H_u\cdot H_d|^2+\left[\frac{\lambda_5}{2}(H_u\cdot
H_d)^2+(\lambda_6|H_u|^2+\lambda_7|H_d|^2)H_u\cdot H_d+h.c.\right]\non\\
&+&m_S^2|S|^2+\kappa^2|S|^4+\lambda_P^u|S|^2|H_u|^2+\lambda_P^d|S|^2|H_d|^2
+\left ( \lambda_M|S|^2H_u\cdot H_d+h.c.\right)
\end{eqnarray}\noi 
\item $\{Q_{H_u}+Q_{H_d}\neq0, Q_{S}=0\}$ constrains severely the doublet
sector, as well as the mixing terms, while leaving the pure-singlet potential
untouched:
\begin{eqnarray}
{\cal V}^{S-D}_{PQ}&=&m_{H_u}^2|H_u|^2+m_{H_d}^2|H_d|^2+\frac{\lambda_1}{2}
|H_d|^4+\frac{ \lambda_2}{2}|H_u|^4+\lambda_3|H_u|^2|H_d|^2+\lambda_4|H_u\cdot
H_d|^2 \non \\
&+&m_S^2|S|^2+\kappa^2|S|^4+\left[\lambda_TS+\frac{\mu^2_S}{2}
S^2+\frac{A_S}{3}S^3+\frac{\tilde{A}_S}{3}S|S|^2+\frac{\kappa_S^2}{4}S^4+\frac{
\tilde{\kappa}_S^2}{4}S^2|S|^2+h.c.\right] \non \\
&+&\lambda_P^u|S|^2|H_u|^2+\lambda_P^d|S|^2|H_d|^2+\left[\tilde{\lambda}
_P^uS^2|H_u|^2+\tilde{\lambda} _P^dS^2|H_d|^2+h.c.\right] \non \\
&+&\left[A_{us} S|H_u|^2+A_{ds}S|H_d|^2+h.c\right]
\end{eqnarray}
 \item $\{Q_{H_u}+Q_{H_d}\neq0, Q_{S}=-(Q_{H_u}+Q_{H_d})\}$ is the `usual'
Peccei-Quinn symmetry (e.g.\ \cite{miller03}) and, without loss of generality,
one may choose $(Q_{H_u}=1=Q_{H_d}, Q_{S}=-2)$. It induces a potential of the
same form as that of the $\mathbb{Z}_3$-symmetry (Eq.(\ref{Z3pot})),
with the further requirement that $A_S$ and $\lambda_P^M$ vanish.
\begin{eqnarray}
\label{PQpot}
 {\cal V}^S_{PQ}&=&m_{H_u}^2|H_u|^2+m_{H_d}^2|H_d|^2+\frac{\lambda_1}{2}
|H_d|^4+\frac{ \lambda_2}{2}|H_u|^4+\lambda_3|H_u|^2|H_d|^2+\lambda_4|H_u\cdot
H_d|^2 \non \\
&+&m_S^2|S|^2+\kappa^2|S|^4+\lambda_P^u|S|^2|H_u|^2+\lambda_P^d|S|^2|H_d|^2
+\left [ A_{ud}SH_u\cdot H_d+h.c.\right]
\end{eqnarray}\noi 
 \item $\{Q_{H_u}+Q_{H_d}\neq0, Q_{S}=Q_{H_u}+Q_{H_d}\}$ is equivalent to the
preceding case with the replacement $S\mapsto\tilde{S}=S^*$.
 \item $\{Q_{H_u}+Q_{H_d}\neq0, Q_{S}=\frac{1}{2}(Q_{H_u}+Q_{H_d})\}$ is a
variant,
concerning the singlet-doublet mixing-sector. This is again a subcase of the 
$\mathbb{Z}_3$-potential (Eq.\ref{Z3pot}), with vanishing $A_S$ and 
$A_{ud}$: in a coarse understanding of the term, this may be considered as
the `R-symmetric' potential.
\begin{eqnarray}\label{VPQp}
 {\cal V}^S_{PQ'}&=&m_{H_u}^2|H_u|^2+m_{H_d}^2|H_d|^2+\frac{\lambda_1}{2}
|H_d|^4+\frac{ \lambda_2}{2}|H_u|^4+\lambda_3|H_u|^2|H_d|^2+\lambda_4|H_u\cdot
H_d|^2 \non \\
&+&m_S^2|S|^2+\kappa^2|S|^4+\lambda_P^u|S|^2|H_u|^2+\lambda_P^d|S|^2|H_d|^2
+\left [ \lambda_P^MS^{*\,2}H_u\cdot H_d+h.c.\right]
\end{eqnarray}\noi 
Note that if one is interested in a SUSY-inspired model, this $PQ'$-symmetry 
would {a priori} forbid the $\lambda SH_u\cdot H_d$ term, resulting in
vanishing 
tree-level conditions for most of the parameters of Eq.(\ref{VPQp}): it is 
therefore best understood as a R-symmetry at the SUSY level.
\item $\{Q_{H_u}+Q_{H_d}\neq0, Q_{S}=-\frac{1}{2}(Q_{H_u}+Q_{H_d})\}$ is
equivalent
to the preceding choice, with the replacement $S\mapsto\tilde{S}=S^*$.
\item $\left\{Q_{H_u}+Q_{H_d}\neq0,
Q_{S}\neq\pm\left\{0,\frac{1}{2},1\right\}(Q_{H_u}+Q_{
H_d})\right\}$ forbids all the complex terms, hence leading to another,
more-constrained
subcase of the $\mathbb{Z}_3$-potential:
\begin{eqnarray}
 {\cal V}^{S-C}_{PQ}&=&m_{H_u}^2|H_u|^2+m_{H_d}^2|H_d|^2+\frac{\lambda_1}{2}
|H_d|^4+\frac{ \lambda_2}{2}|H_u|^4+\lambda_3|H_u|^2|H_d|^2+\lambda_4|H_u\cdot
H_d|^2 \non \\
&+&m_S^2|S|^2+\kappa^2|S|^4+\lambda_P^u|S|^2|H_u|^2+\lambda_P^d|S|^2|H_d|^2
\end{eqnarray}
\end{enumerate}
In the following, we shall focus only on ${\cal V}^S_{PQ}$ and ${\cal
V}^S_{PQ'}$, which
both can be viewed as subcases of ${\cal V}^S_{\mathbb{Z}_3}$.
 \item $U(1)'$-gauge symmetries: they can be regarded as the gauged-version of
the P.Q. symmetries,
with the important consequence that the P.Q.-axion is now unphysical. They
emerge naturally 
from $U(1)'$-SUSY models, containing SM-singlets charged under the
additional $U(1)'$-gauge symmetry and breaking it spontaneously while acquiring
v.e.v.'s. The simplest version of such models, with only one singlet, is called
UMSSM 
\cite{UMSSM} and leads back to the $\mathbb{Z}_3$-invariant Higgs potential, but
with vanishing $A_S$ 
and $\lambda_P^M$, {i.e.}\ ${\cal V}^S_{UMSSM}={\cal V}^S_{PQ}$: see
Eq.(\ref{PQpot}). The further
tree-level conditions are shifted from Eq.(\ref{NMSSMmatch})
according to (with $Q_{S,H_u,H_d}$ the Higgs charges under the $U(1)'$-symmetry
and 
$g_{Z'}$, the coupling constant):
\begin{equation}\label{UMSSM}
\lambda_{1,2}\ra\lambda_{1,2}+\frac{g^2_{Z'}}{2}Q_{H_
{u,d}}^2\hspace{0.2cm};\hspace{0.2cm}\lambda_3\ra\lambda_3+g^2_{Z'}Q_{H_u
}Q_{H_d}\hspace{0.2cm};\hspace{0.2cm}\lambda_P^{u,d}\ra\lambda_P^{u,d}
+g^2_{Z'}Q_{H_{u,d}}^2\hspace{0.2cm};\hspace{0.2cm}\kappa^2=\frac{g^2_{Z'}}{2}
Q_S^2
\end{equation}\noi
Note that the SM-fermion sector is also charged under the $U(1)'$-gauge group,
so as to
ensure invariance of the usual Yukawa terms. To avoid a chiral anomaly of the
$U(1)'$
symmetry, an exotic fermion sector will also be necessary.
\end{itemize}

\par\noindent One may also write tree-level conditions of a different form, not
protected by any symmetry: 
this is for instance the case in the nMSSM, where a $\mathbb{Z}_5^R$ or a
$\mathbb{Z}_7^R$ symmetry \cite{nMSSM-tad} is imposed 
at the level of the superpotential, so as to forbid all renormalizable pure
singlet-terms, then broken explicitly by 
gravity effects, in order to arrange for an effective tadpole term (so as to
break the resulting P.Q. symmetry), 
broken also explicitly by the soft-terms. The tree-level potential then differs
from the $\mathbb{Z}_3$ case 
(\ref{Z3pot}) by the requirements:
\begin{equation}\label{nMSSM}
 \l^M_P = \k = A_S = 0\ \ ;\quad \l_T,m_{12}^2 \neq 0
\end{equation}\noi
We hence define:
\begin{eqnarray}
 \label{nMSSMpot}
{\cal V}^S_{T}&=&m_{H_u}^2|H_u|^2+m_{H_d}^2|H_d|^2-\left(m_{12}^2H_u\cdot
H_d+h.c.\right)\non \\&+&\frac{\lambda_1}{2}|H_d|^4+\frac{
\lambda_2}{2} |H_u|^4+\lambda_3|H_u|^2|H_d|^2+ \lambda_4|H_u\cdot H_d|^2 \non \\
&+&m_S^2|S|^2+\left[\l_TS+h.c.\right]
+\lambda_P^u|S|^2|H_u|^2+\lambda_P^d|S|^2|H_d|^2+ \left[A_{ud}SH_u\cdot
H_d+h.c.\right] 
\end{eqnarray}\noi
While the absence of a residual symmetry at low-energy is a deliberate feature
of the nMSSM (in order to 
circumvent both axion and domain-wall problems), the resulting lack of
protection of the tree-level couplings 
at low-energy will lead to sizeable consequences for the parameter
reconstruction at the loop-level, 
as we will see later.

\subsection{Mass matrices}
Spontaneous symmetry breaking is achieved when the scalar fields develop a
v.e.v.,
\begin{equation}
 \left<H_u\right>=\left(\begin{array}{c}0\\ v_u\end{array}\right),\hspace{0.3cm}
\left<H_d\right>=\left(\begin{array}{c}v_d\\0\end{array}\right),\hspace{0.3cm}
\left<S\right>=s
\end{equation}\noi 
Imposing the minimization conditions associated with the most general potential
in Eq.(\ref{genradpot}), 
one may trade the parameters $m_{H_d}^2$, $m_{H_u}^2$, $m_S^2$ for the v.e.v.'s
$v_u$, $v_d$, $s$. Introducing the 
usual definitions $v\equiv\sqrt{v_u^2+v_d^2}\simeq174$~GeV, $\tan \b \equiv
v_u/v_d$,
we can write these relations as\footnote{We use the shorthand notations
$\cosb=\cb$,
$\sinb=\sb$, $\sbt=\sbb$, $\tb=\tan \b$ etc\ldots},
\begin{eqnarray}
\label{mincondgen}
m_{H_d}^2&=&
\left[A_{ud}+{\tilde A}_{ud}+(\lambda_P^M+{\tilde\l}^M_P+\l_M)s\right]s\,\tb
-\left[2 A_{ds}+(\lambda_P^d +2{\tilde\l}_P^d)s\right]s \non \\
&-&\lambda_1v^2\cosb^2-\left(\lambda_3+\lambda_4+\l_5\right)v^2\sinb^2
+\left(\l_6 v^2\sinb^2-m_{12}^2\right)\tb+3v^2 \sbt \l_7\non \\
 m_{H_u}^2&=&
\left[A_{ud}+{\tilde
A}_{ud}+(\lambda_P^M+{\tilde\l}^M_P+\l_M)s\right]s\,\tb^{-1}
-\left[2 A_{us}+(\lambda_P^u +2{\tilde\l}_P^u)s\right]s \non \\
&-&\lambda_2v^2\sinb^2-\left(\lambda_3+\lambda_4+\l_5\right)v^2\cosb^2
+\left(\l_7 v^2\cosb^2-m_{12}^2\right)\tb^{-1}+3v^2 \sbt \l_6\\
 m_S^2&=&
\left[A_{ud}+{\tilde A}_{ud}+2(\lambda_P^M+{\tilde\l}_M^P+\l_M)s\right]\frac{
v^2\sbt}{2s}
-\left[A_S+{\tilde A}_S+(2\kappa^2+\k_S^2 +{\tilde\k}_S^2)s\right]s \non \\
&-&\left[(A_{us}+2{\tilde\l}_P^u s)\sinb^2+(A_{ds}+2{\tilde\l}
^d_P s)\cosb^2\right]\frac{ v^2 } { s}
-\lambda_P^uv^2\sinb^2-\lambda_P^dv^2\cosb^2-\frac{ \l_T}{s}
-\mu_S^2 \non 
\end{eqnarray}\noi 
The quadratic terms in $H_{u,d}^\pm$ provide us with the charged Higgs mass
matrix:
\begin{align}
 {\cal M}_{H^{\pm}}^2&\equiv\left[(A_{ud}+{\tilde A}_{ud}+(\l_P^M+{\tilde\l}
^M_P+\l_M)s)s-\left(\frac{1}{2}
(\l_4+\l_5)\sbt-\l_6\sinb^2-\l_7\cosb^2\right)v^2
-m_{12}^2\right]\begin{bmatrix}\tb^{-1}&1\\1&\tb\end{bmatrix}
\end{align}\noi 
Its diagonalization expectedly delivers (massless) charged Goldstone bosons
$G^\pm\equiv \cb H_d^{\pm}-\sb H_u^{\pm}$ and the physical charged Higgs
$H^\pm\equiv \cb H_u^{\pm}+\sb H_d^{\pm}$, with mass:
\begin{equation}
\label{condcharged}
 m_{H^{\pm}}^2=\frac{2}{
\sbt}\left[(A_{ud }+{\tilde A}_{ud}+(\l_P^M+{\tilde\l}^M_P+\l_M)s)s
-\left(\frac{1}{2}(\l_4+\l_5)\sbt-\l_6\sinb^2-\l_7\cosb^2\right)v^2-m_{
12 }^2\right]
\end{equation}\noi
We turn to the CP-odd squared mass matrix, written in the basis
$(a_d^0,a_u^0,a_s^0)$:
\begin{eqnarray}
 {\cal M}_{P\,11}^2
&=&\left[\bigg(A_{ud}+{\tilde A}_{ud}+(\l^M_P+{\tilde\l}^M_P+\l_M)s\bigg)s
+\left(\l_6\sinb^2+\l_7\cosb^2-\l_5\sbt\right)v^2-m_{12}^2\right]\tb \nonumber\\
 {\cal M}_{P\,22}^2
&=&\left[\bigg(A_{ud}+{\tilde A}_{ud}+(\l^M_P+{\tilde\l}^M_P+\l_M)s\bigg)s
+\left(\l_6\sinb^2+\l_7\cosb^2-\l_5\sbt\right)v^2-m_{12}^2\right]\tb^{-1}
\nonumber\\
{\cal M}_{P\,33}^2
&=&\left[A_{ud}+{\tilde A}_{ud}
+ 4(\l^M_P+{\tilde\l}_P^M) s\right]\frac{v^2 \sbt}{2 s}
-\left[3 A_S+\frac{{\tilde A}_S}{3}+(4\k_S^2 +{\tilde\k}_S^2)s \right] s
 \\
& &-\left[\left(A_{us}+4 {\tilde\l}^u_Ps\right)\sinb^2 
+\left(A_{ds}+4 {\tilde\l}^d_Ps\right)\cosb^2\right]\frac{v^2}{s}
-2\mu_S^2-\frac{\l_T}{s}\nonumber \\
{\cal M}_{P\,12}^2
&=&\left[A_{ud}+{\tilde A}_{ud}+(\l^M_P+{\tilde\l}^M_P+\l_M)s\right]s
+\left(\l_6\sinb^2+\l_7\cosb^2-\l_5\sbt \right)v^2-m_{12}^2  \nonumber\\
{\cal M}_{P\,13}^2
&=&\left[A_{ud}-{\tilde A}_{ud}-2(\l^M_P-{\tilde\l}^M_P)s\right]v\sinb
\nonumber\\
{\cal M}_{P\,23}^2
&=&\left[A_{ud}-{\tilde A}_{ud}-2(\l^M_P-{\tilde\l}^M_P)s\right]v\cosb \nonumber
\end{eqnarray}\noi 
The neutral Goldstone boson $G^0\equiv\cb a_d^0-\sb a_u^0$ can be isolated
through the rotation with angle $\b$
and we are left with the $2\times2$ matrix ${\cal M}_{P'}^2$ in the basis
$(a_D^0,a^0_S)$, with $a_D^0\equiv\cb a_u^0+\sb a_d^0$
\begin{eqnarray}
{\cal M}_{P'\,11}^2
 &=&\frac{2}{\sbt}\left[\bigg(A_{ud}+{\tilde A}_{ud}+
(\l^M_P+{\tilde\l}^M_P+\l_M)s\bigg)s-(\l_5\sbt-\l_6\sinb^2-\l_7\cosb^2)v^2-m_
{12}^2\right]\nonumber \\
{\cal M}_{P'\,22}^2
&=&\left[A_{ud}+{\tilde A}_{ud}
+ 4(\l^M_P+{\tilde\l}_P^M) s\right]\frac{v^2 \sbt}{2 s}
-\left[3 A_S+\frac{{\tilde A}_S}{3}+(4\k_S^2 +{\tilde\k}_S^2)s \right] s \\
& &-\left[\left(A_{us}+4 {\tilde\l}^u_Ps\right)\sinb^2 
+\left(A_{ds}+4 {\tilde\l}^d_Ps\right)\cosb^2\right]\frac{v^2}{s}
-2\mu_S^2-\frac{\l_T}{s}\nonumber\\
{\cal M}_{P'\,12}^2
&=&\left[A_{ud}-{\tilde A}_{ud}-2(\l^M_P-{\tilde\l}_M^P)s\right]v \nonumber
\end{eqnarray}\noindent
${\cal M}_{P'}^2$ is diagonalized in the subblock of the physical states
$(a_D^0,a_S^0)$
by the orthogonal matrix $P'$, to give the two physical
CP-odd squared mass $m^2_{a_1^0}$, $m^2_{a_2^0}$, such that
\begin{equation}
\label{condCPodd}
 \mbox{diag}(m^2_{a_1^0},m^2_{a_2^0})=P'{\cal M}_{P'}^2P'^{-1}
\end{equation}\noi
\par\noi Finally, the CP-even squared mass matrix, in the basis
$(h_d^0,h_u^0,h_S^0)$, reads:
\begin{eqnarray}
 {\cal M}_{S\,11}^2
&=&
\left[\bigg(A_{ud}+{\tilde A}_{ud}+(\l^M_P+{\tilde\l}^M_P+\l_M)s\bigg)s
+(\l_6\sinb^2  -3\l_7\cosb^2)v^2-m_{12}^2\right]\tb +2 \l_1 v^2\cosb^2 
\nonumber\\
 {\cal M}_{S\,22}^2
&=&
\left[\bigg(A_{ud}+{\tilde A}_{ud}+(\l^M_P+{\tilde\l}^M_P+\l_M)s\bigg)s +
(\l_7 \cosb^2 -3\l_6\sinb^2)v^2 -  m_{12}^2)\right]\tb^{-1}+2 \l_2 v^2\sinb^2
\nonumber\\
{\cal M}_{S\,33}^2
&=&\left[A_{ud}+{\tilde A}_{ud}\right]\frac{v^2\sbt}{2s}
+\left[A_S+{\tilde A}_S + 2(2 \k^2+\k_S^2+{\tilde\k}_S^2)s\right]s 
-\left(A_{us}\sinb^2+A_{ds}\cosb^2\right)\frac{v^2}{s}-\frac{\l_T}{s}
\nonumber\\
{\cal M}_{S\,12}^2
&=&-\left[A_{ud}+{\tilde A}_{ud}+(\lambda_P^M+{\tilde\l}^M_P
+\l_M)s\right]s+\left[(\lambda_3+\lambda_4+\l_5)\sbt
-3(\l_6\sinb^2+\l_7\cosb^2)\right]v^2+m_{12}^2 \nonumber\\
{\cal M}_{S\,13}^2
&=&-\left[A_{ud}+{\tilde A}_{ud}+2(\l_P^M+{\tilde\l}^M_P+\l_M)s\right]v
\sinb+2\left[A_{ds}+(\l_P^d+2{\tilde\l}_P^d)s \right]v \cosb \\
{\cal M}_{S\,23}^2
&=&-\left[A_{ud}+{\tilde A}_{ud}+2(\l_P^M+{\tilde\l}^M_P+\l_M)s\right]v
\cosb+2\left[A_{us}+(\l_P^u+2{\tilde\l}_P^u)s \right]v \sinb\nonumber
\end{eqnarray}\noi
which is diagonalized by a $3\times3$ orthogonal matrix $S$, resulting in
three CP-even squared masses $m^2_{h_1^0}$, $m^2_{h_2^0}$, $m^2_{h_3^0}$, such
that 
\begin{equation}
\label{condCPeven}
 \mbox{diag}(m^2_{h_1^0},m^2_{h_2^0},m^2_{h_3^0})=S{\cal M}_{S}^2S^{-1}
\end{equation}\noi 
We are thus finally left with seven physical Higgs particles, once the three
Goldstone bosons $G^0$, $G^\pm$, giving mass to the $W^\pm$ and $Z^0$ bosons, 
have been discarded. In the particular case of the $U(1)'$-gauge symmetry,
however, 
the P.Q.-axion (associated to the vanishing eigenvalue of ${\cal M}_{P'}^2$) is
also 
unphysical (giving mass to the $Z'$-boson, gauge-field of the $U(1)'$ symmetry 
\cite{UMSSM}), so that we are left with only one CP-odd physical mass.

\section{Reconstruction of the effective parameters}
\subsection{Masses and mixing angles as physical input}
From an experimental point of view, the `$\l_i$' parameters are not directly
accessible: they will enter as combinations within the expressions for the Higgs
masses and self-couplings. The latter can hopefully be accessed through the
experimental
measurement of physical quantities. `Inverting the system', we can therefore
trade 
some $\l_i$ parameters for such physical input. In the simplest case, one would
directly 
use the Higgs masses and their mixing angles, assuming these can be measured 
({e.g.}\ from fermion/gauge couplings), as the new, physical input. For the 
$2$-doublet$+1$-singlet system, such quantities provide us with 12 conditions 
(input measurements) on the $\l_i$'s: the masses of the $2$ CP-odd bosons, $3$
CP-even 
and $1$ (complex) charged Higgs; the mixing angles from the CP-even ($3$),
CP-odd ($1$) and the Goldstone ($1$: $\b$)
sectors; finally, the electroweak v.e.v.\ $v = \sqrt{v_u^2 + v_d^2}$ (from $M_W$
for example). Should those twelve relations prove insufficient to determine all
the $\l_i$'s
(as is obviously the case for the most general potential), one would have to
resort to Higgs self-couplings (or input from another sector) in order to fully
determine the parameters.Accessing such self-couplings would
require that double or triple Higgs production are kinematically open. This
task would most comprehensibly done at future linear colliders. In the
meanwhile, the measurements of masses and mixing
angles still allow for a partial inversion.

\par\noindent We will assume in the following that all the Higgs-masses have
been 
measured. Note that this hypothesis is somewhat optimistic since singlet-like 
fields do not couple directly to SM-fermions and gauge-bosons, hence are
essentially 
elusive: only when there is substantial mixing with the doublet states can we
expect
to access them without having to rely on multi-Higgs couplings. As for the
mixing 
angles, assuming all the Higgs states have been observed in SM decay-channels,
one
can derive them from the couplings to fermions (note that leptonic decay
channels are
likely to give cleaner information) and gauge bosons. For a type II model, we have (taken from
\cite{nmssmrev-ellwanger}): 
\begin{eqnarray}
 h^0_i t_L t^c_R &=& -\frac{Y_t}{\sqrt{2}}S_{i2} \nonumber\\
 h^0_i b_L b^c_R &=& \frac{Y_b}{\sqrt{2}}S_{i1}  \nonumber\\
 h^0_i \tau_L \tau^c_R &=& \frac{Y_\tau}{\sqrt{2}}S_{i1}  \nonumber\\
 a^0_i t_L t^c_R &=& -i\frac{Y_t}{\sqrt{2}} c_\beta P'_{i1} \\
 a^0_i b_L b^c_R &=& i\frac{Y_b}{\sqrt{2}} s_\beta P'_{i1} \nonumber\\
 a^0_i \tau_L \tau^c_R &=& i\frac{Y_\tau}{\sqrt{2}} s_\beta P'_{i1}  \nonumber\\
 H^+ b_L t^c_R &=& Y_t c_\beta \nonumber\\
 H^- t_L b^c_R &=& -Y_b s_\beta \nonumber\\
 H^- {\nu_\tau}_L \tau^c_R &=&-Y_\tau s_\beta  \nonumber
\end{eqnarray}\noindent
where,
\begin{equation}
 Y_t = \frac{m_t}{v s_\beta}, \quad Y_b = \frac{m_b}{v c_\beta}, \quad Y_\tau =
\frac{m_\tau}{v c_\beta}
\end{equation}\noindent
and (we mention here only the $1$-Higgs to $2$-gauge couplings; note that,
albeit more difficult to measure, 
$2$-Higgs to $1$-gauge as well as quartic couplings shall play a very important
role for testing the model):
\begin{eqnarray}
 h^0_i Z_\mu Z_\nu &=&g_{\mu\nu}\frac{g'^2 + g^2}{\sqrt{2}}v\left(c_\beta
S_{i1} + s_\beta S_{i2}\right) \nonumber \\
 h^0_i W^+_\mu W^-_\nu &=& g_{\mu\nu}\frac{g^2}{\sqrt{2}}v\left(c_\beta
S_{i1} + s_\beta S_{i2}\right)
\end{eqnarray}\noi
Combining Higgs couplings to the vector bosons with those to up/down fermions,
one can access {e.g.}\ $S_{i1}$/$S_{i2}$. Moreover, one may be 
tempted to use Higgs decays into two photons to extract information about the 
mixing angles: even admitting that such processes are dominated by quark loops,
the corresponding relation of branching ratios to mixing angles is already
non-trivial and would require an involved extraction procedure for exploitation.

\par\noindent Unitarity relations could also prove useful. For example, a
`missing' 
matrix element $S_{ij}$ could be reconstructed from
\begin{equation}
 \sum_{k=0}^3 S_{ik}S_{jk}= \delta_{ij}=\sum_{k=0}^3 S_{ki}S_{kj} \quad
i,j=1,2,3
\end{equation}\noi 
\par\noindent
A possible (naive) strategy to reconstruct the mixing angles would be the
following: having measured the charged Higgs decay into third generation quarks,
one could then deduce $\tb$, 
since the ratio $m_{t,b}/v$ is fixed by SM measurements. Then the (doublet)
elements $S_{i1}$, $S_{i2}$, $P'_{i1}$ could be obtained unambigously from the
decays of neutral higgs states into fermions and gauge-bosons. The unitarity
relations would finally provide the magnitude of the $S_{i3}$ and $P'_{i2}$
(singlet) elements.
\par\noindent Note finally that, while a full experimental determination of the
Higgs mass
matrices may seem over-optimistic in the short run, there exists a practical case where we have
access to such
data: it is that of the output of spectrum generators ({e.g.}\ the publicly
available \texttt{NMSSMTools}, 
\cite{nmssmtools}). We will resort to that practical application in the last
part of the present paper.

\subsection{Partial reconstruction in the general case}
Considering the general potential of Eq.(\ref{genradpot}) and discarding any
assumption as to an underlying model, 
a complete reconstruction of the 29
parameters (28 of which being relevant) cannot succeed with only the twelve
mass/mixing conditions, 
hence calls for the measurement of Higgs self-couplings. Yet, information from
Eqs.(\ref{condcharged},\ref{condCPodd},\ref{condCPeven})
can already be implemented in a partial reconstruction:
\begin{equation}
\label{gencond}
 \begin{cases}
  [(A_{ud}+\l^M_Ps)s-m_{12}^2]\frac{2}{\sbt}
=m^2_{a_i^0}P^{'\,2}_{i1}+\l_P^1\\
 (A_{ud}-2\l^M_Ps)v=m^2_{a_i^0}P'_{i1}P'_{i2}+\l_P^{12}\\
 -3 A_S s +\left(A_{ud}+ 4\l^M_P s\right)\frac{v^2 \sbt}{2
s}-\frac{\l_T}{s}=m^2_{a_i^0}P_{i2}^{'2} +\l_P^{2}\\
\frac{2}{\sbt}\left[(A_{ud }+\l_P^Ms)s-\frac{\l_4}{2}v^2\sbt-m_{12}^2\right]
=m_{H^{\pm}}^2+\l_{\pm}\\
 [(A_{ud}+\l_P^Ms)s-m_{12}^2]\tb+2\l_1v^2\cosb^2=m^2_{h_i^0}S_{i1}^2-\l_S^1\\
 [(A_{ud}+\l_P^Ms)s-m_{12}^2]\tb^{-1}+2\l_2v^2\sinb^2=m^2_{h_i^0}S_{i2}
^2-\l_S^2\\
 A_Ss+4\k^2s^2+A_{ud}\frac{v^2}{2s}\sbt-\frac{\l_T}{s}=m^2_{h_i^0}S_{i3}
^2-\l_S^3\\
-(A_{ud}+\l_P^Ms)s+(\l_3+\l_4)v^2\sbt+m_{12}^2=m^2_{h_i^0}S_{i1}S_{i2}-\l_S^{12}
\\
 -(A_{ud}+2\l_P^Ms)v\sinb+2\l_P^dsv\cosb=m^2_{h_i^0}S_{i1}S_{i3}-\l_S^{13}\\
 -(A_{ud}+2\l_P^Ms)v\cosb+2\l_P^usv\sinb=m^2_{h_i^0}S_{i2}S_{i3}-\l_S^{23}
 \end{cases}
\end{equation}\noi 
where $\l^{1,2,3}_P,\l^{12}_P,\l_{\pm},\l^{1,2,3}_S,\l^{12,13,23}_S$ are given
by
\begin{equation}
\label{qmcond}
\begin{cases}
  \l_P^1 = -\frac{2}{\sbt}\left[2\left({\tilde A}_{ud}+
({\tilde\l}_M^P+\l_M)s\right)s-(\l_5\sbt-\l_6\sinb^2-\l_7\cosb^2)v^2\right] \\
 \l_P^{12} = ({\tilde A}_{ud}-2{\tilde\l}_M^Ps)v\\
 \l_P^{2}=-\left({\tilde A}_{ud}+4{\tilde\l}_P^Ms\right)\frac{v^2 \sbt}{2 s}
+\left[\frac{{\tilde A}_S}{3}+(4\k_S^2 +{\tilde\k}_S^2)s\right]s  
+\left[(A_{us}+4 {\tilde\l}^u_Ps)\sinb^2+(A_{ds}+4 {\tilde\l}^d_Ps)\cosb^2
\right]\frac{v^2}{s}+2 \mu_S^2\\
 \l_{\pm} = -\frac{2}{\sbt}\left[({\tilde A}_{ud}+({\tilde\l}^M_P+\l_M)s)s
-\left(\frac{1}{2}\l_5\sbt-\l_6\sinb^2-\l_7\cosb^2\right)v^2\right]\\
\l_S^1=\left[\bigg({\tilde A}_{ud}+({\tilde\l}^M_P+\l_M)s\bigg)s
+(\l_6\sinb^2  -3\l_7\cosb^2)v^2\right]\tb\\
\l_S^2 =\left[\bigg({\tilde A}_{ud}+({\tilde\l}^M_P+\l_M)s\bigg)s +
(\l_7 \cosb^2 -3\l_6\sinb^2)v^2\right]\tb^{-1}\\
\l_S^3 = {\tilde A}_{ud}\frac{v^2\sbt}{2s}+{\tilde A}_S s +
2(\k_S^2+{\tilde\k}_S^2)s^2-\left(A_{us}\sinb^2+A_{ds}\cosb^2\right)\frac{v^2}{s
} \\
 \l_S^{12}=-({\tilde A}_{ud}+({\tilde\l}^M_P+\l_M)s)s+\left[\l_5\sbt
-3(\l_6\sinb^2+\l_7\cosb^2)\right]v^2\\
 \l_S^{13}=-\left[{\tilde A}_{ud}+2({\tilde\l}^M_P+\l_M)s\right]v
\sinb+2(2{\tilde\l}_P^ds + A_{ds})v\cosb\\
 \l_S^{23}=-\left[{\tilde A}_{ud}+2({\tilde\l}^M_P+\l_M)s\right]v
\cosb+2(2{\tilde\l}_P^us +A_{us})v\sinb 
 \end{cases}
\end{equation}\noi
Our (arbitrary) choice in ordering the parameters within
Eqs.(\ref{gencond},\ref{qmcond}) was guided by the terms 
that are relevant at leading order in the n/NMSSM and the UMSSM potentials:
beyond $m_{H_d}^2$, $m_{H_u}^2$, $m_S^2$, which are 
common to the three models, those are given by
\begin{center}
\begin{tabular}{cl}
 NMSSM &: $\l_{1-4}$, $\l^{u,d}_P$, $\l^M_P$, $\k^2,A_{ud}$, $A_S$  \\
 nMSSM &: $\l_{1-4}$, $m_{12}^2$, $\l^{u,d}_P$, $\l_T$, $A_{ud}$ \\
 UMSSM   &: $\l_{1-4}$, $\l^{u,d}_P$, $\k^2$, $A_{ud}$\\
\end{tabular}
\end{center}\noi 
These parameters were collected on the left-hand side of Eq.(\ref{gencond}), 
while the remaining ones enter the right-hand side through Eq.(\ref{qmcond}). 

\par\noindent Note that the relations of Eq.(\ref{gencond}) hold at any order
(since
Eq.(\ref{genradpot}) is the most general renormalizable potential satisfying the
gauge-symmetry).
A practical use of Eq.(\ref{gencond}) would lie in a model-independent analysis
of a $2$-doublet$+1$singlet
potential (in order to discriminate among models, constrain them through precision tests). Then the twelve mass conditions can be used to simplify twelve
(arbitrarily chosen)
parameters, hence leaving the remaining couplings as the relevant degrees of
freedom intervening in / to be 
determined from the Higgs self-couplings. Not much predictivity should be
expected, however, in this general 
case.

\subsection{Reconstruction at the classical level in the constrained
models}\label{secLinv}
We focus here on the specific cases inspired by the SUSY models: ${\cal
V}^S_{T}$, ${\cal V}^S_{PQ}$, 
${\cal V}^S_{PQ'}$ and ${\cal V}^S_{\mathbb{Z}_3}$. Note that such potentials
are considered at 
the classical order: quantum effects and explicit/spontaneous breaking of the
symmetries in principle 
destabilize those potentials to generate the most general one. At this leading
order, however, the 
Eqs.(\ref{qmcond}) vanish, leaving Eqs.(\ref{gencond}) in a very simple form.
Note additionally the 
further requirements for each potential:
\begin{center}
\begin{tabular}{cl}
 ${\cal V}^S_{\mathbb{Z}_3}$ &: $m_{12}^2=\l_T=0$\\
 ${\cal V}^S_{T}$ &: $A_S=\k^2=\l_M^P=0$ \\
 ${\cal V}^S_{PQ}$ &:  $A_S=\l_M^P=m_{12}^2=\l_T=0$\\
 ${\cal V}^S_{PQ'}$ &:  $A_S=A_{ud}=m_{12}^2=\l_T=0$
\end{tabular}
\end{center}\noi 

\par\noi We end up with eleven classical parameters and eleven
conditions\footnote{The explicit presence of 
a P.Q.-axion, identified as $a_1^0$, leads to one trivial condition in the
CP-odd sector.} for both the 
potentials ${\cal V}^S_{PQ}$ and ${\cal V}^S_{PQ'}$. In these cases, all the
parameters in the Higgs potential 
can thus be reconstructed (at leading order) from Higgs masses and mixings: this
procedure is explicitly 
carried out in appendix \ref{aprec}, Eqs.(\ref{invPQ},\ref{invPQp}). 

\par\noi In the case of ${\cal V}^S_{\mathbb{Z}_3}$, the thirteen classical
parameters cannot be fully 
determined from the twelve conditions. The remaining degree of freedom is
conveniently chosen as the 
singlet v.e.v.\ $s$: the reconstruction is also given in appendix \ref{aprec},
Eqs.(\ref{lambdoub},\ref{lambnew}). 
Several tracks can 
be followed in order to determine this remaining degree of freedom. The first
one, sticking to the Higgs 
potential, would consist in relying on trilinear couplings, such as $h_i^0 H^+
H^-$ or $h_i^0 a_j^0 a^0_j$,
where the neutral Higgs fields would be largely singlet in nature: kinematical
limits and the elusive 
nature of singlets would tend to disfavor this strategy. Another possibility
would be to input information 
from some other sector (if any): measurement of the higgsino masses in the NMSSM
could provide the missing 
information. Finally, a more predictive option would be to enforce some
additional requirement, such as 
relations among the tree-level couplings: the tree-level relations of the
NMSSM, 
$\frac{\lambda_P^u}{\lambda_P^d}=1$ or
$\frac{\kappa^2(a\lambda_P^u+b\lambda_P^d)}{(\lambda_P^M)^2\cdot(a+b)}=1$ 
(where $a,b$ are real numbers), for instance, or a measure of the P.Q. symmetry
breaking, such as
$\frac{(a+b)\lambda_P^M}{a\lambda_P^u+b\lambda_P^d}\sim\frac{\kappa}{\lambda}$,
may be used as guidelines.

\par\noi Finally for ${\cal V}^S_{T}$, we have twelve parameters and twelve
conditions. Yet a full inversion is 
not possible either, because CP-even and CP-odd singlet masses are explicitly
degenerate in this potential, leaving
a bound system. The remaining degree of freedom is again chosen as the singlet
v.e.v.\ $s$ in appendix
\ref{aprec}, Eq.(\ref{invT}), but could be replaced by {e.g.}\ $\lambda_T$, as a
measure of the violation of $\mathbb{Z}_3$,
for example.

\par\noi So far, we have considered only the Higgs potentials separately. Moving
explicitly to the underlying 
SUSY models, however, the $\l_i$'s are further constrained by the tree-level
relations resulting from their 
supersymmetric origins: we count $7$ parameters in the nMSSM Higgs sector
($\lambda_T$, $m_{12}^2$, $m^2_{H_u}$, 
$m^2_{H_d}$, $m^2_{S}$, $\lambda$, $A_{\lambda}$), $7$ in the NMSSM as well
($m^2_{H_u}$, $m^2_{H_d}$, $m^2_{S}$, 
$\lambda$, $A_{\lambda}$, $\kappa$, $A_{\kappa}$) and $6$ in the UMSSM
($m^2_{H_u}$, $m^2_{H_d}$, $m^2_{S}$, 
$\lambda$, $A_{\lambda}$, $g_{Z'}$; note that we regard the Higgs charges under
$U(1)'$ as fixed). Those 
parameters are then over-constrained by Eq.(\ref{gencond}) and one should thus
consider the remaining 
conditions as a measurement of the deviation from tree-level conditions due to
higher orders (we remind here that the tree-level relations induced by the model of 
origin among the parameters 
of the potential are likely to be spoilt by quantum corrections). Depending on the
information at our disposal in the remaining spectrum ({e.g.}\ SUSY masses),
such conditions may be used to 
estimate the missing parameters ({e.g.}\ sfermion masses or trilinear soft
couplings) or regarded as precision tests of the model. Note that if the SUSY
spectrum
is sufficiently documented as well, this measurement of the Higgs parameters at
leading order, would allow for
a (perturbative) computation of all the $\l_i$'s within the specific models at
higher orders.

\subsection{Reconstruction at the loop level: NMSSM vs.\ nMSSM}
Now we want to apply this formalism to higher order effects. The purpose is
simple:
it has been shown that, in the MSSM, the bulk of the corrections in
Higgs-to-Higgs 
couplings could be absorbed in writing such couplings in terms of the corrected
masses (see for example \cite{brignole-3h,hollik-3h} and the third reference in
\cite{1lpotMSSM}); could a similar recipe apply to the $2$-doublet$+1$singlet
setup? A first strategy is the one presented at the end of the previous
subsection: in a definite model, the Higgs spectrum may allow for a
determination of the Higgs parameters at leading order; then, provided
sufficient information from the other sectors stand at our disposal,
reconstructing all the $\l_i$'s at higher order is simply a matter of
perturbative calculations. Yet, this approach relies on a heavy machinery and on
input which is external to the Higgs sector. We would like to consider cases
where input from the Higgs sector only (or almost only) would already improve on
the simple tree-level expression for the Higgs self-couplings. 

\par\noindent In principle, whatever the potential looked like at the classical
level, 
quantum corrections will generate contributions to all the parameters in the
general 
potential -- Eq.(\ref{genradpot}) -- (unless a symmetry protects certain parameters, but
we 
have seen that such symmetries are spontaneously broken by the Higgs v.e.v.'s
anyway). 
Therefore, while the partial-inversion of the general case
(Eqs.(\ref{gencond},\ref{qmcond}))
is still possible, little predictivity or practical use is to be expected from
such 
relations, because the number of undetermined parameters is high. To extract
meaningful 
information, beyond the leading order, from the Higgs spectrum, one would need
the 
corrected potential to retain a sufficiently simple form beyond the classical
order.

\par\noindent To be more specific, we consider a tree-level potential of the
form ($H$
representing any of the Higgs fields, $\mu^2$, a bilinear, $A$, a trilinear, and
$\l_i$, a quartic
coupling):
\begin{equation}
 {\cal V}_{\rm tree} = \mu^2 H^2 + A H^3 + \l H^4 
\end{equation}\noi 
We now include the radiative corrections, which shift the potential as:
\begin{equation}
 {\cal V}_{\rm eff} = (\mu^2 + \delta \mu^2) H^2 + (A + \delta A) H^3 + (\l +
\delta \l) H^4 + \delta\tilde\mu^2 H^2 + \delta\tilde A H^3 + \delta\tilde \l
H^4
\end{equation}\noi 
where $\delta \mu^2$, $\delta A$ and $\delta \l$ represent corrections to
parameters 
existing at tree-level, while $\delta\tilde\mu^2$, $\delta\tilde A$ and
$\delta\tilde \l$
denote new couplings which were forbidden by symmetries at tree-level and emerge
only at the radiative level. Neglecting numerical coefficients,
the corrected Higgs mass $m^2$ and the trilinear self-coupling $g$ will read
(schematically):
\begin{equation}
\label{masscoupcorr}\begin{cases}
 m^2 \simeq \mu^2 + \delta \mu^2 + \delta\tilde\mu^2 +(A+\delta A +
\delta\tilde A)\langle H \rangle+ (\l +\delta\l + \delta\tilde\l)\langle H
\rangle^2=m^2_{\rm tree}+O(\delta,\tilde\delta)\\
 g \simeq A + \delta A +\delta\tilde A + (\l +\delta\l+\delta\tilde\l)\langle H
\rangle=g_{\rm tree}+O(\delta,\tilde\delta)\end{cases}
\end{equation}\noi 
(with the short-hand notation $\delta/\tilde\delta$ for loop induced corrections
to parameters
present/absent at tree level.) We now assume that we have access to the mass
$m^2$, either from 
experimental data or from a spectrum generator. Using $g_{\rm tree}$ in the
computation of 
physical quantities (branching ratios, cross-sections) will result in an error
of order 
$O\left(\frac{\delta,\tilde\delta}{g}\right)$. If we use the expression for the 
corrected mass to inverse (partially) the relation between mass and tree level
parameters,
we obtain: $\delta = \delta_{m^2} +\mathcal{O}(\tilde\delta)$, where
$\delta_{m^2}$ symbolises 
the result of the inversion procedure. The trilinear couplings then provide:
$g_{m^2} = g +
\mathcal{O}(\tilde\delta)$, resulting in an error of
$\mathcal{O}\left(\frac{\tilde\delta}{g}
\right)$ at the level of cross sections/branching ratios. Claiming that the
inversion 
procedure carries any improvement with respect to a simple tree-level evaluation
holds
at the sole condition that radiative corrections $\delta$ to tree-level
parameters are more 
important, in magnitude, than the contributions $\tilde\delta$ to other
operators.
Otherwise, even if we identify the parameters subject to large contributions, it
is unlikely 
that the mass-matrices would suffice in determining both these parameters and
those intervening
at tree-level, unless we input some additional tree-level relations, as in the
case of the 
matching conditions in Eq.(\ref{NMSSMmatch}), (\ref{UMSSM}) or (\ref{nMSSM}).

\par\noindent This discussion shows that, to extract some benefits -- beyond the
leading order -- 
from the conditions relating masses to effective parameters, we need to identify
which terms 
are potentially subject to large radiative corrections. A simple criterion can
be invoked at the 
one-loop level: it is that of the leading logarithms. To identify those, we
simply resort to the 
Coleman-Weinberg \cite{colemanweinberg} one-loop effective potential and analyse
the outcome for the special case of
the SUSY-inspired models under scrutiny. This method has long been employed for
the computation 
of corrections to the Higgs masses, both in the MSSM \cite{1lpotMSSM} and in the
NMSSM 
\cite{nmssmrev-ellwanger,nmssmtools} (and references therein). In this approach,
the effective corrections to the scalar potential at a scale $\Lambda$ are
determined by the field-dependent tree-level mass matrices
$M_{\Phi}^2(S,H_d,H_u,\ldots)$ of the various fields $\Phi$ entering the 
spectrum, according to (in the $\overline{DR}$-scheme, but note that we shall be
interested in the logarithms only):
\begin{equation}\label{colW}
 \Delta {\cal V}_{\rm eff}^\Lambda(S,H_d,H_u,\ldots) = \frac{1}{64 \pi^2}
\sum_{\Phi}
C_{\Phi} M^4_{\Phi}
\left[\ln\left(\frac{M_{\Phi}^2}{\Lambda^2}\right)-\frac{3}{2}\right]
\end{equation}\noi
Here $C_\Phi$, which counts the degrees of freedom, takes the values $1$ for
real scalar fields,
$2$ for complex ones, $-2$ for Majorana fermions, $-4$ for Dirac fermions and
$3$ for real
gauge-fields. Note that we are interested in the Higgs potential solely, so 
that we will retain dependence on $S,H_d,H_u$ only, within $M_{\Phi}^2$.
Moreover, we consider no EW-violating effects so that we will
not expand the doublet fields $H_d$, $H_u$ around their v.e.v.'s
(except within logarithms). Additionally, the $SU(2)_L$-symmetry can then be
invoked to retain only the neutral Higgs fields $S,H_d^0,H_u^0$ (the dependence
on the charged Higgs fields can then be restored afterwards in virtue of
$SU(2)_L$: only the $\lambda_3$ and $\lambda_4$ parameters cannot be 
disentangled in this fashion, but both parameters being present at tree-level in
the models we consider, this will be of little consequence for our analysis). We
then determine the contributions to the parameters of Eq.(\ref{genradpot}) by
letting the singlet take its v.e.v., $S=s+\tilde{S}$, then truncating
Eq.(\ref{colW}) to renormalizable terms, finally projecting on the couplings of
Eq.(\ref{genradpot}).

\par\noindent The results of our analysis of the large logarithms, in the cases
of the NMSSM and 
nMSSM, are provided in appendix \ref{apCW}. The situation of the NMSSM is quite
simple: leading 
logarithms favor $\mathbb{Z}_3$-conserving terms. We can thus claim, for this
model, that the
inversion procedure for the $\mathbb{Z}_3$-conserving potential, presented in
the previous subsection and Eqs.(\ref{lambdoub},\ref{lambnew}),
improves on the tree-level implementation of the couplings and actually includes
leading-logarithms automatically. Note that, as defined in
Eqs.(\ref{lambdoub},\ref{lambnew}), the effective $\mathbb{Z}_3$-conserving
parameters are directly determined in terms of physical quantities, meaning
that they do not depend on the renormalization scale $\Lambda$ : they are
simply the parameters of the effective $\mathbb{Z}_3$-conserving potential
associated with the physical Higgs spectrum. What we checked explicitly in the
Coleman-Weinberg approach (which depends on the renormalization scale
$\Lambda$) is that this constrained form of an effective potential was
legitimate at least up to leading logarithms. Beyond, the effect of the
$\mathbb{Z}_3$-violating terms (due to the truncation of the potential to 
operators of mass-dimension $\leq4$) cannot be neglected.
In the case of the nMSSM, however, potentially large logarithms affect
non-classical terms. 
In fact, all the sectors contribute to the $\mathbb{Z}_3$-conserving parameters
(including those 
vanishing at tree level in this model). Additionally, logarithms originating from the nMSSM
Higgs sector 
(the only sector which is sensitive to the breakdown of $\mathbb{Z}_3$ at
tree-level) also affect
$\mathbb{Z}_3$-violating terms. Inclusion of the leading higher-order effects
from the inversion procedure of 
subsection \ref{secLinv}, Eq.(\ref{invT}), thus seems dubious in this case. It
seems natural to ascribe this difference of 
behavior, between nMSSM and NMSSM, to the protection of the parameters by the
$\mathbb{Z}_3$-symmetry, 
which albeit spontaneously broken by the singlet v.e.v., continues to
favor $\mathbb{Z}_3$-conserving terms within the NMSSM. We should thus expect a
similar property, whatever the $\mathbb{Z}_3$-symmetric model is 
(SUSY or not), and, beyond the $\mathbb{Z}_3$-symmetry, in any model retaining a
symmetry (or approximate symmetry) at low-energy, {e.g.} $PQ$ or $PQ'$.

\section{Phenomenological consequences for the NMSSM}
We now explain how, with the formalism derived above, we can
improve the computation of some observables in the NMSSM.
As we have already highlighted, there is one practical case where the Higgs
spectrum is fully available: it is 
that of the spectrum generators. The Higgs masses are often, in such a case,
corrected, while couplings
are typically taken at tree-level. In the case of the NMSSM, we have shown that
leading quantum corrections could be absorbed within the tree-level parameters
of the Higgs potential. This allows us, at a very cheap cost, to improve the
accuracy of the Higgs self-couplings by reexpressing them in terms of the Higgs
masses and mixing angles provided by the spectrum generator. We refer to
appendix \ref{hhhcoup} for the explicit expressions of the couplings in terms of
the effective parameters. We implement this recipe both within
\texttt{NMSSMTools\_3.2.0} \cite{nmssmtools,nmhdecay} and within
\texttt{SloopS} \cite{boudjema05,baro08}, and investigate phenomenological
consequences. 

\subsection{Impact on Higgs-constraints within \texttt{NMSSMTools\_3.2.0}}
\texttt{NMSSMTools\_3.2.0} includes several phenomenological constraints on the
NMSSM parameter space,
originating {e.g.}\ from LEP \cite{aleph06}, TeVatron \cite{gutierrez10}, $B$- 
and $\Upsilon$-physics \cite{lowenNMSSM} as well as early
(now outdated but in the process of getting updated) LHC-data
\cite{toolsLHC}. The Higgs-sector evidently
plays a central part in the interplay of these experimental limits and we would
like to investigate whether our analysis could have meaningful consequences 
at this level.

\par\noi The basic routine \texttt{mhiggs.f} of the \texttt{NMSSMTools} Package
computes the corrections to the 
Higgs mass matrices, incorporating typically leading-logarithmic effects
(although leading two-loop contributions 
from the fermion sector are also implemented) at a scale determined by the stops
and sbottoms, then rescaling 
the fields at the EW scale, finally adding pole corrections (this whole
procedure is more precisely described in the 
appendix C.3 of \cite{nmssmrev-ellwanger}). In this subsection, we shall be
relying on this routine for the 
calculation of the Higgs masses and mixing angles from the NMSSM parameter
input. Note however that 
\texttt{NMSSMTools} offers a second possibility, which consists in the
evaluation of the Higgs masses according
to \cite{slavichNMSSM}, including the full one-loop corrections as well as the
two loop ${\cal O}(\alpha_t \alpha_s 
+ \alpha_b \alpha_s)$ (with $\alpha_{t,b}=Y_{t,b}^2/4\pi$) effects in the
effective potential approach. This option 
will be used in the next subsections. Whatever the source
of the masses and mixing matrices however, we will treat the latter as input for
the physical Higgs matrices, allowing
us to compute the $\lambda_i$'s.

\par\noi The Higgs couplings implemented within \texttt{NMSSMTools\_3.2.0} are
actually not purely tree-level
couplings: possibly large radiative corrections from the quarks of the third
generation are included as well,
as explained in the last paragraph of the appendix A.2 of
\cite{nmssmrev-ellwanger}. One can check that these
corrections arise from (s)fermion contributions to $\lambda_{1,2}$ (as one can
recover considering our results for
the Coleman-Weinberg analysis in appendix \ref{apCW}): such effects are thus in
principle automatically incorporated
within our procedure (given that the corresponding contributions to the Higgs
mass are also implemented within 
\texttt{NMSSMTools}). The choice of the singlet v.e.v.\ $s$
deserves an additional comment, since it cannot be extracted from the masses.
After comparing with a few variants leading to minor deviations (a few percent)
at the numerical level, we settled for the simple definition $s=\mu_{\rm
eff}/\l$, with $\mu_{\rm eff}$ and $\l$ the parameters inputed in
\texttt{NMSSMTools}. Note that this choice is coherent with the recurrent use
of $\mu_{\rm eff}/\l$ as the singlet v.e.v.\ within the routines of
\texttt{NMSSMTools}.

\par\noindent To perform the comparison, we simply implement our corrected
$\l_i$'s (see Eqs.(\ref{lambdoub},\ref{lambnew})) 
and the ensuing triple Higgs couplings
(Eq.(\ref{hHHmass},\ref{haamass},\ref{hhhmass})) within the routine
\texttt{decay.f} of \texttt{NMSSMTools}, computing the Higgs decays. A flag
enables 
us to choose between the original setup of \texttt{NMSSMTools} and our modified
version, which we denote in the 
following as \texttt{NMSSMTools*}.
\begin{figure}[h]
 \begin{center}
\includegraphics[width=8.2cm]{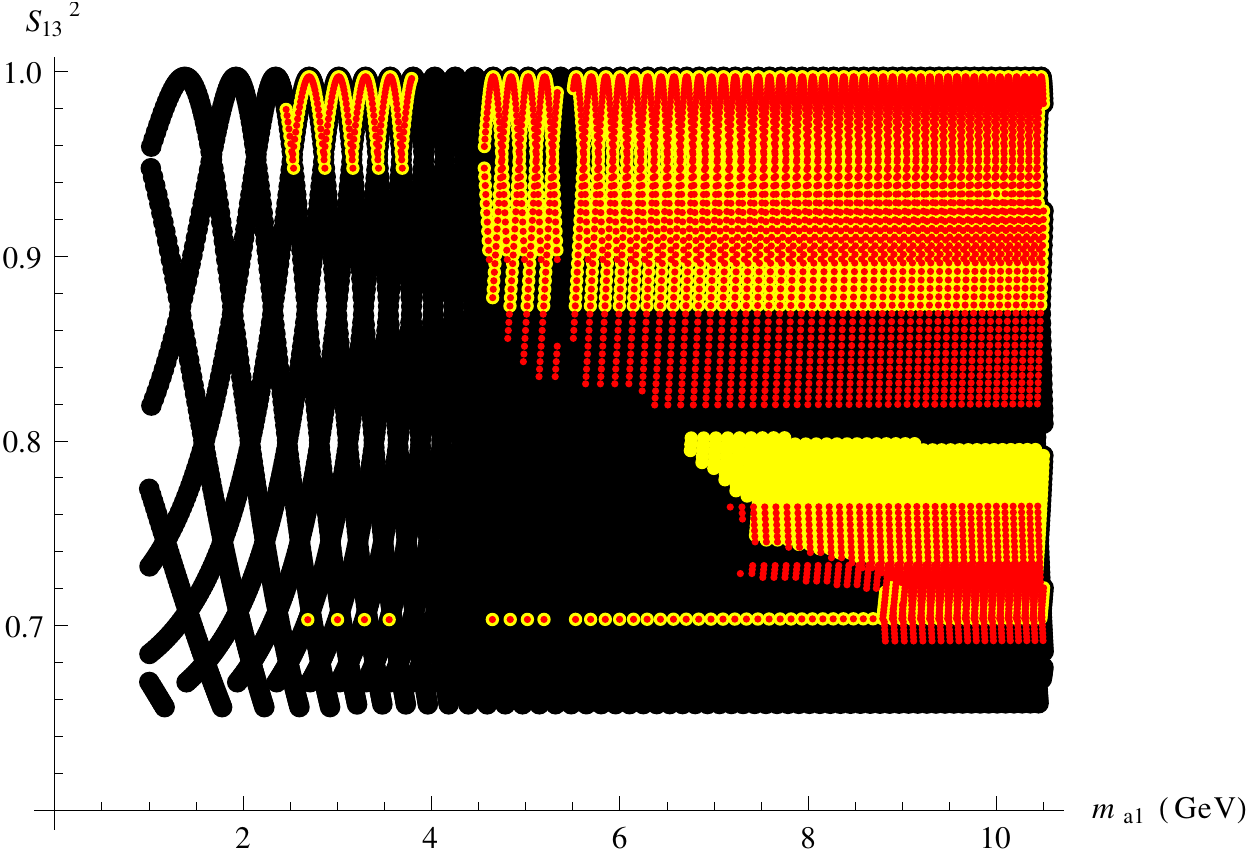}\hspace{0.5cm}
\includegraphics[width=8.2cm]{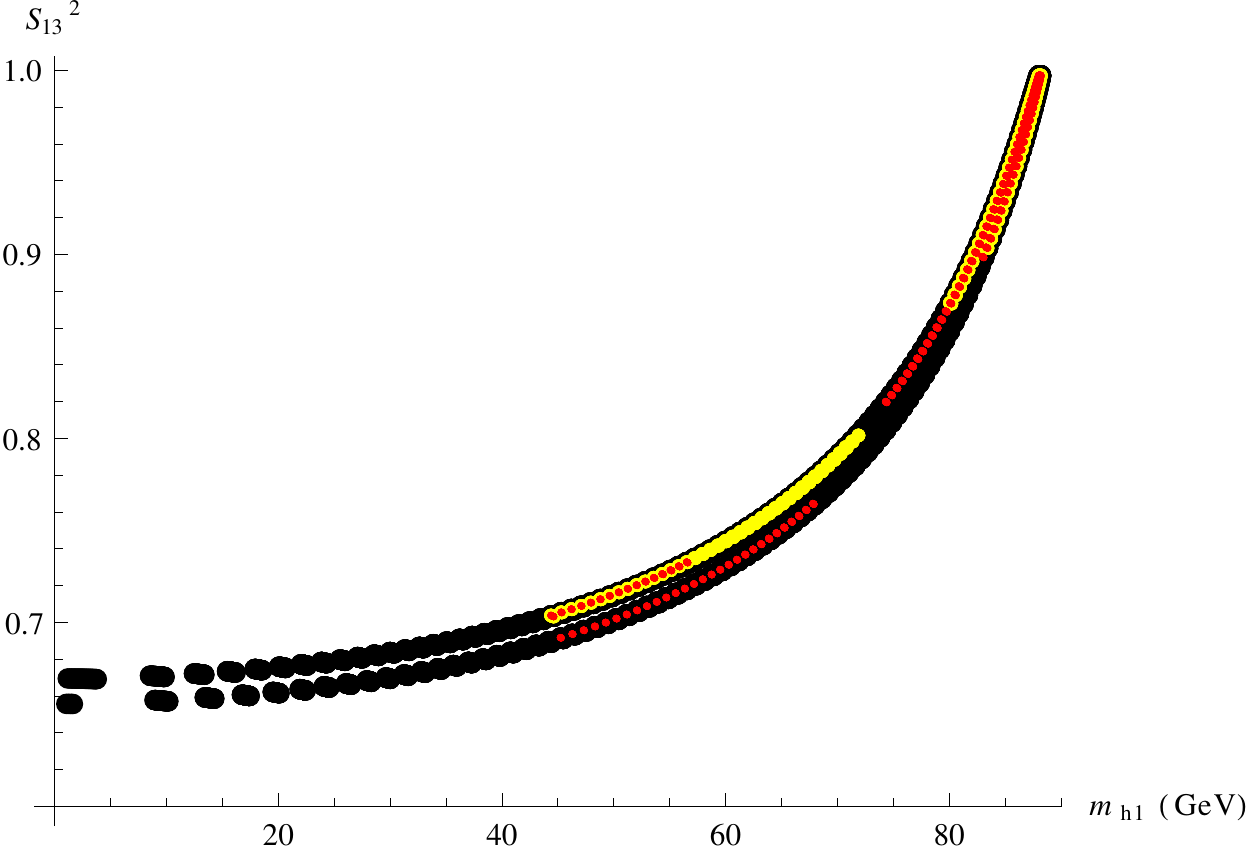}
 \end{center}
\caption{Constraints in the planes defined by $(m_{a^0_1},S_{13}^2)$ (left-hand
side) and $(m_{h_1^0},S_{13}^2)$ 
(right-hand side) for $\mu_{\mbox{\tiny eff}}=300$~GeV. Black dots correspond to
the points on which we perform the 
scan (without collider constraints); yellow dots are allowed by
\texttt{NMSSMTools\_3.2.0} while red dots signal
points allowed with self-Higgs couplings defined as in our procedure
\texttt{NMSSMTools*}.}
\label{S13vsmh1}
\end{figure}

\par\noindent Admittedly, the modification is essentially a fringe effect and
one needs to go to a region of the 
parameter space where the Higgs self-couplings intervene very finely to discover
substantial deviation between 
the two approaches. We thus consider a specific region in the NMSSM parameter
space, characterized by a light 
CP-even Higgs with mass typically under $100$~GeV, sizeable singlet-doublet
mixing $S_{13}^2\sim0-100\%$ and a 
light CP-odd Higgs with mass $<10.5$~GeV, allowing for $h^0_1\to2a^0_1$ decays.
Such a scenario is possible {e.g.}\ in 
an approximate Peccei-Quinn limit $\frac{\kappa}{\lambda}\ll1$, with the
parameters (\texttt{NMSSMTools} input: refer 
to \cite{nmssmtools,nmhdecay}; we use the index `sferm' to denote any of the
sfermions): 
$\tan\beta=5$, $\lambda=0.5$, $\kappa=0.05$,
$6M_1=3M_2=M_3=1.2~\mbox{TeV}=m_{\rm sferm.}=-A_{\rm sferm.}$, 
$\mu_{\mbox{\tiny eff}}\in[100;900]$~GeV, $|A_{\kappa}|<30$~GeV and
$M_A\in[0;4]$~TeV. Incidentally for those points, 
the doublet-like state $h_i^0$ ($i=1$ or $2$, depending on the specific point)
reaches a mass of $\sim125$~GeV in the 
limit of singlet-doublet decoupling  $S_{i3}^2\sim0$. Note however that we did
not specifically attempt to preserve 
this feature of a Higgs state at $125$~GeV (so that for significant mixing, we
may have typically $m_{h_1^0}^2
\sim90$~GeV while $m_{h_2^0}^2\sim150$~GeV): we simply mean to show that our
procedure is liable to affect the 
output of \texttt{NMSSMTools}. Note finally that, for simplicity, we discard
constraints from $(g-2)_{\mu}$, which 
may be taken care of separately, by tuning the slepton sector. All other
collider constraints -- from LEP, 
$B/\Upsilon$-physics, TeVatron or early LHC data -- within
\texttt{NMSSMTools\_3.2.0} are kept.
\begin{figure}[h]
 \begin{center}
\includegraphics[width=8.2cm]{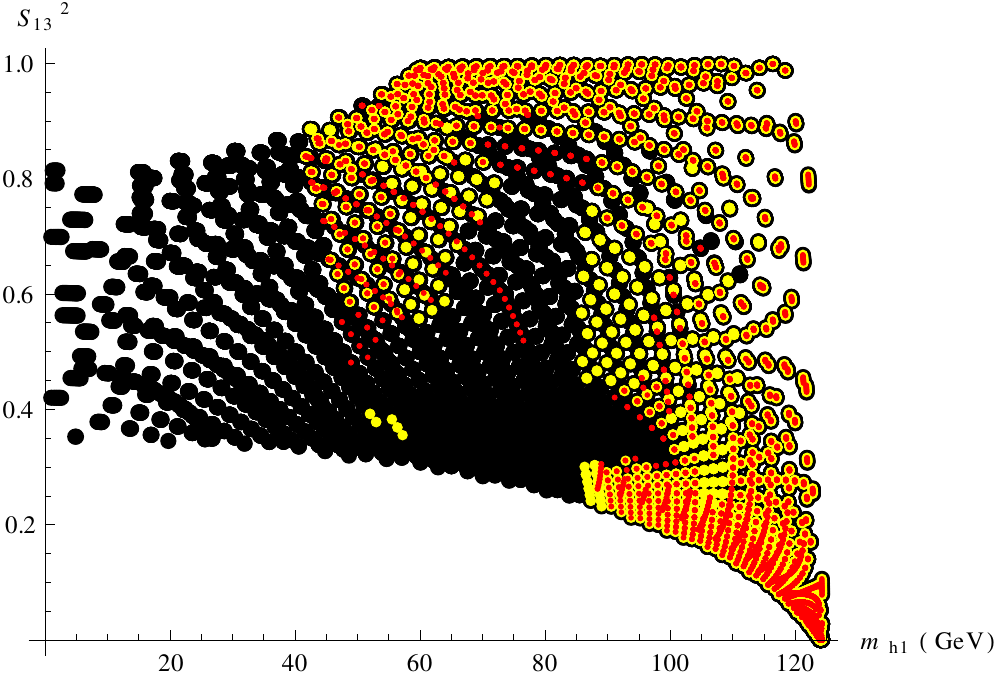}\hspace{0.5cm}
\includegraphics[width=8.2cm]{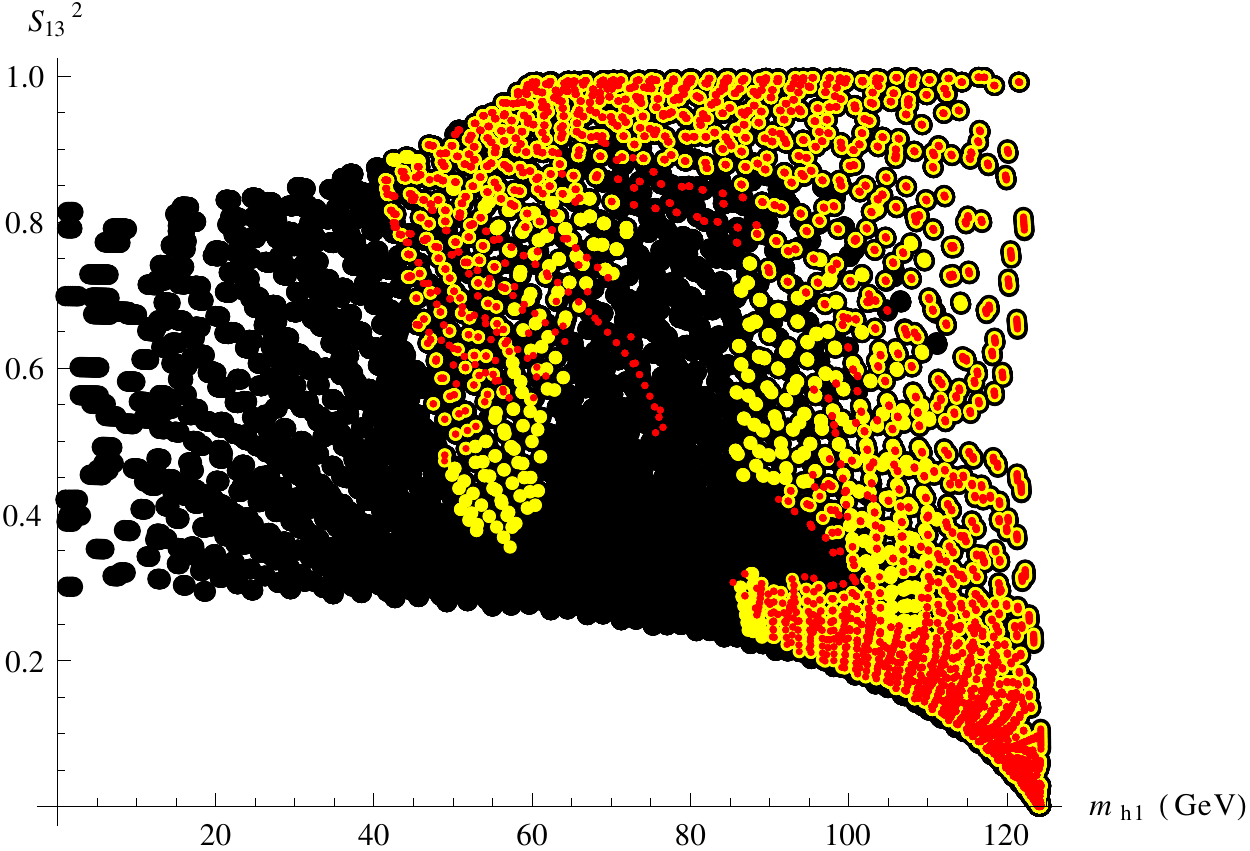}
 \end{center}
\caption{Constraints in the plane defined by $(m_{h_1^0},S_{13}^2)$ for
$\mu_{\mbox{\tiny eff}}\in[100;900]$~GeV. 
On the left-hand side, the scan with the yellow dots uses tree-level Higgs
couplings, while the corresponding one 
on the right-end side is obtained with the couplings implemented in
\texttt{NMSSMTools}, adding fermion 
corrections. The color code is otherwise similar to that of the previous
figure.}
\label{S13vsmS1}
\end{figure}

\noindent We first specialize to the case $\mu_{\mbox{\tiny eff}}=300$~GeV: the
lightest CP-even Higgs is then 
dominantly singlet ($S_{13}^2\sim70-100\%$). We display our results for this
scenario in Fig.\ref{S13vsmh1}. 
The black dots represent points on which we scanned (with no collider
constraints applied; note that their distribution 
is an artifact of the scan and should not be paid particular attention), the
yellow dots, the output of \texttt{NMSSMTools} with the 
original Higgs couplings. Our results (\texttt{NMSSMTools*}), with corrected
Higgs couplings, are depicted by the red dots. Since we scan over 
two variables, the output is two-dimensional in the
$(m_{a^0_1},S_{13}^2)$-plane. In the plane $(m_{h_1^0},S_{13}^2)$, the 
constraint $m_{a^0_1}<10.5~\mbox{GeV}\ll m_{h_1^0}$ reduces the apparent
dimensionality to $1$. To investigate the whole 
$(m_{h_1^0},S_{13}^2)$-plane, one may additionally scan on the parameter
$\mu_{\mbox{\tiny eff}}$, which we show on 
Fig.\ref{S13vsmS1}. There, however, the plot on the left-hand side corresponds
to the case of purely tree-level couplings 
(for the yellow dots), which we obtained by removing the fermion-corrections
from the original couplings implemented 
in \texttt{NMSSMTools}. The obvious conclusion is that, although partially
compatible, the yellow and red dots do not 
exactly coincide, so that the details of the constraints are affected by our
procedure. Note that both points admitted by 
\texttt{NMSSMTools} while excluded by \texttt{NMSSMTools*} and points admitted
by \texttt{NMSSMTools*} while excluded by \texttt{NMSSMTools} are to be found.

\par\noindent We insist however on the fact that such displacement effects 
in the acceptable points of the parameter space are noticeable only because we considered a region
where phenomenological constraints on the Higgs spectrum and decays are particularly severe, due to
the presence of very light Higgs-states which need to be sufficiently `invisible' to escape experimental
limits. A slight perturbation of the $\lambda_i$'s is then liable to result into insufficient $h_i^0\to2a_1^0$
(`invisible') branching ratio, excessive decays into {e.g.}\ SM-fermions and/or excessive $a_1^0$-signals
in $B$-physics: in such cases, an increased `invisibility' of the light Higgs states, that is increased 
singlet components, is required. If, on the contrary, the perturbation of the $\lambda_i$'s stabilizes the
invisibility of the light Higgs states, additional parameter space becomes
available.Given the sensitivity
of such regions to perturbations and the complex interplay of constraints at
stake there, it is very difficult to predict to which extent the allowed
parameter space would be shifted or not. In any case,  a detailed new analysis
on the NMSSM parameter space is beyond the scope of this work. Our procedure
simply ensures the consistency of
the calculation at leading-log order,
the couplings being adequately related to the spectrum.

\subsection{Implementation in \texttt{SloopS}}
In \texttt{SloopS}, the complete spectrum and set of vertices are generated at
tree-level from the NMSSM -- SUSY 
and soft -- parameters through the \texttt{LanHEP} package\cite{lanhep}. There,
$g$, $g'$, $v_u$, $v_d$ are determined by
the physical input $M_Z$, $M_W$, $v$ and $\tan \b$. 
Then the radiative part of the Higgs potential needs to be implemented: the
tree-level Higgs parameters 
$\l_i^0$, given in Eq.(\ref{NMSSMmatch}), are thus shifted as $\l_i = \l_i^0 +
\Delta \l_i$, $\Delta \l_i$ defining the radiative 
corrections to the parameters $\l_i$ of Eq.(\ref{Z3pot}). Yet, the corrected
Higgs masses are not computed 
within \texttt{SloopS}, but imported from \texttt{NMSSMTools} through the
\texttt{SLHA} interface. Applying 
the inversion procedure (Eqs.(\ref{lambdoub}.\ref{lambnew})), we obtain the
$\mathbb{Z}_3$-invariant $\Delta \l_i$'s
from the inputed masses, diagonalizing angles and tree level $\l_i^0$'s.
From now on we will call this procedure the `effective physical
potential approach' and refer to it through the acronym `PhA'. The
complete set of Feynman rules is then derived automatically in the
\texttt{FormCalc}\cite{formcalc} conventions, the latter performing the
calculation of the decay width.

\par\noi A powerful feature of \texttt{SloopS} is the ability to check gauge 
invariance of results through a generalized non-linear gauge fixing, which was 
adapted to the NMSSM \cite{chalons11}. The gauge-fixing Lagrangian can be
written in a general form
\begin{equation}
\label{gaugefixing}
{\mathcal L}_{GF} = -\frac{1}{\xi_W} F^+ F^- - \frac{1}{2  {{ \xi_Z}}  }|
F^Z|^2 - \frac{1}{2 {{ \xi_A}} } | F^A|^2
\end{equation}\noi 
where the non-linear functions of the fields $F$ are given by
\begin{eqnarray}
F^+ & = & \bigg(\partial_\mu - ie {\tilde{\alpha}}  A_\mu - igc_W 
{\tilde{\beta}} Z_\mu\bigg) W^{\mu \, +}\non \\
&  &+ i{{ \xi_W}}  \frac{g}{2}\bigg(v +  {\tilde{\delta}}_1
H_1 + {\tilde{\delta}_2} H_2  +{\tilde{\delta}_3} H_3 + i({\tilde{\kappa}} G^0+
{\tilde{\rho}}_1 A_1 + {\tilde{\rho}}_2 A_2)\bigg)G^+\non\\
F^Z & = & \partial_\mu Z^\mu + {{ \xi_Z}}  \frac{g}{2c_W}\bigg(v +
    {\tilde{\epsilon}}_1  H_1 +  {\tilde{\epsilon}}_2 H_2 +  
{\tilde{\epsilon}}_3 H_3  \bigg)G^0 \\
F^A & = & \partial_\mu A^\mu \non
\end{eqnarray}\noi
The parameters $\tilde \alpha,\tilde \beta, \cdots, \tilde \epsilon_3$ are
generalized gauge-fixing parameters. We also set $\xi_{A,Z,W} =
1$ to keep a simple form for the gauge boson propagators. 
 \par\noi
The ghost Lagrangian $\mathcal{L}_{Gh}$ is established by requiring that the
full
effective Lagrangian is invariant under BRST transformations. This implies 
that the full quantum Lagrangian, with $\Lag_C$ the classical
Lagrangian,
\begin{equation}
 \Lag_Q  = \Lag_C + \Lag_{GF}+ \Lag_{Gh}
\end{equation}\noi 
be such that $\delta_{\mbox{\tiny BRS}} \Lag_Q = 0$ and hence
$\delta_{\mbox{\tiny BRS}}\Lag_{GF} = - \delta_{\mbox{\tiny BRS}}\Lag_{Gh}$
\cite{grace-1loop}. The BRST transformation for the gauge fields can be found
for example in \cite{grace-1loop}. The NMSSM specific transformations for the
scalar fields can be found in \cite{chalons11}. For the decay $h_i^0 \ra \g\g$
not all the parameters are relevant: only $\tilde\alpha$ and $\tilde\delta_i$
are.

\subsection{The decay \texorpdfstring{$H\ra \g\g$}{Htogaga}}
The diphoton decay is an interesting process to investigate due to the
relevance of this channel in the recent discovery at LHC and because the gauge
invariance is fully at play there. Indeed, in the SM, the
W-boson loop, together with the top-quark one (the latter being of course gauge
invariant, as the remaining contributions), dominate the decay. The calculation
of the diphoton rate in the non-linear gauge was originally performed in
\cite{gavela81} in order to simplify the calculation of the
Higgs decay into two photons in the SM. In short details, with the specific
choice $\tilde\alpha = -1$, the $W^\pm G^\mp \g$ coupling vanishes and the 
gauge-boson loop calculation is made easier. In our calculation we will refrain 
from adopting this choice in order to preserve the ability of checking the
cancellation of the unphysical gauge-dependent part in the gauge loops. We will
discuss this effect only at one-loop order, meaning that we consider only the LO
decay width: this will be sufficient for our purposes since we do not aim at a
more precise evaluation. Nevertheless it is worth reminding that the full
two-loop EW+QCD corrections for the SM-like Higgs decay into $\g\g$ are known
and under 2\% \cite{2lhgg-passarino} below the WW threshold. The full two-loop
SUSY corrections are as yet unknown and anyway, our procedure would not be
suited for such a calculation as the renormalization of the tree level Higgs
potential would be mandatory whereas our renormalization is effective and
explicitly breaks SUSY. We
only aim at showing how one may consistently use the radiatively corrected Higgs
masses for an improved LO calculation of this decay.
\par\noi
Once we trade the ``$\l_i$'' parameters for the masses, using
Eqs.(\ref{lambdoub},\ref{lambnew}), we can re-express the
Higgs self-couplings,  obtained from the restricted
$\mathbb{Z}_3$-invariant potential, in terms of them. From now
we will call ``$\l$-representation'' of the trilinear Higgs couplings their
expression in terms of the $\l_i$'s. Moreover, when the $\l_i$'s are
explicitly replaced by Higgs masses, v.e.v's and mixing angles, we will speak
of ``mass-representation''. As
far as the diphoton signal is concerned, the relevant couplings are those
connecting the CP-even Higgs with the charged ones but also with the charged
Goldstones. In the mass representation they are given by\footnote{For the sake
of clarity we reproduce the expression of the charged Higgs coupling. It
can also be found in appendix \ref{hhhcoup}, together with its general
expression in the $\l$-representation.}
\begin{eqnarray}
\label{chargedcoup}
 g_{h_i^0 H^+ H^-} &=&
\frac{1}{v\sqrt{2}}\left\{m^2_{h^0_i}\left(\frac{\sb^2}{\cb}S_{i1}+\frac{
\cb^2 } { \sb }S_{i2}\right)+2 m_{H^\pm}^2(\cb S_{i1}+\sb S_{i2}) \non \right.\\
& &\left. -\frac{2 m_{a^0_j}^2 P'_{j1} P'_{j2}S_{i3}}{3 \sin 2
\b}-m_{a^0_j}^2P^{'\,2}_{j1}\left(\frac{S_{i1}}{\cb}+\frac{S_{i2}}{\sb}
-\frac{4}{3}\frac { v }{s}S_{i3}\right)\right\}\\
 g_{h_i^0 G^+ G^-} &=& \frac{1}{v \sqrt{2}}\left\{m^2_{h^0_i}(\cb
S_{i1}+\sb S_{i2})+2 M_W^2 {\tilde{\delta}_i}\right\}\nonumber
\end{eqnarray}\noi 
where $m^2_{h^0_i}$, $m^2_{a^0_i}$ and $m^2_{H^\pm}$ are the \textit{physical}
masses and the mixing elements $S_{ij},P^{'}_{ij}$ form the matrices
diagonalizing the effective mass matrices
Eqs.(\ref{condCPodd},\ref{condCPeven}).
Note also that in the non-linear gauge the $h_i^0 G^+ G^-$ couplings
depend explicitly on the gauge through the parameters $\tilde\delta_i$. These
parameters also appear within the ghost sector in the couplings $h_i^0 \bar
c^\pm c^\pm$, where $\bar c^\pm, c^\pm$ are the charged ghost fields. The
non-linear gauge parameter $\tilde\alpha$ also appears in the course of the
calculation. It originates from couplings with physical fields like $W^\pm W^\mp
\g$, $W^\pm W^\mp \g\g$ and unphysical ones: $G^\pm W^\mp\g$, $\bar c^\pm c^\mp
\g$, $\bar c^\pm c^\mp \g\g$ (see for example \cite{grace-1loop}). The latter
quartic coupling emerges from the non-linear gauge condition only. All these
couplings arise purely from the ghost and Goldstone part of the gauge sector and
are not modified by the effective potential of the Higgs sector.
\par\noi 
For the numerical evaluation, as explained before, we obtain the Higgs masses
and the mixing elements from \texttt{NMSSMTools} and the values are fed into
\texttt{SloopS} through the \texttt{SLHA} interface. Here we use the second
possibility offered by \texttt{NMSSMTools}, to compute the Higgs masses
following\cite{slavichNMSSM}. There pole-mass corrections can be taken into
account or not. In both cases the mixing matrices are computed in the effective
potential approximation (\textit{i.e} at $p^2 = 0$ where $p$ is the external
momentum entering the Higgs self-energies).
\par\noi
 As an illustration of the gauge invariance of the parameter reconstruction
(Eqs.(\ref{lambdoub},\ref{lambnew})), we considered two benchmark points from
\cite{bmp-nmssm-125}, named NMP2 and NMP5 after the conventions in
\cite{bmp-nmssm-125}. Their respective Higgs sector parameters are recalled in
Table \ref{bmphgg}, together with the soft SUSY breaking masses of the stop
sector $M_{\tilde Q_L^3}$, $M_{\tilde t_R}$ and of the gluino sector $M_3$. All
remaining soft masses and trilinear parameters that are not given in Table
\ref{bmphgg} are set at 1 TeV.
\begin{table}[htbp]
 \begin{center}
  \begin{tabular}{|c|c|c|}
   \hline
   \hline
  Parameter & NMP2 & NMP5 \\
  \hline
  $\tan \b$ & 2 &3  \\
  $\l$ & 0.6 & 0.66\\
  $\k$ & 0.18 & 0.12\\
  $\mu_{\rm eff.}$ [GeV] & 200 & 200\\
  $A_\l$ [GeV] & 405 & 650\\
  $A_\k$ [GeV] & -10 & -10\\
  $M_{\tilde Q^3_L}$ [GeV] & 700 & 600\\
  $M_{\tilde t_R}$ [GeV] & 700 & 600\\
  $M_3$ [GeV] & 600 & 600\\
  \hline 
  \hline
  \end{tabular}
 \end{center}
\caption{Benchmark points taken from \cite{bmp-nmssm-125} for the diphoton
decay width.\label{bmphgg}}
\end{table}
The resulting Higgs spectrum is summarized in Table \ref{higgspec} and computed
within \texttt{NMSSMTools} according to the procedure \cite{slavichNMSSM}
with and without the pole-mass corrections.
\begin{table}[htbp]
 \begin{center}
  \begin{tabular}{|c|c|c|c|c|}
   \hline
   \hline 
   Mass  & \multicolumn{2}{c|}{NMP2}&  \multicolumn{2}{|c|}{NMP5} \\
   \cline{2-5}
  $\rm [GeV]$ & no pole & pole & no pole & pole \\
   \hline
   $m_{h_1^0}$ &129.4 &126.5&96.1 &95.6 \\
   $m_{h_2^0}$ &133.1 &132.4&128.9 &126.5\\
   $m_{h_3^0}$ &470.8 &464.5&  659.9&655.8\\
   $m_{a_1^0}$ & 116.4&115.7& 93.9 &93.2\\
   $m_{a_2^0}$ & 468.7&462.8& 660.1&656.5\\
   $m_{H^\pm}$ & 454.4&454.5& 644.8&644.9 \\
   \hline
   \hline
  \end{tabular}
 \end{center}
\caption{Higgs spectrum of the benchmark points
considered. For the NMP2 point the lightest CP-even Higgs is
SM-like and for NMP5 it is the second-to-lightest one.\label{higgspec}}
\end{table}\noi
\par\noi
In addition to the masses given in the \texttt{SLHA} output we also need the
mixing elements $S_{ij}$ and $P'_{ij}$ to obtain the couplings entering the
diphoton decay width. In the effective potential approach used in
\cite{slavichNMSSM}, that we henceforth denote as EPA, they are
obtained by diagonalizing the radiatively corrected Higgs mass matrices in the
$\drbar$-scheme anew. However the definition of the diagonalizing matrices is
ambiguous since the self energies entering the radiatively corrected mass
matrices depend on the external momentum. In \cite{slavichNMSSM} the rotations
matrices are defined as those that diagonalize the mass matrices at $p^2=0$.
Then, whether pole-mass corrections are taken into account or not leads to the
same diagonalization matrices $S$ and $P'$: This can be seen
formally as a missing higher-order correction, but, in this fashion, the
\textit{physical} Higgs mass matrix would not correspond to the $\drbar$ one in
the EPA. This inconvenience is circumvented in the physical
effective approach (PhA) as we \textit{force} the
radiatively corrected mass matrices to be the \textit{physical} ones by imposing
Eqs.(\ref{lambdoub},\ref{lambnew}).
\par\noi
 To reproduce the EPA method using pole-masses within \texttt{SloopS}, we take
the $\drbar$-masses for the reconstruction of the potential of Eq.(\ref{Z3pot}) but
the
pole masses for the kinematics of the process. Therefore the Higgs mass
appearing in the coupling of Eq.(\ref{chargedcoup}) is the $\drbar$-mass, which
differs from the energy at which the decay is evaluated. This mismatch in the
EPA with pole-masses will lead to a violation of gauge invariance within our
generalized non-linear gauge, as we will show numerically. On the other hand, in the
PhA, we reconstruct the potential of
Eq.(\ref{Z3pot}) (and consequently the Higgs mass matrices as well) directly from
the pole-masses, which are also still used in the kinematics. This procedure
will guarantee that gauge invariance is maintained because
Eqs.(\ref{lambdoub},\ref{lambnew}) are fulfilled. At the numerical level we
will vary the parameters $\tilde\alpha,\tilde\delta_i$ within \texttt{SloopS} to
exemplify the gauge-invariance of the calculation in the PhA method. The
results are presented in Table \ref{hggcomp}. They were cross-checked
with the standard \texttt{NMSSMTools} version and with \texttt{NMSSMTools*}.
The corresponding output of \texttt{NMSSMTools}, whether in the standard or
modified version, does not check internally any consistency requirement, such as
gauge-invariance, and simply uses an analytic, pre-computed expression for the
effective $h_i^0 \g\g$ couplings.
\begin{table}[htbp]
 \begin{center}
  \begin{tabular}{|c|c|c|}
\hline
\hline
    \multicolumn{3}{|c|}{NMP2}  \\
\hline
$\Gamma(h_1^0 \ra \g\g)$   & $\tilde\alpha=\tilde\delta_1 = 0$ &
$\tilde\alpha=\tilde\delta_1 = 10$ \\
\hline
\texttt{SloopS} (EPA)  &1.138108952362$.\,10^{-5}$
&4.490893854783$.\,10^{-5}$ \\
 \hline 
\texttt{SloopS} (PhA)  &1.125710969262$.\,10^{-5}$
&1.125710969261$.\,10^{-5}$ \\
  \hline
\texttt{NMSSMTools\_3.2.0} &\multicolumn{2}{c|}{1.12699441$.\,10^{-5}$}\\
  \hline
\texttt{NMSSMTools*} &\multicolumn{2}{c|}{1.12737737$.\,10^{-5}$}\\
 \hline
     \multicolumn{3}{|c|}{NMP5}  \\
\hline
$\Gamma(h_2^0 \ra \g\g)$   & $\tilde\alpha=\tilde\delta_2 = 0$ &
$\tilde\alpha=\tilde\delta_2 = 10$ \\
\hline
\texttt{SloopS} (EPA) &1.053756232511$.\,10^{-5}$
&3.628709516521$.\,10^{-5}$ \\
 \hline 
\texttt{SloopS}(PhA) &1.044860481657$.\,10^{-5}$
&1.044860481613$.\,10^{-5}$ \\
\hline
\texttt{NMSSMTools\_3.2.0} &
\multicolumn{2}{c|}{1.04342526$.\,10^{-5}$}\\
\hline
\texttt{NMSSMTools*} &
\multicolumn{2}{c|}{1.04361857$.\,10^{-5}$}\\
\hline
\hline
  \end{tabular}
 \end{center}\caption{Gauge invariance test for the
computation of $\Gamma(h_{1,2}^0 \ra \g\g)$ (in GeV) in the EPA and PhA
procedures. Only the PhA approach passes the gauge invariance
test within \texttt{SloopS}. There is no such test available with
\texttt{NMSSMTools}, whether it is the modified version or
not.\label{hggcomp}}
\end{table}\noi 
The excellent agreement among the results of \texttt{NMSSMTools}
and the computation of \texttt{SloopS} for vanishing non-linear gauge parameters
({i.e.}\ in a linear gauge) is therefore a welcomed feature. The sources for
possible discrepancies between \texttt{SloopS} and \texttt{NMSSMTools} lie in
the treatment of the sfermion sector: in \texttt{SloopS} it is treated purely
at tree-level whereas \texttt{NMSSMTools} includes several corrections to the
spectrum and couplings\cite{nmhdecay}. For the two benchmark points
investigated this difference is almost invisible because the sfermion sector is
essentially decoupled.
\par\noi
These illustrative
examples show however evidently that the EPA calculation is
not gauge-invariant. The origin of this breakdown can be traced back to
the observation that Eqs.(\ref{lambdoub},\ref{lambnew}) are not satisfied. In
more restricted gauges the gauge dependence would be seen at higher orders only.
As stated above, setting $\tilde\alpha = -1$ removes the $G^\pm W^\mp \g$
coupling and varying $\tilde\delta_{1,2}$ then gives gauge-invariant results in
both procedures: they differ only by a finite and gauge-independent piece. This
is due to the fact that the $\tilde\delta_{1,2}$ gauge-dependent parts only
appear proportionally to $M_W^2/v\sqrt{2}$. On the contrary, in the general
case, $\tilde\alpha$-dependent parts are
proportional to the Higgs mass $m_{h_i^0}$, originating from the
kinematics (i.e. the center of mass energy $\sqrt{s}$), and the $g_{h_i^0
G^+ G^-}$ coupling, see Eq.(\ref{chargedcoup}). Recall that in the EPA the mass
appearing in the coupling $g_{h_i^0 G^+ G^-}$ is \textit{not} equal to the pole-mass, 
which is used for the on-shell decay. The gauge-dependence of the EPA is
precisely caused by this mismatch between the ``kinematical'' mass and the mass
appearing in the coupling of Eq.(\ref{chargedcoup}) \textit{if} pole-mass
corrections are applied. In the opposite case, that of the PhA, the
``kinematical'' mass and the one appearing in the coupling are the same and the
gauge dependent part $\propto \tilde\alpha$ vanishes. We can render the
EPA approach gauge-invariant if no pole-mass corrections are applied 
(this would then be a `PhA with $\overline{DR}$-masses'). However,
in this last case some precision is lost since, looking back to
Table \ref{higgspec}, there is a 2-3 GeV mass difference between pole and
running masses. Within our
reconstruction, the charged Higgs contribution is modified through the $g_{h_i^0
H^+ H^-}$ coupling: we do not expect significant modifications with respect
to previous calculations, in the MSSM-limit of the NMSSM, since it is known that
the charged Higgs contribution to the diphoton decay width in the MSSM
remains small (see for example \cite{Djouadi_Susy_Higgs}). In the 
NMSSM with large $\l$ (a form of the so-called `$\l$-SUSY' models
\cite{barbieri06}, which typically leads to a Landau pole below the GUT scale)
and a relatively light charged Higgs mass, one could modify significantly
$g_{h_i^0, H^+ H^-}$, without requiring large doublet-singlet mixing. This was
explicitly shown in \cite{schmidt12}. 
\par\noi 
Note however that a serious issue would arise with the gauge-dependent
calculation of $h_i^0 \ra \g\g$ if one would choose to use it in order to
derive some fundamental parameters at the Lagrangian level, which should
preferably be determined from gauge-independent observables. As a final remark
concerning this section, beyond maintaining gauge invariance, the PhA is clearly
advantageous as it enables us to use the pole-masses easily in the calculation
of the decay width, in a consistent way, without resorting to the technical task
of computing the ``pole-corrected'' mixing elements $S_{ij}, P'_{ij}$.

\subsection{Comparison with \texttt{micrOMEGAs}}
A comparable approach, based on an effective potential approach, had been
carried out in \texttt{micrOMEGAs} \cite{micromegas,belanger-nmssm}, a code
computing the DM relic density $\Omegah$ in ({e.g.})\ the NMSSM. Since light
Higgs states can be present, annihilation channels into $h_1^0 h_1^0$, $h_1^0
a_1^0$, $a_1^0 a_1^0$ can contribute significantly to $\Omegah$ and such
channels are affected by radiative corrections in the Higgs sector. The
effective scalar potential was implemented in this code as (see
\cite{belanger-nmssm}, where a slightly different version was proposed),
\begin{eqnarray}\label{micropot}
 {\cal V}_{\rm rad} & = & 
\l_1^M |H_u|^2 + \l_2^M |H_d|^2 + \l_3^M |H_u|^2 |H_d|^2 + \l_4^M |H_u \cdot
H_d|^2 + \frac{\l_5^M}{2}\left[  (H_u \cdot H_d)^2 + h.c \right]\non \\
& + & \l_1^s |S|^2 |H_u|^2 + \l_2^s |S|^2 |H_d|^2 +
\left[\frac{\l_5^s}{2}S^2 (H_u \cdot H_d) +\l^s_p S^4 + h.c \right]+
\frac{\l^s_s}{2} |S|^4 
\end{eqnarray}\noi 
This potential is to be understood as a radiative potential, which means that
all the parameters are loop-induced. To make a connection with our conventions
we have
\begin{equation}
  \Delta\lambda_1 =\lambda_2^M\,, \Delta\lambda_2 =
\lambda_1^M\,,\Delta\lambda_3 = \lambda_3^M\,,\Delta\lambda_4 =
\lambda_4^M\,,\Delta\lambda_P^u = \lambda_1^s\,,\Delta\lambda_P^d =
\lambda_2^s\,,\Delta\kappa^2 =\lambda_s^s
\end{equation}\noi 
where the $\Delta \l_i$ are the loop-induced part of the $\l_i$ parameters
appearing in Eq.(\ref{Z3pot}) when we split them as $\l_i = \l_i^0 +
\Delta\l_i$.
The remaining parameters $\l^s_p$, $\lambda_5^M$, and $\lambda_5^s$ have no
equivalent in our
\textit{restricted} potential of Eq.(\ref{Z3pot}), but correspond, in our
conventions 
for the general potential (Eq.\ref{genradpot}), to
\begin{equation}
 \l_5^M = \l_5,\, \l^s_p = {\tilde\k}^2_S/4\,,\tilde\lambda^M_P =
\lambda_5^s/2
\end{equation}\noi
Conversely, our parameters $A_{ud}$, $A_S$ and $\lambda_P^M$ receive no
correction 
in the \texttt{micrOMEGAs} approach, which, obviously, does not rely on the
$\mathbb{Z}_3$ 
symmetry. As a consequence, while an inversion procedure is also possible with
the potential of 
Eq.(\ref{micropot}), the radiative corrections to the masses will be distributed in
a different way among the $\l_i$'s, leading to differences at the level of the
Higgs self-couplings. 
\par\noi
We remind here, that if one aims at improving on the 
tree-level couplings, as is obviously the purpose of a radiative potential, {i.e.}\ of
Eq.(\ref{micropot}), it becomes crucial to identify the $\lambda_i$'s that are subject to
large quantum corrections: that was our discussion in subsection \ref{secLinv}. An arbitrary
truncation of the potential, albeit allowing for an inversion in terms of the Higgs masses provided it is sufficiently 
simple, is not a receivable option because the accuracy contained within the couplings brings
no improvement with respect to the tree-level evaluation. Our study of the large logarithms within the
Coleman-Weinberg approach tends to convince us that our choice of a
$\mathbb{Z}_3$-invariant potential should be preferred, while the choice in
Eq.(\ref{micropot}) seems arbitrary. Possible reasons for this choice within \texttt{micrOMEGAs} could lie on the
facts that the loop-corrections to $\l_5$ are sizeable in the MSSM, and one
could have expected the same behavior in the NMSSM, while the
parameters $\l^s_p$ and $\lambda_5^s$ (or rather $\tilde\k_S^2$ and
$\tilde\l_P^M$ in our conventions) appear in the trilinear couplings $h_i^0
h_j^0 h_k^0$ and $h_i^0 a_j^0 a_k^0$ with a factor $s$ (questionably an enhancement factor in the MSSM limit): see
Eqs.(\ref{haalamb},\ref{hhhlamb}). However other parameters in the general
potential of Eq.(\ref{genradpot}) share this latter property and are still
arbitrarily absent. We will see in a numerical example that ensuing deviations
between our implementation and that in \texttt{micrOMEGAs} could be significant.

\par\noi 
 We considered a point in the NMSSM parameter space where the DM relic
density $\Omegah$ is in the correct experimental range (at the $2\sigma$ level:
$0.1 \lesssim \Omegah \lesssim 0.124$ \cite{PDG12}),
when computed with \texttt{micrOMEGAs\_2.4.1} and the Higgs radiative
potential of Eq.(\ref{micropot}). This specific point passes warnings
from \texttt{NMSSMTools\_3.2.0} as well, and features a SM-like CP-even Higgs
mass around 125 GeV. The \texttt{NMSSMTools} input for this point is given in
Table \ref{RDpoint}. 
\begin{table}[t]
 \begin{center}
  \begin{tabular}{|c|c||c|c|}
  \hline
  \hline
   Parameter & Value & Parameter & Value\\
  \hline
   $M_1$ [GeV] &84.49 & $\tb$ &2 \\
   $M_2$ [GeV] &359 & $\l$ & 0.63\\
   $M_3$ [GeV] & 1200& $\k$ &0.05\\
   $A_{f}$ [GeV]& -1500 & $A_\l$ [GeV] & 694\\ 
   $M_{\tilde l}$ [GeV] & 200 & $A_\k$ [GeV] & 0\\ 
   $M_{\tilde q}$ [GeV] & 600 & $\mu_{\rm eff.}$ [GeV]& 300   \\
  \hline 
  \hline
  \end{tabular}
 \end{center}
\caption{ {SUSY point for the comparison between
\texttt{micrOMEGAs, SloopS} and \texttt{NMSSMTools}. $M_{\tilde
l}$ and $M_{\tilde q}$ are common sleptons and squarks soft
masses.}\label{RDpoint}}
\end{table}\noi 
The main channel contributing to $\Omegah$ is $\neuto\neuto \ra a_1^0 a_1^0$
(at 72\%) and the rest of the contributions involve fermions
in the final state, dominated by the $b\bar b$ final state. The process
$\neuto\neuto \ra a_1^0 a_1^0$ is dominated by the s-channel exchange of the
SM-like Higgs $h_2^0$ close to its mass shell, as
can be seen on Table \ref{RDspec}, where the Higgs spectrum,
the lightest neutralino mass $m_{\neuto}$ and the resulting relic density are
provided.
\begin{table}[htbp]
 \begin{center}
  \begin{tabular}{|c|c|}
\hline 
\hline
 \multicolumn{2}{|c|}{Spectrum} \\
\hline 
$m_{\neuto}$ [GeV] & 63.2 \\
$m_{h_1^0}$ [GeV] & 110.9 \\
$m_{h_2^0}$ [GeV] & 126.4 \\
$m_{h_3^0}$ [GeV] & 727.8 \\
$m_{a_1^0}$ [GeV] & 59.7 \\
$m_{a_2^0}$ [GeV] & 732.4 \\
$m_{H^\pm}$ [GeV] & 721.7 \\
\hline
\multicolumn{2}{|c|}{Relic density}\\
\hline
$\Omegah$ & 0.103 \\
\hline
\hline 
\end{tabular}
 \end{center}
\caption{ {Resulting spectrum and $\Omegah$ from the data point presented
in Table \ref{RDpoint}.}\label{RDspec}}
\end{table}\noi 

\par\noi
The decay modes $h_{2,3}^0 \ra 2 a_1^0$ are kinematically open.
Let us compare these decay widths within the three codes:
\texttt{SloopS}, with our effective implementation
(Eq.(\ref{lambdoub},\ref{lambnew})), \texttt{micrOMEGAs} using
the radiative potential in Eq.(\ref{micropot}) and \texttt{NMSSMTools\_3.2.0},
where only the leading logarithms in top/bottom corrections are taken into
account. The output of \texttt{NMSSMTools*} is also considered. The results are
displayed in Table \ref{RDdecay}.
\begin{table}[htbp]
 \begin{center}
  \begin{tabular}{|c|c|c|}
 \hline
\hline
Decay [GeV] & $\Gamma(h_2^0 \ra a_1^0 a_1^0)$ & $\Gamma(h_3^0 \ra a_1^0
a_1^0)$ \\
\hline 
\texttt{SloopS} & $3.566\, 10^{-2}$& $1.900\,10^{-4}$\\
\texttt{micrOMEGAs\_2.4.1} & $2.960\,10^{-2}$&$4.665\,10^{-5}$ \\
\texttt{NMSSMTools\_3.2.0} & $2.730\,10^{-2}$& $1.233\,10^{-4}$\\
\texttt{NMSSMTools*\_3.2.0}& $3.566\, 10^{-2}$&$1.900\,10^{-4}$\\
\hline
\hline 
\end{tabular}
 \end{center}
\caption{ {Comparison of the decay widths $h_{2,3}^0\ra 2
a_1^0$.}\label{RDdecay}}
\end{table}\noi 
We observe significant discrepancies between
\texttt{SloopS}/\texttt{NMSSMTools*\_3.2.0} (which are in remarkable
agreement), on one side, \texttt{micrOMEGAS\_2.4.1/NMSSMTools\_3.2.0} (which
also show some disagreement between them), on the other. Our calculation
for the main channel $h_2^0\ra 2 a_1^0$ (that we denote henceforth as
$\Gamma^S(h_2^0 \ra a_1^0 a_1^0)$) is about a factor 1.2 larger than the
\texttt{micrOMEGAs} result (labeled as $\Gamma^M(h_2^0 \ra a_1^0 a_1^0)$).

\par\noi
Giving the modified prediction of $\Omegah$ within our procedure is
beyond the scope of this work, but we can nevertheless make a rough estimation
of this quantity. As the process $\neuto\neuto \ra a_1^0 a_1^0$ is dominated by
the $h_2^0$ resonance, and only the $h_2^0 a_1^0 a_1^0$ coupling is modified, we
can reasonably approximate,
\begin{equation}
\frac{ \sigma^S(\neuto\neuto \ra a_1^0 a_1^0)}{\sigma^M(\neuto\neuto \ra
a_1^0 a_1^0)} \sim \frac{\Gamma^S(h_2^0 \ra a_1^0 a_1^0)}{\Gamma^M(h_2^0 \ra
a_1^0 a_1^0)}\sim 1.2
\end{equation}\noi 
Denoting $\sigma^M_r$ as the contribution of the rest of the processes
to the cross-sections involved in $\Omega_\chi^M h^2$, the relic density
computed within \texttt{micrOMEGAs}, we can write the sum of all contributions
$\sigma_{\rm tot}^M(\neuto\neuto\ra X)$ as, $X$ standing for any relevant final
state,
\begin{equation}
\sigma_{\rm tot}^M(\neuto\neuto\ra X)=\sum_X \sigma^M(\neuto\neuto\ra X) =
\sigma^M(\neuto\neuto \ra a_1^0 a_1^0) + \sigma^M_r
\end{equation}\noi 
The \texttt{micrOMEGAs} calculation gives $\sigma^M(\neuto\neuto \ra
a_1^0 a_1^0)/\sigma^M_{\rm tot}(\neuto\neuto\ra X)= 72 \%$, as we already
mentioned, and the ratio of relic densities in both approaches is approximately
determined by $\Omega_\chi^M h^2 / \Omega_\chi^S h^2 \simeq \sigma^S_{\rm
tot}/\sigma^M_{\rm tot}$, with $\Omega_\chi^S
h^2$ the relic abundance in our calculation, and where
\begin{equation}
 \sigma^S_{\rm tot} = \sigma^S(\neuto\neuto \ra a_1^0 a_1^0) + \sigma^S_r
\end{equation}\noi 
Moreover we have $\sigma^S_r = \sigma^M_r$, since the remaining relevant
contributions are annihilations into light fermions and hence unaffected by
corrections in the Higgs sector. Thus we obtain the following estimate,
\begin{equation}
 \Omega^S_\chi h^2 \simeq  0.090
\end{equation}\noi 
A reduction of the relic density with respect to the \texttt{micrOMEGAs}
calculation was to be anticipated since in our computation the annihilation into
light pseudoscalars is enhanced, thus depleting the abundance
of relic neutralinos more efficiently. In turn, and contrarily to the
prediction of \texttt{micrOMEGAs}, this point would actually lie outside the
cosmologically
interesting region if one relies on our estimate. Of course the derived value
of $\Omegah$ also depends crucially on the precision of the evaluation of
$m_{h_2^0}$ (and $m_{\neuto}$) since the annihilation $\neuto\neuto \ra a_1^0
a_1^0$ occurs at the $h_2^0$ resonance. These considerations are of particular
significance when one considers that the
\texttt{PLANCK} satellite\cite{planck} should improve the
experimental determination of cosmological parameters\cite{planckforecast} soon.
For a discussion concerning the accuracies required from colliders to match the
precision of the relic density measurement, see
for example \cite{wmaplhclc-requirements}.

\section*{Conclusions}
This study of the Higgs potential with two Higgs
doublets and one gauge singlet has
put forward several points of interests
that we would like to summarize briefly here.

\par\noi The most general effective renormalizable Lagrangian of
the $2$-doublet$+1$-singlet setup, contains $28$ (plus
one superfluous) parameters, far beyond the $11$
ones of the 2HDM, even 
after complex phases have been discarded. Therefore, if future experimental
measurements should point towards such a rich Higgs sector, a full reconstruction of its
potential through experimental data in the Higgs sector could succeed only after
an exhaustive measurement of the Higgs self-couplings, beyond that of the masses
and mixing angles: if the purpose for such a reconstruction is sound from the
point of view of model identification and precision tests, it is also probably
condemned to a very long delay, as far as the experimental phase is concerned.
This situation is eventually that of most models, albeit constrained, once considered at
the radiative level, for symmetries are spontaneously broken by the
Higgs v.e.v.'s and loop corrections end up contributing to all possible terms in
the potential. We emphasize, however, that a precise determination, in a general parametrical
form, of the potential at future (linear) colliders, shall help discriminate
among such models and constrain their parameters: in turn, the predictions of specific models for the parameters of the
Higgs potential should be known at the radiative level so as
to allow for comparison/precision tests.

\par\noi Requirements for additional symmetries, beyond the EW-invariance, or
matching conditions originating from more-elaborate
models may constrain the effective potential at the classical order. Provided
its form is simple enough, an identification
at leading order of the parameters of the underlying model is achievable from
the Higgs spectrum solely. Then, assuming
the remaining sectors of the model are sufficiently documented as well, a full
determination of the
effective potential
within the more-fundamental model is essentially a matter of perturbative
calculation. We lent particular attention to the
$\mathbb{Z}_3$-invariant and PQ-conserving potentials, which could both be
embedded within a SUSY extension of the SM,
respectively, the NMSSM or the UMSSM, the PQ'-conserving potential (R-symmetric limit of the NMSSM) or the
potential driven by an underlying nMSSM. A
reconstruction of the classical parameters was explicitly carried out, at
leading order, for those models.

\par\noi Further achievements seemed within reach in models ensuring a residual
symmetry at the EW scale. Our test-model
here was the NMSSM, and the study of the large logarithms within the
Coleman-Weinberg approach confirmed that the
leading-logarithmic effects would not spoil the $\mathbb{Z}_3$-symmetry,
extending the validity of our parameter
reconstruction in terms of the Higgs spectrum to this order. By contrast, 
in the nMSSM, where no residual symmetry
is present at low-energy, logarithms do not observe the classical form, spoiling
a reconstruction beyond LO.

\par\noi We finally considered a few phenomenological consequences of this
parameter reconstruction at the leading-logarithmic
order in the NMSSM. We based our discussion on the Higgs spectrum computed in
the public code \texttt{NMSSMTools} and 
implemented the reconstruction both within \texttt{NMSSMTools}, directly, and
within \texttt{SloopS}. The latter allowed
us to visit the diphoton
decay of the SM-like CP-even scalar again and
clarified the conditions for a gauge-independent implementation. Comparison
with the previous implementation of an effective Higgs potential within
\texttt{micrOMEGAs} was also carried out: different
choices in the radiative potential result in different Higgs-to-Higgs couplings
at the order of leading logarithms, as the
radiative effects encoded within the masses are distributed differently among
the parameters of Eq.(\ref{genradpot}); while the 
form in Eq.(\ref{micropot}) is seemingly arbitrary, our choice
(Eq.(\ref{Z3pot})) is justified by the analysis of the 
logarithms appearing in the Coleman-Weinberg approach, and should thence prove
{a priori} more reliable. As far as the 
phenomenology of the NMSSM is concerned, we found fine effects in
collider-constraints or the calculation of the DM relic 
density, appearing essentially for points of the parameter space which rely
heavily on Higgs-to-Higgs couplings, such as those 
entering the processes $h_1^0\to a_1^0a_1^0$ or $\neuto\neuto\to a_1^0a_1^0$,
mediated by a CP-even Higgs in the s-channel.

\par\noi Finally, let us mention that, although the state discovered at LHC is
in a favourable mass-range for singlet-extensions
of the MSSM, a long stage of experimental measurements and identifications of
additional Higgs states lies ahead of us, should 
the $2$-doublet$+1$-singlet setup be realized at all in Nature.

\section*{Acknowledgements}
This work has been supported by the BMBF grant 05H12VKF. G.C. would like to
thank the IKTP TU Dresden, where parts of this work
was realized, for the warm hospitality. The authors would like to
thank U.~Nierste and D.~Stöckinger, for useful discussion and comments on the
manuscript. The authors also thank M.~M.~M\"{u}hlleitner for clarifying
discussion about the benchmark points and G.~Bélanger for help with the
\texttt{micrOMEGAS}/\texttt{NMSSMTools} interface. U.~Ellwanger and F.~Boudjema
are also thanked for reviewing the manuscript.

\clearpage
\appendix
\input{apInversion}

\input{apCoupHHH}\clearpage 
\input{apColemanWeinberg}
\clearpage
\newpage
\bibliographystyle{unsrt}

\end{document}

%% file: apInversion.tex
\section{Parameter reconstruction for simple classical potentials}\label{aprec}
We provide here the results of the inversion procedure described in section \ref{secLinv} for a few classical potentials. Note 
that, for completeness, one should also replace the parameters within Eq.(\ref{mincondgen}) to fully determine the potential.

{\bf $\mathbb{Z}_3$-invariant potential ${\cal V}^S_{\mathbb{Z}_3}$:}\newline
The quartic doublet couplings are entirely determined by the Higgs mass-matrices:
\begin{equation}
 \label{lambdoub}\begin{cases}
 \lambda_1=\frac{1}{2v^2}\left[\frac{m^2_{h_i^0}S_{i1}^2}{\cos^2\beta}-
m^2_{a_i^0}P'^2_{i1}\tan^2\beta\right]\\
 \lambda_2=\frac{1}{2v^2}\left[\frac{m^2_{h_i^0}S_{i2}^2}{\sin^2\beta}-\frac{
m^2_{a_i^0}P'^2_{i1}}{\tan^2\beta}\right]\\
 \lambda_3=\frac{1}{2v^2}\left[2m_{H^{\pm}}^2+\frac{2m^2_{h_i^0}S_{i1}S_{i2}}{
\sin2\beta}- m^2_{a_i^0}P'^2_{i1}\right]\\
 \lambda_4=\frac{1}{v^2}\left[m^2_{a_i^0}P'^2_{i1}-m_{H^{\pm}}^2\right]
 \end{cases}
\end{equation}\noi
One degree of freedom remains, which can be chosen conveniently as the singlet v.e.v.\ $s$. One then obtains for the remaining 
parameters:
\begin{equation}
 \begin{cases}
 A_{ud}=\frac{1}{3}\left[\frac{\sin2\beta}{s}m^2_{a_i^0}P'^2_{i1}+\frac{1}{v}
m^2_{a_i^0}P'_{i1}P'_{i2}\right]\\
 \lambda_P^M=\frac{1}{3s}\left[\frac{\sin2\beta}{2s}m^2_{a_i^0}P'^2_{i1}-\frac{1
}{v}m^2_{a_i^0}P'_{i1}P'_{i2}\right]\\
 A_S=\frac{1}{3s}\left[\frac{v^2\sin^22\beta}{2s^2}m^2_{a_i^0}P'^2_{i1}-m^2_{
a_i^0}P'^2_{i2}-\frac{v\sin2\beta}{2s}m^2_{a_i^0}P'_{i1}P'_{i2}\right]\\
 \kappa^2=\frac{1}{4s^2}\left[m_{h_i^0}^2S_{i3}^2+\frac{1}{3}m^2_{a_i^0}P'^2_{i2
}-\frac{v^2\sin^22\beta}{3s^2}m^2_{a_i^0}P'^2_{i1}\right]\\
 \lambda_P^u=\frac{m_{h_i^0}^2S_{i2}S_{i3}}{2sv\sin\beta}+\frac{1}{3s\tan\beta}
\left[\frac{\sin2\beta}{s}m^2_{a_i^0}P'^2_{i1}-\frac{1}{2v}m^2_{a_i^0}P'_{i1}P'_
{i2}\right]\\
 \lambda_P^d=\frac{m_{h_i^0}^2S_{i1}S_{i3}}{2sv\cos\beta}+\frac{\tan\beta}{3s}
\left[\frac{\sin2\beta}{s}m^2_{a_i^0}P'^2_{i1}-\frac{1}{2v}m^2_{a_i^0}P'_{i1}P'_
{i2}\right]
 \end{cases}
\label{lambnew}\end{equation}\noi
In replacement of $s$, one may use any combination of these latter equations to define a new parameter. For instance,
\begin{multline}
 \varepsilon_{(a,b)}\equiv\frac{\lambda_{P}^M(a+b)}{a\lambda_P^u+b\lambda_P^d}\hspace{1cm}\Leftrightarrow\\
s=\frac{v\sin2\beta\,m^2_{a_i^0}P'^2_{i1}\left[1-2\varepsilon_{(a,b)}\left(\frac
{a}{\tan\beta}+b\tan\beta\right)\right]}{2m^2_{a_i^0}P'_{i1}P'_{i2}\left[1-\frac
{\varepsilon_{(a,b)}}{2}\left(\frac{a}{\tan\beta}+b\tan\beta\right)\right]
+3\varepsilon_{(a,b)}m_{h_i^0}^2S_{i3}\left(a\frac{S_{i1}}{\sin\beta}+b\frac{S_{
i2}}{\cos\beta}\right)}
\end{multline}\noi
$\varepsilon_{(a,b)}$ coincides with $\kappa/\lambda$ in the NMSSM at tree-level and may be regarded as a measurement of the 
breakdown of the Peccei-Quinn symmetry. Alternatively,
\begin{multline}
 \rho_{ud}\equiv\frac{\lambda_P^u}{\lambda_P^d}-1\hspace{1cm}\Leftrightarrow\\
s=\frac{2v\sin2\beta\,m_{a_i^0}^2P'^2_{i1}\left[1-(1+\rho_{ud})\tan^2\beta\right
]}{m_{a_i^0}^2P'_{i1}P'_{i2}\left[1-(1+\rho_{ud})\tan^2\beta\right]-3m_{h_i^0}
^2S_{i3}\left[\frac{S_{i1}}{\cos\beta}-(1+\rho_{ud})\tan^2\beta\frac{S_{i2}}{
\sin\beta}\right]}
\end{multline}
$\rho_{ud}$ vanishes at tree-level in the NMSSM and may represent another possibility.

{\bf Peccei-Quinn-invariant potential ${\cal V}^S_{PQ}$:}\newline
The system is fully determined by the mass matrices:
\begin{equation}
 \label{invPQ}\begin{cases}
 \lambda_1=\frac{1}{2v^2}\left[\frac{m^2_{h_i^0}S_{i1}^2}{\cos^2\beta}-
m^2_{a_2^0}P'^2_{21}\tan^2\beta\right]\\
 \lambda_2=\frac{1}{2v^2}\left[\frac{m^2_{h_i^0}S_{i2}^2}{\sin^2\beta}-\frac{
m^2_{a_2^0}P'^2_{21}}{\tan^2\beta}\right]\\
 \lambda_3=\frac{1}{2v^2}\left[2m_{H^{\pm}}^2+\frac{2m^2_{h_i^0}S_{i1}S_{i2}}{
\sin2\beta}- m^2_{a_2^0}P'^2_{21}\right]\\
 \lambda_4=\frac{1}{v^2}\left[m^2_{a_2^0}P'^2_{21}-m_{H^{\pm}}^2\right]\\
 A_{ud}=\frac{m^2_{a_2^0}P'_{21}P'_{22}}{v}\\
 s=\frac{v}{2}\sin{2\beta}\frac{P'_{21}}{P'_{22}}\\
 \kappa^2=\frac{(P'_{22}/P'_{21})^2}{v^2\sin^22\beta}\left[m_{h_i^0}^2S_{i3}
^2-m^2_{a_2^0}P'^2_{22}\right]\\
 \lambda_P^u=\frac{1}{2v^2\sin^2\beta}\frac{P'_{22}}{P'_{21}}\left[\frac{m_{
h_i^0}^2S_{i2}S_{i3}}{\cos{\beta}}+m^2_{a_2^0}P'_{21}P'_{22}\right]\\
 \lambda_P^d=\frac{1}{2v^2\cos^2\beta}\frac{P'_{22}}{P'_{21}}\left[\frac{m_{
h_i^0}^2S_{i1}S_{i3}}{\sin{\beta}}+m^2_{a_2^0}P'_{21}P'_{22}\right]
 \end{cases}
\end{equation}\noi

{\bf Peccei-Quinn'-invariant potential ${\cal V}^S_{PQ'}$:}\newline
The system is fully determined by the mass matrices:
\begin{equation}
 \label{invPQp}\begin{cases}
 \lambda_1=\frac{1}{2v^2}\left[\frac{m^2_{h_i^0}S_{i1}^2}{\cos^2\beta}-
m^2_{a_2^0}P'^2_{21}\tan^2\beta\right]\\
 \lambda_2=\frac{1}{2v^2}\left[\frac{m^2_{h_i^0}S_{i2}^2}{\sin^2\beta}-\frac{
m^2_{a_2^0}P'^2_{21}}{\tan^2\beta}\right]\\
 \lambda_3=\frac{1}{2v^2}\left[2m_{H^{\pm}}^2+\frac{2m^2_{h_i^0}S_{i1}S_{i2}}{
\sin2\beta}- m^2_{a_2^0}P'^2_{21}\right]\\
 \lambda_4=\frac{1}{v^2}\left[m^2_{a_2^0}P'^2_{21}-m_{H^{\pm}}^2\right]\\
 \lambda_P^M=\frac{m^2_{a_2^0}P'^2_{22}}{2v^2\sin{2\beta}}\\
 s=-v\sin{2\beta}\frac{P'_{21}}{P'_{22}}\\
 \kappa^2=\frac{m_{h_i^0}^2S_{i3}^2}{4v^2\sin^22\beta}\left(\frac{P'_{22}}{P'_{
21 }}\right)^2\\
 \lambda_P^u=\frac{1}{4v^2\sin^2\beta}\frac{P'_{22}}{P'_{21}}\left[m^2_{a_2^0}
P'_{21}P'_{22}-\frac{m_{h_i^0}^2S_{i2}S_{i3}}{\cos{\beta}}\right]\\
 \lambda_P^d=\frac{1}{4v^2\cos^2\beta}\frac{P'_{22}}{P'_{21}}\left[m^2_{a_2^0}
P'_{21}P'_{22}-\frac{m_{h_i^0}^2S_{i2}S_{i3}}{\sin{\beta}}\right]
 \end{cases}
\end{equation}\noi

{\bf nMSSM-inspired potential ${\cal V}^S_{T}$:}\newline
Although only twelve parameters are to be determined within the potential, application of the constraints
of Eq.(\ref{gencond}) leave one degree of freedom, due to the degenerescence of the CP-even and CP-odd 
singlet in this model: $m_{h_i^0}^2S_{i3}^2=m_{a_i^0}^2P'^2_{i2}$. We again choose $s$ to be this degree of 
freedom. The remaining parameters read:
\begin{equation}
 \label{invT}\begin{cases}
 \lambda_1=\frac{1}{2v^2}\left[\frac{m^2_{h_i^0}S_{i1}^2}{\cos^2\beta}-
m^2_{a_i^0}P'^2_{i1}\tan^2\beta\right]\\
 \lambda_2=\frac{1}{2v^2}\left[\frac{m^2_{h_i^0}S_{i2}^2}{\sin^2\beta}-\frac{
m^2_{a_i^0}P'^2_{i1}}{\tan^2\beta}\right]\\
 \lambda_3=\frac{1}{2v^2}\left[2m_{H^{\pm}}^2+\frac{2m^2_{h_i^0}S_{i1}S_{i2}}{
\sin2\beta}- m^2_{a_i^0}P'^2_{i1}\right]\\
 \lambda_4=\frac{1}{v^2}\left[m^2_{a_i^0}P'^2_{i1}-m_{H^{\pm}}^2\right]\\
 \lambda_T=\frac{v}{2}\sin{2\beta}m^2_{a_i^0}P'_{i1}P'_{i2}-sm^2_{a_i^0}P'^2_{i2
}\\
 m_{12}^2=m^2_{a_i^0}P'_{i1}P'_{i2}\frac{s}{v}-m^2_{a_i^0}P'^2_{i1}\frac{\sin{2\beta}}{2}\\
 A_{ud}=\frac{m^2_{a_i^0}P'_{i1}P'_{i2}}{v}\\
 \lambda_P^u=\frac{1}{2vs}\left[\frac{m_{h_i^0}^2S_{i2}S_{i3}}{\sin{\beta}}
+\frac{m^2_{a_i^0}P'_{i1}P'_{i2}}{\tan\beta}\right]\\
 \lambda_P^d=\frac{1}{2vs}\left[\frac{m_{h_i^0}^2S_{i1}S_{i3}}{\cos{\beta}}+m^2_
{a_i^0}P'_{i1}P'_{i2}\tan\beta\right]
 \end{cases}
\end{equation}\noi

%% file: apCoupHHH.tex
\section{Trilinear Higgs-to-Higgs couplings}\label{hhhcoup}
In this appendix we give the physical trilinear Higgs-to-Higgs couplings $h_i^0
H^+ H^-$, $h_i^0 a_j^0 a_k^0$ and $h_i^0 h_j^0 h_k^0$ in the $\l$-representation
obtained from the general potential Eq.(\ref{genradpot}) and in the mass
representation from the restricted $\mathbb{Z}_3$ potential Eq.(\ref{Z3pot})
only (as in the general potential the results are cumbersome). In the following
the matrix $P_{ij}$ is defined as the $3\times3$ diagonalization matrix which
rotates the gauge eigenstates $(a_d^0, a_u^0, a_s^0)$ directly to the physical
basis $(a_1^0, a_2^0, G^0)$ such that,
\par
\begin{equation}
  \left(\begin{array}{c}a_1\\a_2\\G^0\end{array}\right) = 
\begin{bmatrix}
 P'_{11} \sinb & P'_{11} \cosb & P'_{12} \\
 P'_{21} \sinb & P'_{21} \cosb & P'_{22} \\
 \cosb & -\sinb & 0 
\end{bmatrix}
\left(\begin{array}{c}
a_d^0\\a_u^0\\a_S^0\end{array}\right) =
 \left[\begin{array}{ccc}
 P_{11} & P_{12} & P_{13}\\
 P_{21} & P_{22} & P_{23}\\
 P_{31} & P_{32} & P_{33}\\
\end{array}\right]\left(\begin{array}{c} a_d^0 \\ a_u^0 \\
a_S^0\end{array}\right)
\end{equation}\noi
with $P'$ defined from Eq.(\ref{condCPodd}). To cast the couplings in a more
compact form we also define the following mixing elements combinations,
\begin{eqnarray}
  (\Pi^A)^{a,b,c}_{i,j,k} &=& S_{ia}\left[P_{jb}P_{kc}+P_{jc}P_{kb} \right]\\
(\Pi^S)^{a,b,c}_{i,j,k}&=&S_{ia} S_{jb}S_{kc}+S_{ia}S_{jc}S_{kb}+S_{ib}
S_{ja}S_{kc}+S_{ib}S_{jc}S_{ka}+S_{ic} S_{ja}S_{kb}+S_{ic}S_{jb}S_{ka}
\end{eqnarray}\noi 

\subsection{Trilinear couplings in the \texorpdfstring{$\l$}{l}-representation}
\paragraph*{CP-even Higgs to charged Higgses \texorpdfstring{$h_i^0 H^+
H^-$}{hi0H+H-} coupling}
\begin{eqnarray}
\label{hHHlamb}
 g_{h_i^0 H^+ H^-} &=& \frac{\l_1 v \sinb \sbt S_{i1}}{\sqrt{2}}
+\frac{\l_2 v \cosb \sbt S_{i2}}{\sqrt{2}}
+\sqrt{2}\l_3 v [ \cosb^3 S_{i1} + \sinb^3 S_{i2}]
-\frac{(\l_4 + \l_5)v \sbt}{\sqrt{2}}\left [\sinb S_{i1}+\cosb S_{i2}
\right]\non \\
&-&\frac{\l_6 v \cosb}{\sqrt{2}}\left[\sbt S_{i1}-(1-3 c_{2\b})S_{i2} \right]
+\frac{\l_7 v \sinb}{\sqrt{2}}\left[(1+3 c_{2\b})S_{i1}-\sbt
S_{i2}\right]\non \\
&+&\frac{\sbt}{\sqrt{2}}\left[A_{ud}+{\tilde A}_{ud}
+4(\l^M_P+\lt^M_P+\l_M)s)\right]S_{i3}\non \\
&+&\sqrt{2}\left[\left(A_{ds}+(\l^d_P+2\lt^d_P)s\right)\sinb^2
+\left(A_{us}+(\l^u_P+2\lt^u_P)s\right)\cosb^2\right] S_{i3}
\end{eqnarray}\noi
\paragraph*{CP-even Higgs to 2 CP-odd Higgs \texorpdfstring{$h_i^0 a_i^0
a_j^0$}{hi0aj0ak0}}
\begin{eqnarray}
\label{haalamb}
 g_{h^0_i a_j^0 a_k^0} &=& \frac{\l_1 v
\cosb}{\sqrt{2}}(\Pi^A)^{1,1,1}_{i,j,k} 
+\frac{\l_2 v \sinb}{\sqrt{2}}(\Pi^A)^{2,2,2}_{i,j,k}
+\frac{(\l_3+\l_4)v}{\sqrt{2}}\left[\cosb(\Pi^A)^{1,2,2}_{i,j,k}+\sinb(\Pi^A)^{2
,1,1}_{ i,j,k}\right] \non \\
&-&\frac{\l_7 v \sinb}{\sqrt{2}}(\Pi^A)^{1,1,1}_{i,j,k}
-\frac{\l_6 v \cosb}{\sqrt{2}}(\Pi^A)^{2,2,2}_{i,j,k}
-\frac{v(\l_5 \cosb -\l_6\sinb)}{\sqrt{2}}(\Pi^A)^{1,2,2}_{i,j,k}
-\frac{v(\l_5 \sinb - \l_7\cosb)}{\sqrt{2}}(\Pi^A)^{2,1,1}_{i,j,k}\non \\
&-&\sqrt{2}v\left[(\l_5 \cosb+\l_6\sinb)(\Pi^A)^{2,1,2}_{i,j,k}
(\l_5 \sinb+\l_7\cosb)(\Pi^A)^{1,2,1}_{i,j,k}\right] \non \\
&+&\frac{A_{ud}-2\l^M_P s}{\sqrt{2}}(\Pi^A)^{1,2,3}_{i,j,k}
+\frac{A_{ud}-2\l^M_P s}{\sqrt{2}}(\Pi^A)^{2,1,3}_{i,j,k}
+\frac{A_{ud}+2\l^M_P s}{\sqrt{2}}(\Pi^A)^{3,1,2}_{i,j,k} \non \\
&-&\frac{\tilde A_{ud}-2\lt^M_P s}{\sqrt{2}}(\Pi^A)^{1,2,3}_{i,j,k}
-\frac{\tilde A_{ud}-2\lt^M_P s}{\sqrt{2}}(\Pi^A)^{2,1,3}_{i,j,k}
+\frac{\tilde A_{ud}+2(\lt^M_P+\l_M) s}{\sqrt{2}}(\Pi^A)^{3,1,2}_{i,j,k} \\
&+&\frac{A_{ds}+2\lt^d_P s}{\sqrt{2}}(\Pi^A)^{3,1,1}_{i,j,k}
+\frac{A_{us}+2\lt^u_P s}{\sqrt{2}}(\Pi^A)^{3,2,2}_{i,j,k}
-\frac{3A_S-\tilde A_S-3(2\k^2-3{\tilde\k}_S^2)
s}{3\sqrt{2}}(\Pi^A)^{3,3,3}_{i,j,k}\non \\
&-&\sqrt{2}(\l_P^M -\lt^M_P)v \left[\sinb(\Pi^A)^{3,1,3}_{i,j,k}
+\cosb(\Pi^A)^{3,2,3}_{i,j,k} \right]
+\frac{\l^d_P s}{\sqrt{2}}(\Pi^A)^{3,1,1}_{i,j,k}
+\frac{\l^u_P s}{\sqrt{2}}(\Pi^A)^{3,2,2}_{i,j,k}\non \\
&+&\frac{[(\lt^M_P-\l_M)\sinb-2\lt^d_P\cosb]v}{\sqrt{2}}(\Pi^A)^{1,3,3}_{i,j,k}
+\frac{[(\lt^M_P-\l_M)\cosb-2\lt^u_P\sinb]v}{\sqrt{2}}(\Pi^A)^{2,3,3}_{i,j,k}
\non 
\end{eqnarray}\noi
\paragraph*{Triple CP-even Higgs coupling \texorpdfstring{$h_i^0 h_j^0
h_k^0$}{hi0hj0hk0}}
\begin{eqnarray}
\label{hhhlamb}
 g_{h_i^0 h_j^0 h_k^0} &=&
\frac{\l_1 v\cosb}{\sqrt{2}}(\Pi^S)^{1,1,1}_{i,j,k}+
\frac{\l_2 v \sinb}{\sqrt{2}}(\Pi^S)^{2,2,2}_{i,j,k}+
\frac{(\l_3+\l_4)v}{\sqrt{2}}\left[\cosb(\Pi^S)^{1,2,2}_{i,j,k}+\sinb(\Pi^S)^{2,
1,1 }_{ i,j,k}\right] \non \\
&-&\frac{\l_6v}{\sqrt{2}}\left[\cosb(\Pi^S)^{2,2,2}_{i,j,k}+3\sinb(\Pi^S)^{1,2,2
} _{ijk}\right]-\frac{\l_7v}{\sqrt{2}}\left[\sinb(\Pi^S)^{1,1,1}_{i,j,k}
+3\cosb(\Pi^S)^{2,1,1}_{ijk}\right]\non \\
&+&\frac{(\l^d_P \cosb - \l^M_P\sinb)v}{\sqrt{2}}(\Pi^S)^{1,3,3}_{i,j,k}
+\frac{(\l^u_P \sinb  - \l^M_P\cosb)v}{\sqrt{2}}(\Pi^S)^{2,3,3}_{i,j,k}
+\frac{\l^d_P s}{\sqrt{2}}(\Pi^S)^{3,1,1}_{i,j,k}
+\frac{\l^u_P s}{\sqrt{2}}(\Pi^S)^{3,2,2}_{i,j,k}\non \\
&+&\frac{(2 {\tilde\l}^d_P\cosb-(\l_M+{\tilde\l}^M_P)\sinb)v}{\sqrt{2}}
(\Pi^S)^{1,3,3}_{i,j,k}
+\frac{(2 {\tilde\l}^u_P\sinb-(\l_M+{\tilde\l}^M_P)\cosb)v}{\sqrt{2}}
(\Pi^S)^{2,3,3}_{i,j,k} \\
&+&\sqrt{2}\left({\tilde\l}^d_P s(\Pi^S)^{3,1,1}_{i,j,k}+
{\tilde\l}^u_P s(\Pi^S)^{3,2,2}_{i,j,k}\right)
-\frac{A_{ud}+2\l^M_P s}{\sqrt{2}}(\Pi^S)^{3,1,2}_{i,j,k}
+\frac{A_S+6\k^2 s}{3\sqrt{2}}(\Pi^S)^{3,3,3}_{i,j,k}\non \\
&-&\frac{{\tilde A}_{ud}+2(\l_M+{\tilde\l}^M_P)
s}{\sqrt{2}}(\Pi^S)^{3,1,2}_{i,j,k}
+\frac{{\tilde A}_S+3(\k_S^2+{\tilde\k}_S^2)
s}{3\sqrt{2}}(\Pi^S)^{3,3,3}_{i,j,k}\non 
\end{eqnarray}\noi
\subsection{Trilinear couplings in the mass-representation for the
\texorpdfstring{$\mathbb{Z}_3$}{Z3}-conserving potential}
To obtain the mass representation we trade the $\mathbb{Z}_3$-conserving
$\l_i$'s of the couplings in the $\l$-representation (see previous
subsection) against the masses, mixing angles and
v.e.v.'s (Eq.(\ref{lambdoub},\ref{lambnew})) and set the remaining ones to zero.
\paragraph*{CP-even Higgs to charged Higgses
\texorpdfstring{$h_i^0 H^+
H^-$}{hi0H+H-} coupling}
\begin{eqnarray}
\label{hHHmass}
 g_{h_i^0 H^+ H^-} &=&
\frac{1}{v\sqrt{2}}\left\{m^2_{h^0_i}\left(\frac{\sb^2}{\cb}S_{i1}+\frac{
\cb^2 } { \sb }S_{i2}\right)+2 m_{H^\pm}^2(\cb S_{i1}+\sb S_{i2}) \non \right.\\
& &\left. -\frac{2 m_{a^0_j}^2 P'_{j1} P'_{j2}S_{i3}}{3 \sin 2
\b}-m_{a^0_j}^2P^{'\,2}_{j1}\left(\frac{S_{i1}}{\cb}+\frac{S_{i2}}{\sb}
-\frac{4}{3}\frac { v }{s}S_{i3}\right)\right\}
\end{eqnarray}\noi
\paragraph*{CP-even Higgs to 2 CP-odd Higgs \texorpdfstring{$h_i^0 a_i^0
a_j^0$}{hi0aj0ak0}}
\begin{eqnarray}
\label{haamass}
g_{h^0_i a_j^0 a_k^0} =\sum_{l=1}^3 \frac{m^2_{h^0_l}}{2\sqrt{2}} 
&\bigg \{&\frac{S_{l1}^2}{v\cosb}(\Pi^A)^{1,1,1}_{i,j,k}
+\frac{S_{l2}^2}{v\sinb}(\Pi^A)^{2,2,2}_{i,j,k}
+\frac{S_{l3}^2}{s}(\Pi^A)^{3,3,3}_{i,j,k}
+S_{l1}S_{l2}\bigg[
\frac{(\Pi^A)^{1,2,2}_{i,j,k}}{v\sinb}+\frac{(\Pi^A)^{2,1,1}_{i,j,k}}{v\cosb}
\bigg ]\non \\
&+ &
S_{l3}\bigg[S_{l1}\left(\frac{(\Pi^A)^{3,1,1}_{i,j,k}}{v\cosb}
+\frac{(\Pi^A)^{1,3,3}_{i,j,k}}{s}\right)
+S_{l2}\left(\frac{(\Pi^A)^{3,2,2} _{i,j,k}}{v\sinb}
+ \frac{(\Pi^A)^{2,3,3}_{i,j,k}}{s}\right) \bigg ] \non \\
+\sum_{l=1}^{2}\frac{m^2_{a^0_l}}{\sqrt{2}}&\bigg \{&
P_{l1}P_{l2}\bigg [
\frac{(\Pi^A)^{1,2,2}_{i,j,k}-\tb(\Pi^A)^{1,1,1}_{i,j,k}}{2v \cosb}
+\frac{(\Pi^A)^{2,1,1}_{i,j,k}-\tb^{-1}(\Pi^A)^{2,2,2}_{i,j,k}}{2v \sinb} \non\\
&+&\frac{4}{3 s} \bigg (
\tb(\Pi^A)^{3,1,1}_{i,j,k}
+\tb^{-1}(\Pi^A)^{3,2,2}_{i,j,k}
+(\Pi^A)^{3,1,2}_{i,j,k}\bigg ) \non \\
&+&\frac{v}{3s^2}\bigg(\sinb (\Pi^A)^{1,3,3}_{i,j,k}
+\cosb(\Pi^A)^{2,3,3}_{i,j,k}
-4(\sinb(\Pi^A)^{3,1,3}_{i,j,k}+\cosb(\Pi^A)^{3,2,3}_{i,j,k})
-\frac{2 v\sbt}{s}(\Pi^A)^{3,3,3}_{i,j,k} \bigg ]\non \\
&+&P_{l1}P_{l3}\bigg [\frac{1}{2v}\left(
\frac{(\Pi^A)^{1,2,3}_{i,j,k}}{\sinb}
+\frac{(\Pi^A)^{2,1,3}_{i,j,k}}{\sinb}
-\frac{(\Pi^A)^{3,1,2}_{i,j,k}}{3\sinb}
-\frac{(\Pi^A)^{3,1,1}_{i,j,k}}{3\cosb}\right) \\
&+&\frac{1}{3 s}\bigg ((\Pi^A)^{3,1,3}_{i,j,k}+\tb^{-1}(\Pi^A)^{3,2,3}_{i,j,k}
-(\Pi^A)^{1,3,3}_{i,j,k}
-\frac{\tb^{-1}(\Pi^A)^{2,3,3}_{i,j,k}}{2}
+\frac{v\cosb(\Pi^A)^{3,3,3}_{i,j,k}}{2s}\bigg )\bigg ]\non \\
&+&P_{l2}P_{l3}\bigg [\frac{1}{2v}\left(
\frac{(\Pi^A)^{1,2,3}_{i,j,k}}{\cosb}
+\frac{(\Pi^A)^{2,1,3}_{i,j,k}}{\cosb}
-\frac{(\Pi^A)^{3,1,2}_{i,j,k}}{3\cosb}
-\frac{(\Pi^A)^{3,2,2}_{i,j,k}}{3\sinb}\right)
\non \\
&+&\frac{1}{3 s}\bigg ((\Pi^A)^{3,2,3}_{i,j,k}+t_\b(\Pi^A)^{3,1,3}_{i,j,k}
-(\Pi^A)^{2,3,3}_{i,j,k}
-\frac{t_\b(\Pi^A)^{2,3,3}_{i,j,k}}{2}+\frac{v\sinb(\Pi^A)^{3,3,3}_{i,j,k}}{2s}
\bigg )\bigg ]\non \\
&+&\frac{P^{2}_{l3}}{2s}(\Pi^A)^{3,3,3}_{i,j,k}\bigg \}\non 
\end{eqnarray}\noi 
\paragraph*{Triple CP-even Higgs coupling \texorpdfstring{$h_i^0 h_j^0
h_k^0$}{hi0hj0hk0}}
\begin{eqnarray}
\label{hhhmass}
g_{h_i^0 h_j^0 h_k^0} &=& 
\sum_{l=1}^3 \frac{m^2_{h^0_l}}{2\sqrt{2}}\bigg\{\frac{S_{l1}^2
(\Pi^S)^{1,1,1}_{i,j,k}}{v
\cosb} +\frac{S_{l2}^2(\Pi^S)^{2,2,2}_{i,j,k}}{v\sinb}
+S_{l1}S_{l_2}\left[\frac{(\Pi^S)^{2,1,1}_{i,j,k}}{v\cosb}
+\frac{(\Pi^S)^{1,2,2}_{i,j,k}}{v\sinb}\right] \non \\
& &+S_{l1}S_{l_3}\left[\frac{(\Pi^S)^{3,1,1}_{i,j,k}}{v\cosb}
+\frac{(\Pi^S)^{1,3,3}_{i,j,k}}{s}\right]
+S_{l2}S_{l_3}\left[\frac{(\Pi^S)^{3,2,2}_{i,j,k}}{v\sinb}
+\frac{(\Pi^S)^{2,3,3}_{i,j,k}}{s}\right]
+\frac{S_{l3}^2(\Pi^S)^{3,3,3}_{i,j,k}}{s}\bigg\}\non \\
&+&\sum_{l=1}^2 \frac{m^2_{a^0_l}}{2\sqrt{2}}\bigg\{
P_{l1}^2 \bigg[
\frac{\tb^{-1}(\Pi^S)^{2,1,1}_{i,j,k}-(\Pi^S)^{1,1,1}_{i,j,k}}{v\cosb}\non \\
& &\qquad+\frac{4}{3s}\bigg(
(\Pi^S)^{3,1,1}_{i,j,k}
+\tb^{-1}(\Pi^S)^{3,1,2}_{i,j,k}
+\frac{v\cosb^2}{s}\bigg(\frac{(\Pi^S)^{1,3,3}_{i,j,k}}{2\sinb}
-\frac{v (\Pi^S)^{3,3,3}_{i,j,k}}{6s}\bigg)\bigg)\bigg] \\
&+&
P_{l2}^2 \bigg[
\frac{\tb(\Pi^S)^{1,2,2}_{i,j,k}-(\Pi^S)^{2,2,2}_{i,j,k}}{v\sinb}\non \\
& &\qquad+\frac{4}{3s}\bigg(
(\Pi^S)^{3,2,2}_{i,j,k}
+\tb(\Pi^S)^{3,1,2}_{i,j,k}
+\frac{v\sinb^2}{s}\bigg(\frac{(\Pi^S)^{2,3,3}_{i,j,k}}{2\cosb}
-\frac{v(\Pi^S)^{3,3,3}_{i,j,k}}{6s}\bigg)\bigg)\bigg]\non \\
&-&
\frac{P_{l1}P_{l3}}{3}
\bigg[\frac{(\Pi^S)^{1,1,1}_{i,j,k}}{v \cosb}
+\frac{(\Pi^S)^{3,1,2}_{i,j,k}}{v\sinb}
-\frac{\tb^{-1}(\Pi^S)^{1,3,3}_{i,j,k}}{s}
+\frac{v\cosb}{6s^2}(\Pi^S)^{3,3,3}_{i,j,k}\bigg]\non \\
&-&
\frac{P_{l2}P_{l3}}{3}
\bigg[\frac{(\Pi^S)^{2,2,2}_{i,j,k}}{v \sinb}
+\frac{(\Pi^S)^{3,1,2}_{i,j,k}}{v\cosb}
-\frac{\tb(\Pi^S)^{2,3,3}_{i,j,k}}{s}
+\frac{v\sinb}{6s^2}(\Pi^S)^{3,3,3}_{i,j,k}\bigg]+ 
\frac{P^{2}_{l3}(\Pi^S)^{3,3,3}_{i,j,k}}{9 s} \bigg\}\non
\end{eqnarray}\noi

%% file: apColemanWeinberg.tex
\section{Coleman-Weinberg analysis of the Higgs potential in the NMSSM and the nMSSM}\label{apCW}
The two models under consideration essentially differ, at tree-level, by their Higgs sectors.
Additionally, one should require the limit $\kappa^2\to0$ in the nMSSM neutralino sector, 
with respect to that of the NMSSM.

{\bf SM-fermion contributions:}\newline\noi
In the base of Dirac-fermions $(u,d,\nu_e,e)$, the squared mass-matrix of SM-fermions in terms of neutral Higgs fields 
reads (we omit color and generation indices):
\begin{equation}
 {\cal M}_{f}^2(H_{u,d}^0)=\begin{bmatrix}
  Y_u^2\left|H_u^0\right|^2&0&0&0\\0&Y_d^2\left|H_d^0\right|^2&0&0\\0&0&0&0\\0&0&0&Y_e^2\left|H_d^0\right|^2
 \end{bmatrix}
\end{equation}\noi
leading to the potential:
{\begin{align}
 \delta{\cal V}_{\mbox{\small eff}}^{\Lambda,f}(H_u^0,H_d^0,S)&=-\frac{1}{16\pi^2}\sum_{f}{Y_f^4\left|H_f^0\right|^4\left[\ln\left(\frac{Y_f^2\left|H_f^0\right|^2}{\Lambda^2}\right)-\frac{3}{2}\right]}\nonumber\\
&\simeq-\frac{1}{16\pi^2}\sum_{f}Y_f^4\left|H_f^0\right|^4\ln\left(\frac{Y_f^2\mbox{\small \em v}_f^2}{\Lambda^2}\right)+\ldots
\end{align}}\noi
where we have kept only the leading, $SU(2)_L$-invariant, logarithmic terms. We deduce the corresponding contributions to the Higgs potential:
\begin{equation}\begin{cases}
\lambda_1^f\simeq-\frac{1}{8\pi^2}\sum_{f=d,e}Y_f^4\ln\left(\frac{m_f^2}{\Lambda^2}\right)\\
\lambda_2^f\simeq-\frac{1}{8\pi^2}\sum_{u}Y_u^4\ln\left(\frac{m_u^2}{\Lambda^2}\right)
\end{cases}\end{equation}

{\bf SM-Gauge-boson contributions:}\newline\noi
In the base of real vector fields $(\gamma^0,W^1,W2,Z^0)$, the squared mass-matrix of SM-Gauge-bosons in terms of neutral Higgs fields reads:
\begin{equation}
 {\cal M}_{G}^2(H_{u,d}^0)=\frac{1}{2}\begin{bmatrix}
  0&0&0&0\\0&g^2&0&0\\0&0&g^2&0\\0&0&0&g^2+g'^2
 \end{bmatrix}\left(\left|H_u^0\right|^2+\left|H_d^0\right|^2\right)
\end{equation}\noi
leading to the potential (note that in the $SU(2)_L$-conserving limit, these fields are massless):
{\small\begin{align}
 \delta{\cal V}_{\mbox{\small eff}}^{\Lambda,f}(H_u^0,H_d^0,S)&=\frac{3}{64\pi^2}\left\{\frac{g^4}{2}\left(\left|H_u^0\right|^2+\left|H_d^0\right|^2\right)^2\left[\ln\left(\frac{g^2\left(\left|H_u^0\right|^2+\left|H_d^0\right|^2\right)}{2\Lambda^2}\right)-\frac{3}{2}\right]\right.\nonumber\\
& \left.+\frac{(g^2+g'^2)^2}{4}\left(\left|H_u^0\right|^2+\left|H_d^0\right|^2\right)^2\left[\ln\left(\frac{(g^2+g'^2)\left(\left|H_u^0\right|^2+\left|H_d^0\right|^2\right)}{2\Lambda^2}\right)-\frac{3}{2}\right]\right\}\\
&\simeq\frac{3}{256\pi^2}\left[2g^4\ln\left(\frac{M_W^2}{\Lambda^2}\right)+(g^2+g'^2)^2\ln\left(\frac{M_Z^2}{\Lambda^2}\right)\right]\left(\left|H_u^0\right|^2+\left|H_d^0\right|^2\right)^2\ldots\nonumber
\end{align}}\noi
providing us with the couplings:
\begin{equation}\begin{cases}
\lambda_1^G\simeq\frac{3}{128\pi^2}\left[2g^4\ln\left(\frac{M_W^2}{\Lambda^2}\right)+(g^2+g'^2)^2\ln\left(\frac{M_Z^2}{\Lambda^2}\right)\right]\\
\lambda_2^G\simeq\frac{3}{128\pi^2}\left[2g^4\ln\left(\frac{M_W^2}{\Lambda^2}\right)+(g^2+g'^2)^2\ln\left(\frac{M_Z^2}{\Lambda^2}\right)\right]\\
\lambda_3^G+\lambda_4^G\simeq\frac{3}{128\pi^2}\left[2g^4\ln\left(\frac{M_W^2}{\Lambda^2}\right)+(g^2+g'^2)^2\ln\left(\frac{M_Z^2}{\Lambda^2}\right)\right]\\
\end{cases}\end{equation}

{\bf Sfermion contributions:}\newline\noi
The Sfermion squared mass-matrix, in the base $(\tilde{F}_L,\tilde{F}_R^{c\,*})$ for a flavour $f$, is given by:
{\small \begin{equation}
 {\cal M}^2_{\tilde{F}}(S,H_u^0,H_d^0)=\begin{bmatrix}
m^2_{\tilde{F}_L}+|Y_fH_f^0|^2+\frac{g'^2{\cal Y}_L^f-2g^2I_3^f}{4}(|H_u^0|^2-|H_d^0|^2)&Y_f(A_fH_f^0-\lambda S^*H_{\tilde{f}}^{0\,*})\\
Y_f(A_fH_f^{0\,*}-\lambda SH_{\tilde{f}}^{0})&m^2_{\tilde{F}_R}+|Y_fH_f^0|^2+\frac{g'^2{\cal Y}_R^f}{4}(|H_u^0|^2-|H_d^0|^2)
                        \end{bmatrix}
\end{equation}}\noi
Defining $T_{\tilde{F}}=({\cal M}^2_{\tilde{F}})_{11}+({\cal M}^2_{\tilde{F}})_{22}$ and $R_{\tilde{F}}^2=\left[({\cal M}^2_{\tilde{F}})_{11}-({\cal M}^2_{\tilde{F}})_{22}\right]^2+4|{\cal M}^2_{\tilde{F}}|^2_{12}$, 
we obtain the eigenvalues $m_{\tilde{F}_{\pm}}^2=\frac{1}{2}\left[T_{\tilde{F}}\pm\sqrt{R_{\tilde{F}}^2}\right]$ and the Higgs potential:
\begin{equation}
 \delta{\cal V}_{\mbox{\small eff}}^{\Lambda,\tilde{F}}(H_u^0,H_d^0,S)=\frac{1}{128\pi^2}\sum_{f}\left\{(T_{\tilde{F}}^2+R_{\tilde{F}}^2)\left[\ln\left(\frac{T_{\tilde{F}}^2-R_{\tilde{F}}^2}{4\Lambda^4}\right)-3\right]+2\,T_{\tilde{F}}\cdot R_{\tilde{F}}\ln\left(\frac{T_{\tilde{F}}+R_{\tilde{F}}}{T_{\tilde{F}}-R_{\tilde{F}}}\right)\right\}
\end{equation}\noi
One derives the couplings:
{\small\begin{equation}\begin{cases}
\left(m_{H_{u}}^2\right)^{\tilde{F}}=\frac{1}{16\pi^2}\sum_f\left\{\delta_{fu}Y_f^2\left[{\cal A}^{\tilde{F}}_L+{\cal A}^{\tilde{F}}_R+A_f^2{\cal B}^{\tilde{F}}_{L,R}\right]+\frac{g'^2{\cal Y}_L^f-2g^2I_3^f}{4}{\cal A}^{\tilde{F}}_L+\frac{g'^2{\cal Y}_R^f}{4}{\cal A}^{\tilde{F}}_R\right\}\\
\left(m_{H_{d}}^2\right)^{\tilde{F}}=\frac{1}{16\pi^2}\sum_f\left\{\delta_{f(d,e)}Y_f^2\left[{\cal A}^{\tilde{F}}_L+{\cal A}^{\tilde{F}}_R+A_f^2{\cal B}^{\tilde{F}}_{L,R}\right]+\frac{g'^2{\cal Y}_R^f}{4}{\cal A}^{\tilde{F}}_L+\frac{g'^2{\cal Y}_L^f-2g^2I_3^f}{4}{\cal A}^{\tilde{F}}_R\right\}\\
A_{ud}^{\tilde{F}}=\frac{1}{16\pi^2}\sum_f\lambda Y_f^2A_f{\cal B}^{\tilde{F}}_{L,R}\\
\lambda_1^{\tilde{F}}=\frac{1}{16\pi^2}\sum_f\left\{\delta_{f(d,e)}\left[\left(Y_f^2+\frac{g'^2{\cal Y}^f_R}{4}\right)^2\ln\left(\frac{m_{\tilde{F}_L}^2}{\Lambda^2}\right)+\left(Y_f^2+\frac{g'^2{\cal Y}^f_L-2g^2I_3^f}{4}\right)^2\ln\left(\frac{m_{\tilde{F}_R}^2}{\Lambda^2}\right)\right.\right.\\
\null\hspace{1.8cm}\left.\left.+Y_f^2A_f^2\left[\left(2Y_f^2+\frac{g'^2({\cal Y}^f_L+{\cal Y}_R^f)-2g^2I_3^f}{4}\right){\cal C}^{\tilde{F}}_{L,R}-\left(Y_f^2A_f^2-\frac{g'^2({\cal Y}^f_L-{\cal Y}_R^f)-2g^2I_3^f}{4}\Delta m^{2\,\tilde{F}}_{L,R}\right){\cal D}^{\tilde{F}}_{L,R}\right]\right]\right.\\
\null\hspace{1.7cm}\left.+\delta_{fu}\left[\left(\frac{g'^2{\cal Y}^f_R}{4}\right)^2\ln\left(\frac{m_{\tilde{F}_L}^2}{\Lambda^2}\right)+\left(\frac{g'^2{\cal Y}^f_L-2g^2I_3^f}{4}\right)^2\ln\left(\frac{m_{\tilde{F}_R}^2}{\Lambda^2}\right)\right.\right.\\
\null\hspace{2cm}\left.\left.+Y_f^2\lambda^2s^2\left[\frac{g'^2({\cal Y}^f_L+{\cal Y}_R^f)-2g^2I_3^f}{4}{\cal C}^{\tilde{F}}_{L,R}-\left(Y_f^2\lambda^2s^2-\frac{g'^2({\cal Y}^f_L-{\cal Y}_R^f)-2g^2I_3^f}{4}\Delta m^{2\,\tilde{F}}_{L,R}\right){\cal D}^{\tilde{F}}_{L,R}\right]\right]\right\}\\
\lambda_2^{\tilde{F}}=\frac{1}{16\pi^2}\sum_f\left\{\delta_{fu}\left[\left(Y_f^2+\frac{g'^2{\cal Y}^f_L-2g^2I_3^f}{4}\right)^2\ln\left(\frac{m_{\tilde{F}_L}^2}{\Lambda^2}\right)+\left(Y_f^2+\frac{g'^2{\cal Y}^f_R}{4}\right)^2\ln\left(\frac{m_{\tilde{F}_R}^2}{\Lambda^2}\right)\right.\right.\\
\null\hspace{1.8cm}\left.\left.+Y_f^2A_f^2\left[\left(2Y_f^2+\frac{g'^2({\cal Y}^f_L+{\cal Y}_R^f)-2g^2I_3^f}{4}\right){\cal C}^{\tilde{F}}_{L,R}-\left(Y_f^2A_f^2+\frac{g'^2({\cal Y}^f_L-{\cal Y}_R^f)-2g^2I_3^f}{4}\Delta m^{2\,\tilde{F}}_{L,R}\right){\cal D}^{\tilde{F}}_{L,R}\right]\right]\right.\\
\null\hspace{1.7cm}\left.+\delta_{f(d,e)}\left[\left(\frac{g'^2{\cal Y}^f_L-2g^2I_3^f}{4}\right)^2\ln\left(\frac{m_{\tilde{F}_L}^2}{\Lambda^2}\right)+\left(\frac{g'^2{\cal Y}^f_R}{4}\right)^2\ln\left(\frac{m_{\tilde{F}_R}^2}{\Lambda^2}\right)\right.\right.\\
\null\hspace{2cm}\left.\left.+Y_f^2\lambda^2s^2\left[\frac{g'^2({\cal Y}^f_L+{\cal Y}_R^f)-2g^2I_3^f}{4}{\cal C}^{\tilde{F}}_{L,R}-\left(Y_f^2\lambda^2s^2+\frac{g'^2({\cal Y}^f_L-{\cal Y}_R^f)-2g^2I_3^f}{4}\Delta m^{2\,\tilde{F}}_{L,R}\right){\cal D}^{\tilde{F}}_{L,R}\right]\right]\right\}\\
(\lambda_3+\lambda_4)^{\tilde{F}}=\frac{1}{16\pi^2}\sum_f\left\{\delta_{fu}\left[\frac{g'^2{\cal Y}^f_R}{4}\left(Y_f^2+\frac{g'^2{\cal Y}^f_L-2g^2I_3^f}{4}\right)\ln\left(\frac{m_{\tilde{F}_L}^2}{\Lambda^2}\right)+\frac{g'^2{\cal Y}^f_L-2g^2I_3^f}{4}\left(Y_f^2+\frac{g'^2{\cal Y}^f_R}{4}\right)\ln\left(\frac{m_{\tilde{F}_L}^2}{\Lambda^2}\right)\right.\right.\\
\null\hspace{0.5cm}\left.\left.+Y_f^2\lambda^2s^2\left[\left(Y_f^2+\frac{g'^2({\cal Y}^f_L+{\cal Y}_R^f)-2g^2I_3^f}{8}\left(1+\frac{A_f^2}{\lambda^2s^2}\right)\right){\cal C}^{\tilde{F}}_{L,R}-\left(2A_f^2+\frac{g'^2({\cal Y}^f_L-{\cal Y}_R^f)-2g^2I_3^f}{8}(1-\frac{A_f^2}{\lambda^2s^2})\Delta m^{2\,\tilde{F}}_{L,R}\right){\cal D}^{\tilde{F}}_{L,R}\right]\right]\right.\\
\null\hspace{3cm}\left.+\delta_{f(d,e)}\left[\frac{g'^2{\cal Y}^f_L-2g^2I_3^f}{4}\left(Y_f^2+\frac{g'^2{\cal Y}^f_R}{4}\right)\ln\left(\frac{m_{\tilde{F}_L}^2}{\Lambda^2}\right)+\frac{g'^2{\cal Y}^f_R}{4}\left(Y_f^2+\frac{g'^2{\cal Y}^f_L-2g^2I_3^f}{4}\right)\ln\left(\frac{m_{\tilde{F}_L}^2}{\Lambda^2}\right)\right.\right.\\
\null\hspace{0.5cm}\left.\left.+Y_f^2\lambda^2s^2\left[\left(Y_f^2+\frac{g'^2({\cal Y}^f_L+{\cal Y}_R^f)-2g^2I_3^f}{8}\left(1+\frac{A_f^2}{\lambda^2s^2}\right)\right){\cal C}^{\tilde{F}}_{L,R}-\left(2A_f^2+\frac{g'^2({\cal Y}^f_L-{\cal Y}_R^f)-2g^2I_3^f}{8}(\frac{A_f^2}{\lambda^2s^2}-1)\Delta m^{2\,\tilde{F}}_{L,R}\right){\cal D}^{\tilde{F}}_{L,R}\right]\right]\right\}\\
\lambda_5^{\tilde{F}}=-\frac{1}{16\pi^2}\sum_{f}Y_f^4A_f^2\lambda^2s^2{\cal D}^{\tilde{F}}_{L,R}\\
\lambda_6^{\tilde{F}}=\frac{1}{16\pi^2}\sum_f\left\{\delta_{fu}Y_f^2A_f\lambda s\left[\left(Y_f^2+\frac{g'^2({\cal Y}^f_L+{\cal Y}_R^f)-2g^2I_3^f}{8}\right){\cal C}^{\tilde{F}}_{L,R}-\left(Y_f^2A_f^2+\frac{g'^2({\cal Y}^f_L-{\cal Y}_R^f)-2g^2I_3^f}{8}\Delta m^{2\,\tilde{F}}_{L,R}\right){\cal D}^{\tilde{F}}_{L,R}\right]\right.\\
\null\hspace{2.3cm}\left.+\delta_{f(d,e)}Y_f^2A_f\lambda s\left[\frac{g'^2({\cal Y}^f_L+{\cal Y}_R^f)-2g^2I_3^f}{8}{\cal C}^{\tilde{F}}_{L,R}-\left(Y_f^2\lambda^2s^2+\frac{g'^2({\cal Y}^f_L-{\cal Y}_R^f)-2g^2I_3^f}{8}\Delta m^{2\,\tilde{F}}_{L,R}\right){\cal D}^{\tilde{F}}_{L,R}\right]\right\}\\
\lambda_7^{\tilde{F}}=\frac{1}{16\pi^2}\sum_f\left\{\delta_{f(d,e)}Y_f^2A_f\lambda s\left[\left(Y_f^2+\frac{g'^2({\cal Y}^f_L+{\cal Y}_R^f)-2g^2I_3^f}{8}\right){\cal C}^{\tilde{F}}_{L,R}-\left(Y_f^2A_f^2-\frac{g'^2({\cal Y}^f_L-{\cal Y}_R^f)-2g^2I_3^f}{8}\Delta m^{2\,\tilde{F}}_{L,R}\right){\cal D}{\tilde{F}}_{L,R}\right]\right.\\
\null\hspace{2.3cm}\left.+\delta_{fu}Y_f^2A_f\lambda s\left[\frac{g'^2({\cal Y}^f_L+{\cal Y}_R^f)-2g^2I_3^f}{8}{\cal C}^{\tilde{F}}_{L,R}-\left(Y_f^2\lambda^2s^2-\frac{g'^2({\cal Y}^f_L-{\cal Y}_R^f)-2g^2I_3^f}{8}\Delta m^{2\,\tilde{F}}_{L,R}\right){\cal D}^{\tilde{F}}_{L,R}\right]\right\}\\
(\lambda_P^u)^{\tilde{F}}=\frac{1}{16\pi^2}\sum_{f=d,e}Y_f^2\lambda^2{\cal B}^{\tilde{F}}_{L,R}\\
(\lambda_P^d)^{\tilde{F}}=\frac{1}{16\pi^2}\sum_{f=u}Y_f^2\lambda^2{\cal B}^{\tilde{F}}_{LR}
\end{cases}\end{equation}}\noi
where we have used the notations:
{\small\begin{align}
 &{\cal A}^{\tilde{F}}_L\equiv m_{\tilde{F}_L}^2\left[\ln\left(\frac{m_{\tilde{F}_L}^2}{\Lambda^2}\right)-1\right]\hspace{0.5cm};\hspace{0.5cm}{\cal A}^{\tilde{F}}_R\equiv m_{\tilde{F}_R}^2\left[\ln\left(\frac{m_{\tilde{F}_R}^2}{\Lambda^2}\right)-1\right]\hspace{0.5cm};\hspace{0.5cm}\Delta m^{2\,\tilde{F}}_{L,R}\equiv m_{\tilde{F}_L}^2-m_{\tilde{F}_R}^2\nonumber\\
 &{\cal B}^{\tilde{F}}_{L,R}\equiv \frac{1}{m_{\tilde{F}_L}^2-m_{\tilde{F}_R}^2}\left[{\cal A}^{\tilde{F}}_L-{\cal A}^{\tilde{F}}_R\right]\xrightarrow{m_{\tilde{F}_{L,R}}^2\rightarrow m^2}\ln\left(\frac{m^2}{\Lambda^2}\right)\\
 &{\cal C}^{\tilde{F}}_{L,R}\equiv \frac{1}{m_{\tilde{F}_L}^2-m_{\tilde{F}_R}^2}\ln\left(\frac{m_{\tilde{F}_L}^2}{m_{\tilde{F}_R}^2}\right)\xrightarrow{m_{\tilde{F}_{L,R}}^2\rightarrow m^2}\frac{1}{m^2}\nonumber\\
 &{\cal D}^{\tilde{F}}_{L,R}\equiv \frac{1}{(m_{\tilde{F}_L}^2-m_{\tilde{F}_R}^2)^3}\left[m_{\tilde{F}_L}^2\left(\ln\left(\frac{m_{\tilde{F}_L}^2}{m_{\tilde{F}_R}^2}\right)-2\right)+m_{\tilde{F}_R}^2\left(\ln\left(\frac{m_{\tilde{F}_L}^2}{m_{\tilde{F}_R}^2}\right)+2\right)\right]\xrightarrow{m_{\tilde{F}_{L,R}}^2\rightarrow m^2}\frac{1}{6m^4}\nonumber
\end{align}}

{\bf Chargino contributions:}\newline
The chargino squared mass-matrix, in a base of Dirac (winos,higgsinos), is given by:
\begin{equation}
 {\cal M}^2_{\chi^{\pm}}(S,H_u^0,H_d^0)=\begin{bmatrix}
M_2^2+g^2|H_d^0|^2&g(M_2H_u^{0\,*}+\lambda S H_d^0)\\
g(M_2H_u^{0}+\lambda S^* H_d^{0\,*})&\lambda^2|S|^2+g^2|H_u^0|^2
                        \end{bmatrix}
\end{equation}\noi
Defining $T_{\chi^{\pm}}=({\cal M}^2_{\chi^{\pm}})_{11}+({\cal M}^2_{\chi^{\pm}})_{22}$ and $R_{\tilde{F}}^2=\left[({\cal M}^2_{\chi^{\pm}})_{11}-({\cal M}^2_{\chi^{\pm}})_{22}\right]^2+4|{\cal M}^2_{\chi^{\pm}}|^2_{12}$, 
we obtain the eigenvalues $m_{(\chi^{\pm})_{\pm}}^2=\frac{1}{2}\left[T_{\chi^{\pm}}\pm\sqrt{R_{\chi^{\pm}}^2}\right]$ and the Higgs potential:
\begin{equation}
 \delta{\cal V}_{\mbox{\small eff}}^{\Lambda,\chi^{\pm}}(H_u^0,H_d^0,S)=-\frac{1}{64\pi^2}\left\{(T_{\chi^{\pm}}^2+R_{\chi^{\pm}}^2)\left[\ln\left(\frac{T_{\chi^{\pm}}^2-R_{\chi^{\pm}}^2}{4\Lambda^4}\right)-3\right]+2\,T_{\chi^{\pm}}\cdot R_{\chi^{\pm}}\ln\left(\frac{T_{\chi^{\pm}}+R_{\chi^{\pm}}}{T_{\chi^{\pm}}-R_{\chi^{\pm}}}\right)\right\}
\end{equation}\noi
One can derive the couplings:
{\small\begin{equation}\begin{cases}
\left(m_{H_{u,d}^0}^2\right)^{\chi^{\pm}}=-\frac{g^2}{8\pi^2(M_2^2-\lambda^2s^2)^3}\left\{M_2^8\left[\ln\left(\frac{M_2^2}{\Lambda^2}\right)-1\right]-3M_2^6\lambda^2s^2\left[\ln\left(\frac{M_2^2}{\Lambda^2}\right)-1\right]\right.\\
\hspace{3.5cm}\left.+3M_2^4\lambda^4s^4\left[2\ln\left(\frac{M_2^2}{\Lambda^2}\right)-\ln\left(\frac{\lambda^2s^2}{\Lambda^2}\right)-\frac{8}{3}\right]-M_2^2\lambda^6s^6\left[\ln\left(\frac{\lambda^2s^2}{\Lambda^2}\right)-7\right]-\lambda^8s^8\right\}\\
\left(m_{12}^2\right)^{\chi^{\pm}}=-\frac{g^2M_2\lambda^3s^3}{8\pi^2(M_2^2-\lambda^2s^2)^3}\left\{M_2^4\left[\ln\left(\frac{M_2^2}{\lambda^2s^2}\right)-3\right]+3M_2^2\lambda^2s^2\left[\ln\left(\frac{M_2^2}{\lambda^2s^2}\right)+\frac{2}{3}\right]+\lambda^4s^4\right\}\\
\left(A_{(u,d)s}\right)^{\chi^{\pm}}=\frac{g^2\lambda^4s^3}{4\pi^2(M_2^2-\lambda^2s^2)^3}\left\{2M_2^4\left[\ln\left(\frac{M_2^2}{\lambda^2s^2}\right)-\frac{3}{2}\right]+4M_2^2\lambda^2s^2-\lambda^4s^4\right\}\\
\left(\tilde{A}_{ud}\right)^{\chi^{\pm}}=-\frac{g^2M_2\lambda^3s^2}{8\pi^2(M_2^2-\lambda^2s^2)^3}\left\{M_2^4\left[\ln\left(\frac{M_2^2}{\lambda^2s^2}\right)-3\right]+3M_2^2\lambda^2s^2\left[\ln\left(\frac{M_2^2}{\lambda^2s^2}\right)+\frac{2}{3}\right]+\lambda^4s^4\right\}\\
\left(A_{ud}\right)^{\chi^{\pm}}=\frac{g^2\lambda M_2}{8\pi^2(M_2^2-\lambda^2s^2)^3}\left\{M_2^6\left[\ln\left(\frac{M_2^2}{\Lambda^2}\right)-1\right]-M_2^4\lambda^2s^2\left[5\ln\left(\frac{M_2^2}{\Lambda^2}\right)-2\ln\left(\frac{\lambda^2s^2}{\Lambda^2}\right)-8\right]\right.\\
\null\hspace{3.5cm}\left.+3M_2^2\lambda^4s^4\left[\ln\left(\frac{\lambda^2s^2}{\Lambda^2}\right)-3\right]-\lambda^6s^6\left[\ln\left(\frac{\lambda^2s^2}{\Lambda^2}\right)-2\right]\right\}\\
\left(\lambda_{1,2}\right)^{\chi^{\pm}}=-\frac{g^4}{8\pi^2(M_2^2-\lambda^2s^2)^3}\left\{M_2^6\ln\left(\frac{M_2^2}{\Lambda^2}\right)-3M_2^4\lambda^2s^2\left[\ln\left(\frac{M_2^2}{\Lambda^2}\right)-\frac{2}{3}\right]+3M_2^2\lambda^4s^4\left[\ln\left(\frac{\lambda^2s^2}{\Lambda^2}\right)-\frac{2}{3}\right]-\lambda^6s^6\ln\left(\frac{\lambda^2s^2}{\Lambda^2}\right)\right\}\\
\left(\lambda_{3}+\lambda_4\right)^{\chi^{\pm}}=-\frac{g^4}{8\pi^2(M_2^2-\lambda^2s^2)^3}\left\{M_2^6-2M_2^4\lambda^2s^2\left[\ln\left(\frac{M_2^2}{\lambda^2s^2}\right)-\frac{1}{2}\right]-2M_2^2\lambda^4s^4\left[\ln\left(\frac{M_2^2}{\lambda^2s^2}\right)+\frac{1}{2}\right]-\lambda^6s^6\right\}\\
\left(\lambda_{5}\right)^{\chi^{\pm}}=\frac{g^4M_2^2\lambda^2s^2}{8\pi^2(M_2^2-\lambda^2s^2)^3}\left\{M_2^2\left[\ln\left(\frac{M_2^2}{\lambda^2s^2}\right)-2\right]+\lambda^2s^2\left[\ln\left(\frac{M_2^2}{\lambda^2s^2}\right)+2\right]\right\}\\
\left(\lambda_{M}\right)^{\chi^{\pm}}=\frac{g^2\lambda^3sM_2}{8\pi^2(M_2^2-\lambda^2s^2)^3}\left\{2M_2^4\left[\ln\left(\frac{M_2^2}{\lambda^2s^2}\right)-\frac{3}{2}\right]+4M_2^2\lambda^2s^2-\lambda^4s^4\right\}\\
\left(\lambda^P_{M}\right)^{\chi^{\pm}}=-\frac{g^2\lambda^3sM_2}{16\pi^2(M_2^2-\lambda^2s^2)^3}\left\{M_2^4-2M_2^2\lambda^2s^2\ln\left(\frac{M_2^2}{\lambda^2s^2}\right)-\lambda^4s^4\right\}\\
\left(\tilde{\lambda}^P_{M}\right)^{\chi^{\pm}}=\frac{g^2\lambda^3sM_2}{16\pi^2(M_2^2-\lambda^2s^2)^3}\left\{2M_2^4\left[\ln\left(\frac{M_2^2}{\lambda^2s^2}\right)-\frac{3}{2}\right]+4M_2^2\lambda^2s^2-\lambda^4s^4\right\}\\
\left(\lambda^P_{u,d}\right)^{\chi^{\pm}}=-\frac{g^2\lambda^2}{8\pi^2(M_2^2-\lambda^2s^2)^3}\left\{M^6_2\left[\ln\left(\frac{M_2^2}{\Lambda^2}\right)-1\right]+M_2^4\lambda^2s^2\left[\ln\left(\frac{M_2^2}{\Lambda^2}\right)+2\ln\left(\frac{\lambda^2s^2}{\Lambda^2}\right)-1\right]\right.\\
\null\hspace{3.5cm}\left.+3M_2^2\lambda^4s^4\left[\ln\left(\frac{\lambda^2s^2}{\Lambda^2}\right)+1\right]-\lambda^6s^6\left[\ln\left(\frac{\lambda^2s^2}{\Lambda^2}\right)+1\right]\right\}\\
\left(\tilde{\lambda}^P_{u,d}\right)^{\chi^{\pm}}=-\frac{g^2\lambda^3sM_2}{16\pi^2(M_2^2-\lambda^2s^2)^3}\left\{2M_2^4\left[\ln\left(\frac{M_2^2}{\lambda^2s^2}\right)-\frac{3}{2}\right]+4M_2^2\lambda^2s^2-\lambda^4s^4\right\}\\
\end{cases}\end{equation}\noi
\begin{displaymath}
 \begin{cases}
  \left(\lambda_T\right)^{\chi^{\pm}}=\frac{\lambda^4s^3}{12\pi^2}\hspace{1.cm};\hspace{1.cm}
\left(m_{S}^2\right)^{\chi^{\pm}}=-\frac{\lambda^4s^2}{4\pi^2}\hspace{1.cm};\hspace{1.cm}
\left(\mu_{S}^2\right)^{\chi^{\pm}}=-\frac{\lambda^4s^2}{8\pi^2}\hspace{1.cm};\hspace{1.cm}
\left(\tilde{A}_{S}^2\right)^{\chi^{\pm}}=\frac{3\lambda^4s}{4\pi^2}\\
 \left(\kappa^2\right)^{\chi^{\pm}}=-\frac{\lambda^4}{16\pi^2}\left[\ln\left(\frac{M_2^2}{\Lambda^2}\right)-\frac{3}{2}\right]\hspace{1.2cm};\hspace{1.2cm}
\left(\kappa_S^2\right)^{\chi^{\pm}}=\frac{\lambda^4}{48\pi^2}\hspace{1.2cm};\hspace{1.2cm}
\left(\tilde{\kappa}_S^2\right)^{\chi^{\pm}}=-\frac{\lambda^4}{6\pi^2}
\end{cases}
\end{displaymath}
}

{\bf Neutralino contributions:}\newline
The (hermitian) neutralino squared mass-matrix is determined by its entries in the base of Weyl spinors $(-\imath\tilde{b}^0,
-\imath\tilde{w}^0_3,\tilde{h}^0_u,\tilde{h}^0_d,\tilde{h}^0_s)$:
{\small\begin{align}
 \left({\cal M}^2_{\chi^0}\right)_{11}(S,H_u^0,H_d^0)&=M_1^2+\frac{g'^2}{2}\left(|H_u^0|^2+|H_d^0|^2\right)\nonumber\\
 \left({\cal M}^2_{\chi^0}\right)_{22}(S,H_u^0,H_d^0)&=M_2^2+\frac{g^2}{2}\left(|H_u^0|^2+|H_d^0|^2\right)\nonumber\\
 \left({\cal M}^2_{\chi^0}\right)_{12}(S,H_u^0,H_d^0)&=-\frac{gg'}{2}\left(|H_u^0|^2+|H_d^0|^2\right)\nonumber\\
 \left({\cal M}^2_{\chi^0}\right)_{33}(S,H_u^0,H_d^0)&=\lambda^2(|S|^2+|H_d^0|^2)+\frac{g^2+g'^2}{2}|H_u^0|^2\nonumber\\
 \left({\cal M}^2_{\chi^0}\right)_{44}(S,H_u^0,H_d^0)&=\lambda^2(|S|^2+|H_u^0|^2)+\frac{g^2+g'^2}{2}|H_d^0|^2\nonumber\\
 \left({\cal M}^2_{\chi^0}\right)_{34}(S,H_u^0,H_d^0)&=\left(\lambda^2-\frac{g^2+g'^2}{2}\right)H_u^0H_d^{0\,*}\nonumber\\
 \left({\cal M}^2_{\chi^0}\right)_{55}(S,H_u^0,H_d^0)&=4\kappa^2|S|^2+\lambda^2(|H_u^0|^2+|H_d^0|^2)\\
 \left({\cal M}^2_{\chi^0}\right)_{35}(S,H_u^0,H_d^0)&=\lambda^2S^*H_u^0-2\lambda\kappa SH_d^{0\,*}\nonumber\\
 \left({\cal M}^2_{\chi^0}\right)_{45}(S,H_u^0,H_d^0)&=\lambda^2S^*H_d^0-2\lambda\kappa SH_u^{0\,*}\nonumber\\
 \left({\cal M}^2_{\chi^0}\right)_{13}(S,H_u^0,H_d^0)&=\frac{g'}{\sqrt{2}}(M_1H_u^{0\,*}+\lambda SH_d^0)\nonumber\\
 \left({\cal M}^2_{\chi^0}\right)_{14}(S,H_u^0,H_d^0)&=-\frac{g'}{\sqrt{2}}(M_1H_d^{0\,*}+\lambda SH_u^0)\nonumber\\
 \left({\cal M}^2_{\chi^0}\right)_{23}(S,H_u^0,H_d^0)&=-\frac{g}{\sqrt{2}}(M_2H_u^{0\,*}+\lambda SH_d^0)\nonumber\\
 \left({\cal M}^2_{\chi^0}\right)_{24}(S,H_u^0,H_d^0)&=\frac{g}{\sqrt{2}}(M_1H_d^{0\,*}+\lambda SH_u^0)\nonumber
\end{align}}\noi
One can expand its eigenvalues in terms of doublet fields:
\begin{multline}
 m_i^2(S,H_u^0,H_d^0)=m_i^{2\,(0)}(S)+m_i^{2\,(1)}(S,H_u^0,H_d^0)+m_i^{2\,(2)}(S,H_u^0,H_d^0)+O\left((H^0_{u,d})^5\right)\\
\begin{cases} m_i^{2\,(1)}(S,H_u^0,H_d^0)=m_i^{2\,(1u)}(S)|H_u^0|^2+m_i^{2\,(1d)}(S)|H_d^0|^2+\left(m_i^{2\,(1ud)}(S)H_u^0H_d^0+h.c.\right)\\
m_i^{2\,(2)}(S,H_u^0,H_d^0)=m_i^{2\,(2d)}(S)|H_d^0|^4+m_i^{2\,(2u)}(S)|H_u^0|^4+m_i^{2\,(2ud)}(S)|H_u^0|^2|H_d^0|^2+\ldots\end{cases}
\end{multline}\noi
The associated potential is then given by:
{\small\begin{multline}
 \delta{\cal V}_{\mbox{\small eff}}^{\Lambda,\chi^{0}}(H_u^0,H_d^0,S)=-\frac{1}{32\pi^2}\sum_im_i^4(S,H_u^0,H_d^0)\left[\ln\left(\frac{m_i^2(S,H_u^0,H_d^0)}{\Lambda^2}\right)-\frac{3}{2}\right]\\
=-\frac{1}{32\pi^2}\sum_i\left\{m_i^{4\,(0)}(S)\left[\ln\left(\frac{m_i^{2\,(0)}(S)}{\Lambda^2}\right)-\frac{3}{2}\right]+2m_i^{2\,(0)}(S)m_i^{2\,(1)}(S,H_u^0,H_d^0)\left[\ln\left(\frac{m_i^{2\,(0)}(S)}{\Lambda^2}\right)-1\right]\right.\\\left.+\left(2m_i^{2\,(0)}(S)m_i^{2\,(2)}(S,H_u^0,H_d^0)\left[\ln\left(\frac{m_i^{2\,(0)}(S)}{\Lambda^2}\right)-1\right]+m_i^{4\,(1)}(S,H_u^0,H_d^0)\ln\left(\frac{m_i^{2\,(0)}(S)}{\Lambda^2}\right)\right)+O(H^5_{u,d})\right\}
\end{multline}}

\noi${\cal M}^2_{\chi^0}(S,H_{u,d}^0=0)$ is already diagonal with the eigenstates $\{M_1^2,\left|E_1\right>\}$, $\{M_2^2,\left|E_2\right>\}$, 
$\{\lambda^2|S|^2,\left|E_3\right>,\left|E_4\right>\}$ and $\{4\kappa^2|S|^2,\left|E_5\right>\}$ ($\{\left|E_i\right>\}$ stands for the canonical base of $\mathbb{C}^5$), from which one obtains easily the pure-singlet parameters:
{\small\begin{equation}
 \begin{cases}
  \left(\lambda_T\right)^{\chi^{0}}=\frac{1}{12\pi^2}(\lambda^4+8\kappa^4)s^3\\
\left(m^2_S\right)^{\chi^{0}}=-\frac{1}{4\pi^2}(\lambda^4+8\kappa^4)s^2\\
\left(\mu_S^2\right)^{\chi^{0}}=-\frac{1}{8\pi^2}(\lambda^4+8\kappa^4)s^2\\
\left(\tilde{A}_S\right)^{\chi^{0}}=\frac{3}{4\pi^2}(\lambda^4+8\kappa^4)s\\
\left(\kappa^2\right)^{\chi^{0}}=-\frac{1}{16\pi^2}\left\{\lambda^4\left[\ln\left(\frac{\lambda^2s^2}{\Lambda^2}\right)+\frac{3}{2}\right]+8\kappa^4\left[\ln\left(\frac{4\kappa^2s^2}{\Lambda^2}\right)+\frac{3}{2}\right]\right\}\\
\left(\kappa^2_S\right)^{\chi^{0}}=\frac{1}{48\pi^2}(\lambda^4+8\kappa^4)\\
\left(\tilde{\kappa}^2_S\right)^{\chi^{0}}=-\frac{1}{6\pi^2}(\lambda^4+8\kappa^4)
 \end{cases}
\end{equation}}

\noi The $O(H^2)$ masses also come without much effort:
{\small\begin{align}
 &m_1^{2\,(1)}=\frac{g'^2}{M_1^2-\lambda^2|S|^2}\left[M_1^2(|H_u^0|^2+|H_d^0|^2)+2\lambda M_1\mbox{Re}(SH_u^0H_d^0)\right]\nonumber\\
 &m_2^{2\,(1)}=\frac{g^2}{M_2^2-\lambda^2|S|^2}\left[M_2^2(|H_u^0|^2+|H_d^0|^2)+2\lambda M_2\mbox{Re}(SH_u^0H_d^0)\right]\\
 &m_3^{2\,(1)}+m_4^{2\,(1)}=\frac{g'^2}{\lambda^2|S|^2-M_1^2}\left[\lambda^2|S|^2(|H_u^0|^2+|H_d^0|^2)+2\lambda M_1\mbox{Re}(SH_u^0H_d^0)\right]\nonumber\\
&\null\hspace{2.6cm}+\frac{g^2}{\lambda^2|S|^2-M_2^2}\left[\lambda^2|S|^2(|H_u^0|^2+|H_d^0|^2)+2\lambda M_2\mbox{Re}(SH_u^0H_d^0)\right]\nonumber\\
&\null\hspace{2.6cm}+\frac{2\lambda^2}{(\lambda^2-4\kappa^2)|S|^2}\left[\lambda^2|S|^2(|H_u^0|^2+|H_d^0|^2)-4\kappa\lambda\mbox{Re}(S^{*\,2}H_u^0H_d^0)\right]\nonumber\\
 &m_5^{2\,(1)}=\frac{8\lambda^2}{(4\kappa^2-\lambda^2)|S|^2}\left[\kappa^2|S|^2(|H_u^0|^2+|H_d^0|^2)-\kappa\lambda\mbox{Re}(S^{*\,2}H_u^0H_d^0)\right]\nonumber
\end{align}}\noi
from which we can derive the couplings (we focus on the logarithmic terms):
{\small\begin{equation}
 \begin{cases}
\left(m^2_{H_{u,d}}\right)^{\chi^{0}}=-\frac{g'^2}{16\pi^2(M_1^2-\lambda^2s^2)^3}\left\{M_1^4\left[M_1^4-3M^2_1\lambda^2s^2+6\lambda^4s^4\right]\ln\left(\frac{M_1^2}{\Lambda^2}\right)-\lambda^4s^4\left[3M_1^2+\lambda^2s^2\right]\ln\left(\frac{\lambda^2s^2}{\Lambda^2}\right)\right\}\\
\null\hspace{2cm}-\frac{g^2}{16\pi^2(M_2^2-\lambda^2s^2)^3}\left\{M_2^4\left[M_2^4-3M^2_2\lambda^2s^2+6\lambda^4s^4\right]\ln\left(\frac{M_2^2}{\Lambda^2}\right)-\lambda^4s^4\left[3M_2^2+\lambda^2s^2\right]\ln\left(\frac{\lambda^2s^2}{\Lambda^2}\right)\right\}\\
\left(m^2_{12}\right)^{\chi^{0}}=\frac{g'^2M_1^3\lambda^3s^3}{8\pi^2(M_1^2-\lambda^2s^2)^3}\left[M_1^2+3\lambda^2s^2\right]\ln\left(\frac{M_1^2}{\lambda^2s^2}\right)+\frac{g^2M_2^3\lambda^3s^3}{8\pi^2(M_2^2-\lambda^2s^2)^3}\left[M_2^2+3\lambda^2s^2\right]\ln\left(\frac{M_2^2}{\lambda^2s^2}\right)\\
\left(A_{(u,d)s}\right)^{\chi^{0}}=\frac{g'^2M_1^4\lambda^4s^3}{4\pi^2(M_1^2-\lambda^2s^2)^3}\ln\left(\frac{M_1^2}{\lambda^2s^2}\right)+\frac{g^2M_2^4\lambda^4s^3}{4\pi^2(M_2^2-\lambda^2s^2)^3}\ln\left(\frac{M_2^2}{\lambda^2s^2}\right)\\
\left(A_{ud}\right)^{\chi^{0}}=\frac{g'^2\lambda M_1}{16\pi^2(M_1^2-\lambda^2s^2)^3}\left\{M_1^4\left[M_1^2-5\lambda^2s^2\right]\ln\left(\frac{M_1^2}{\Lambda^2}\right)-\lambda^2s^2\left[2M_1^4+3M_1^2\lambda^2s^2-\lambda^4s^4\right]\ln\left(\frac{\lambda^2s^2}{\Lambda^2}\right)\right\}\\
\null\hspace{1.8cm}+\frac{g^2\lambda M_2}{16\pi^2(M_2^2-\lambda^2s^2)^3}\left\{M_2^4\left[M_2^2-5\lambda^2s^2\right]\ln\left(\frac{M_2^2}{\Lambda^2}\right)-\lambda^2s^2\left[2M_2^4+3M_2^2\lambda^2s^2-\lambda^4s^4\right]\ln\left(\frac{\lambda^2s^2}{\Lambda^2}\right)\right\}\\
\left(\tilde{A}_{ud}\right)^{\chi^{0}}=-\frac{g'^2M_1^3\lambda^3s^2}{16\pi^2(M_1^2-\lambda^2s^2)^3}\left[M_1^2+3\lambda^2s^2\right]\ln\left(\frac{M_1^2}{\lambda^2s^2}\right)-\frac{g^2M_2^3\lambda^3s^2}{16\pi^2(M_2^2-\lambda^2s^2)^3}\left[M_2^2+3\lambda^2s^2\right]\ln\left(\frac{M_2^2}{\lambda^2s^2}\right)\\
\left(\lambda^P_{u,d}\right)^{\chi^{0}}=-\frac{g'^2\lambda^2}{16\pi^2(M_1^2-\lambda^2s^2)^3}\left\{M_1^4\left[M_1^2+\lambda^2s^2\right]\ln\left(\frac{M_1^2}{\Lambda^2}\right)-\lambda^2s^2\left[4M_1^4-3M_1^2\lambda^2s^2+\lambda^4s^4\right]\ln\left(\frac{\lambda^2s^2}{\Lambda^2}\right)\right\}\\
\null\hspace{1.8cm}-\frac{g^2\lambda^2}{16\pi^2(M_2^2-\lambda^2s^2)^3}\left\{M_2^4\left[M_2^2+\lambda^2s^2\right]\ln\left(\frac{M_2^2}{\Lambda^2}\right)-\lambda^2s^2\left[4M_2^4-3M_2^2\lambda^2s^2+\lambda^4s^4\right]\ln\left(\frac{\lambda^2s^2}{\Lambda^2}\right)\right\}\\
\null\hspace{1.8cm}+\frac{\lambda^2}{8\pi^2(4\kappa^2-\lambda^2)}\left\{\lambda^4\ln\left(\frac{\lambda^2s^2}{\Lambda^2}\right)-16\kappa^4\ln\left(\frac{4\kappa^2s^2}{\Lambda^2}\right)\right\}\\
\left(\tilde{\lambda}^P_{u,d}\right)^{\chi^{0}}=-\frac{g'^2M_1^4\lambda^4s^2}{16\pi^2(M_1^2-\lambda^2s^2)^3}\ln\left(\frac{M_1^2}{\lambda^2s^2}\right)-\frac{g^2M_2^4\lambda^4s^2}{16\pi^2(M_2^2-\lambda^2s^2)^3}\ln\left(\frac{M_2^2}{\lambda^2s^2}\right)\\
\left(\lambda^P_M\right)^{\chi^{0}}=\frac{g'^2M_1^3\lambda^5s^3}{16\pi^2(M_1^2-\lambda^2s^2)^3}\ln\left(\frac{M_1^2}{\lambda^2s^2}\right)+\frac{g^2M_2^3\lambda^5s^3}{16\pi^2(M_2^2-\lambda^2s^2)^3}\ln\left(\frac{M_2^2}{\lambda^2s^2}\right)\\
\null\hspace{1.8cm}+\frac{\kappa\lambda^3}{4\pi^2(4\kappa^2-\lambda^2)}\left\{\lambda^2\ln\left(\frac{\lambda^2s^2}{\Lambda^2}\right)-4\kappa^2\ln\left(\frac{4\kappa^2s^2}{\Lambda^2}\right)\right\}\\
\left(\lambda_M\right)^{\chi^{0}}=\frac{g'^2M_1^5\lambda^3s}{8\pi^2(M_1^2-\lambda^2s^2)^3}\ln\left(\frac{M_1^2}{\lambda^2s^2}\right)+\frac{g^2M_2^5\lambda^3s}{8\pi^2(M_2^2-\lambda^2s^2)^3}\ln\left(\frac{M_2^2}{\lambda^2s^2}\right)\\
\left(\tilde{\lambda}^P_M\right)^{\chi^{0}}=\frac{g'^2M_1^5\lambda^3s}{16\pi^2(M_1^2-\lambda^2s^2)^3}\ln\left(\frac{M_1^2}{\lambda^2s^2}\right)+\frac{g^2M_2^5\lambda^3s}{16\pi^2(M_2^2-\lambda^2s^2)^3}\ln\left(\frac{M_2^2}{\lambda^2s^2}\right)\\
 \end{cases}
\end{equation}}\noi
The limit $\kappa^2\to0$ for the nMSSM is straightforward.

{\bf Charged-Higgs contributions -- NMSSM:}\newline
In the base $(H_u^-,H_d^-)$, the hermitian squared mass-matrix of Charged-Higgs bosons in terms of neutral Higgs fields reads (we use the general notation 
of a $\mathbb{Z}_3$-conserving potential; those parameters should be replaced, in practice, by their tree-level value; we also 
define $\lambda_P^.$, replacing $\lambda_P^{u,d}$, which coincide at tree-level; same thing for $\lambda_.$, replacing 
$\lambda_{1,2}$):
{\small\begin{align}
 &({\cal M}_{H^{\pm}}^2)_{11}(S,H_{u,d}^0)=M_{EW}^2+(A_{ud}+\lambda_P^Ms)\frac{s}{\tan\beta}+\lambda_.|H_u^0|^2+\lambda_3|H_d^0|^2+\lambda_P^.(|S|^2-s^2)\nonumber\\
 &({\cal M}_{H^{\pm}}^2)_{22}(S,H_{u,d}^0)=M_{EW}^2+(A_{ud}+\lambda_P^Ms)s\tan\beta+\lambda_.|H_d^0|^2+\lambda_3|H_u^0|^2+\lambda_P^.(|S|^2-s^2)\\
 &({\cal M}_{H^{\pm}}^2)_{12}(S,H_{u,d}^0)=A_{ud}S+\lambda_P^MS^{*\,2}-\lambda_4(H_u^0H_d^0)^*\nonumber
\end{align}}\noi
We have introduced $M_{EW}^2$ to replace constant terms generated by the electroweak v.e.v.\ and regularizing the (otherwise-vanishing)
Goldstone mass (which does not correspond to a Goldstone boson since $SU(2)_L$ is conserved in our approach). In practice, this 
$M_{EW}^2$ should be chosen as $M_W^2$, typically, since it replaces the longitudinal component of $W$-bosons. Now, defining 
$T\equiv\mbox{Tr}{\cal M}_{H^{\pm}}^2(S,H_{u,d}^0)$ and $R^2\equiv T^2-4\mbox{det}{\cal M}_{H^{\pm}}^2(S,H_{u,d}^0)$, we obtain
the two eigenvalues $m^2_{h/H}(S,H_{u,d}^0)=\frac{1}{2}\left[T-/+R\right]$, as well as the potential:
\begin{equation}\label{CHpot}
 \delta{\cal V}_{\mbox{\small eff}}^{\Lambda,H^{\pm}}(H_u^0,H_d^0,S)=\frac{1}{128\pi^2}\left\{(T^2+R^2)\left[\ln\left(\frac{T^2-R^2}{4\Lambda^4}\right)-3\right]+2\,T\cdot R\ln\left(\frac{T+R}{T-R}\right)\right\}
\end{equation}\noi
Then we focus on the logarithms (the notation $\left<\cdot\right>$ means that Higgs fields are replaced by their v.e.v.'s):
\begin{equation}\begin{cases}
 \ln\left(\frac{T^2-R^2}{4\Lambda^2}\right)\simeq\ln\left[\frac{M_{EW}^2}{\Lambda^4}\left(M_{EW}^2+\frac{2(A_{ud}+\lambda_P^M s)s}{\sin{2\beta}}\right)\right]+\ldots\equiv\ln\left<\frac{m_{h}^2m_{H}^2}{\Lambda^{4}}\right>+\ldots\\
\ln\left(\frac{T+R}{T-R}\right)\simeq\ln\left[1+\frac{2(A_{ud}+\lambda_P^M s)s}{\sin{2\beta}M_{EW}^2}\right]+\ldots\equiv-\ln\left<\frac{m_{h}^2}{m_{H}^2}\right>+\ldots
\end{cases}\end{equation}\noi
Expanding their coefficients, we obtain the leading charged-Higgs contributions to the Higgs-potential parameters. Note that the 
coefficients multiplying $\ln\left<\frac{m_{h}^2}{m_{H}^2}\right>$ are in general very complicated. Here, for simplicity, we give only the leading term in 
$\sin{2\beta}\to0$ ($\tan\beta\to\infty$)
{\small\begin{equation}\begin{cases}
\left(\lambda_T\right)^{H^{\pm}}\simeq\frac{\lambda_P^.\lambda_P^Ms^3\sin{2\beta}}{32\pi^2\left(A_{ud}+\lambda_P^Ms\right)}(5A_{ud}+12\lambda_P^Ms)\ln\left<\frac{m_{h}^2}{m_{H}^2}\right>\\
\left(m_{H_{u}}^2\right)^{H^{\pm}}\simeq\frac{1}{32\pi^2}\left\{\left[\frac{2(A_{ud}+\lambda_P^M s)}{\sin{2\beta}}\left(\lambda_.s\cos^2\beta+\lambda_3s\sin^2\beta\right)+(\lambda_.+\lambda_3)(M_{EW}^2-\lambda_P^.s^2)\right]\ln\left<\frac{m_{h}^2m_{H}^2}{\Lambda^{4}}\right>\right.\\
\null\hspace{6cm}\left.+\frac{2\lambda_3s(A_{ud}+\lambda_P^Ms)}{\sin{2\beta}}\ln\left<\frac{m_{h}^2}{m_{H}^2}\right>\right\}\\
\left(m_{H_{d}}^2\right)^{H^{\pm}}\simeq\frac{1}{32\pi^2}\left\{\left[\frac{2(A_{ud}+\lambda_P^M s)}{\sin{2\beta}}\left(\lambda_.s\sin^2\beta+\lambda_3s\cos^2\beta\right)+(\lambda_.+\lambda_3)(M_{EW}^2-\lambda_P^.s^2)\right]\ln\left<\frac{m_{h}^2m_{H}^2}{\Lambda^{4}}\right>\right.\\
\null\hspace{6cm}\left.+\frac{2\lambda_.s(A_{ud}+\lambda_P^Ms)}{\sin{2\beta}}\ln\left<\frac{m_{h}^2}{m_{H}^2}\right>\right\}\\
\left(m_{S}^2\right)^{H^{\pm}}\simeq\frac{1}{32\pi^2}\left\{\left[A_{ud}^2+2\lambda_P^.\left(M_{EW}^2-\lambda_P^.s^2+\frac{s(A_{ud}+\lambda_P^Ms)}{\sin{2\beta}}\right)\right]\ln\left<\frac{m_{h}^2m_{H}^2}{\Lambda^{4}}\right>-\frac{2\lambda_P^.s(A_{ud}+\lambda_P^Ms)}{\sin{2\beta}}\ln\left<\frac{m_{h}^2}{m_{H}^2}\right>\right\}\\
\left(m_{12}^2\right)^{H^{\pm}}\simeq-\frac{\lambda_4\lambda_P^.s^2\sin{2\beta}}{16\pi^2\left(A_{ud}+\lambda_P^Ms\right)}(A_{ud}+3\lambda_P^Ms)\ln\left<\frac{m_{h}^2}{m_{H}^2}\right>\\
\left(\mu_S^2\right)^{H^{\pm}}\simeq-\frac{3\lambda_P^M\lambda_P^.s^2\sin{2\beta}}{16\pi^2\left(A_{ud}+\lambda_P^Ms\right)}(2A_{ud}+3\lambda_P^Ms)\ln\left<\frac{m_{h}^2}{m_{H}^2}\right>\\
\left(A_{ud}\right)^{H^{\pm}}\simeq\frac{\lambda_4A_{ud}}{32\pi^2}\left\{\ln\left<\frac{m_{h}^2m_{H}^2}{\Lambda^{4}}\right>-\ln\left<\frac{m_{h}^2}{m_{H}^2}\right>\right\}\\
\left(\tilde{A}_{ud}\right)^{H^{\pm}}\simeq\frac{\lambda_4\lambda_P^.s\sin{2\beta}}{32\pi^2\left(A_{ud}+\lambda_P^Ms\right)}(A_{ud}+6\lambda_P^Ms)\ln\left<\frac{m_{h}^2}{m_{H}^2}\right>\\
\left(A_{us}\right)^{H^{\pm}}\simeq\frac{\lambda_.\lambda_P^Ms\sin{2\beta}}{32\pi^2\left(A_{ud}+\lambda_P^Ms\right)}(3A_{ud}+4\lambda_P^Ms)\ln\left<\frac{m_{h}^2}{m_{H}^2}\right>\\
\left(A_{ds}\right)^{H^{\pm}}\simeq\frac{\lambda_3\lambda_P^Ms\sin{2\beta}}{32\pi^2\left(A_{ud}+\lambda_P^Ms\right)}(3A_{ud}+4\lambda_P^Ms)\ln\left<\frac{m_{h}^2}{m_{H}^2}\right>\\
\left(A_{S}\right)^{H^{\pm}}\simeq\frac{3\lambda_P^MA_{ud}}{32\pi^2}\left\{\ln\left<\frac{m_{h}^2m_{H}^2}{\Lambda^{4}}\right>-\ln\left<\frac{m_{h}^2}{m_{H}^2}\right>\right\}\\
\left(\tilde{A}_S\right)^{H^{\pm}}\simeq\frac{9\lambda_P^.\lambda_P^Ms\sin{2\beta}}{16\pi^2\left(A_{ud}+\lambda_P^Ms\right)}(A_{ud}+3\lambda_P^Ms)\ln\left<\frac{m_{h}^2}{m_{H}^2}\right>
\end{cases}\end{equation}
\begin{displaymath}\begin{cases}
\left(\lambda_1\right)^{H^{\pm}}\simeq\frac{1}{32\pi^2}\left\{(\lambda_.^2+\lambda_3^2)\ln\left<\frac{m_{h}^2m_{H}^2}{\Lambda^{4}}\right>+(\lambda_3^2-\lambda_.^2)\ln\left<\frac{m_{h}^2}{m_{H}^2}\right>\right\}\\
\left(\lambda_2\right)^{H^{\pm}}\simeq\frac{1}{32\pi^2}\left\{(\lambda_.^2+\lambda_3^2)\ln\left<\frac{m_{h}^2m_{H}^2}{\Lambda^{4}}\right>+(\lambda_.^2-\lambda_3^2)\ln\left<\frac{m_{h}^2}{m_{H}^2}\right>\right\}\\
(\lambda_3+\lambda_4)^{H^{\pm}}\simeq\frac{1}{32\pi^2}\left\{(2\lambda_.\lambda_3+\lambda_4^2)\ln\left<\frac{m_{h}^2m_{H}^2}{\Lambda^{4}}\right>-\lambda_4^2\ln\left<\frac{m_{h}^2}{m_{H}^2}\right>\right\}\\
\left(\lambda_5\right)^{H^{\pm}}\simeq\frac{\lambda^2_4}{32\pi^2}\sin{2\beta}^2\\
\left(\lambda_6\right)^{H^{\pm}}\simeq\frac{\lambda_4\lambda_3}{32\pi^2}\sin{2\beta}\ln\left<\frac{m_{h}^2}{m_{H}^2}\right>\\
\left(\lambda_7\right)^{H^{\pm}}\simeq\frac{\lambda_4\lambda_.}{32\pi^2}\sin{2\beta}\ln\left<\frac{m_{h}^2}{m_{H}^2}\right>\\
(\lambda_P^u)^{H^{\pm}}\simeq\frac{1}{32\pi^2}\left\{\lambda_P^.(\lambda_.+\lambda_3)\ln\left<\frac{m_{h}^2m_{H}^2}{\Lambda^{4}}\right>+\lambda_P^.(\lambda_.-\lambda_3)\ln\left<\frac{m_{h}^2}{m_{H}^2}\right>\right\}\\
(\lambda_P^d)^{H^{\pm}}\simeq\frac{1}{32\pi^2}\left\{\lambda_P^.(\lambda_.+\lambda_3)\ln\left<\frac{m_{h}^2m_{H}^2}{\Lambda^{4}}\right>+\lambda_P^.(\lambda_3-\lambda_.)\ln\left<\frac{m_{h}^2}{m_{H}^2}\right>\right\}\\
(\tilde{\lambda}_P^u)^{H^{\pm}}\simeq-\frac{\lambda_.\lambda_P^M\sin{2\beta}}{32\pi^2\left(A_{ud}+\lambda_P^Ms\right)}(3A_{ud}+\lambda_P^Ms)\ln\left<\frac{m_{h}^2}{m_{H}^2}\right>\\
(\tilde{\lambda}_P^d)^{H^{\pm}}\simeq-\frac{\lambda_3\lambda_P^M\sin{2\beta}}{32\pi^2\left(A_{ud}+\lambda_P^Ms\right)}(3A_{ud}+\lambda_P^Ms)\ln\left<\frac{m_{h}^2}{m_{H}^2}\right>\\
\left(\lambda_M\right)^{H^{\pm}}\simeq-\frac{\lambda_4\lambda_P^.\sin{2\beta}}{32\pi^2(A_{ud}+\lambda_P^Ms)}(2A_{ud}+3\lambda_P^Ms)\ln\left<\frac{m_{h}^2}{m_{H}^2}\right>\\
(\lambda_P^M)^{H^{\pm}}\simeq\frac{\lambda_4\lambda_P^M}{32\pi^2}\left\{\ln\left<\frac{m_{h}^2m_{H}^2}{\Lambda^{4}}\right>-\ln\left<\frac{m_{h}^2}{m_{H}^2}\right>\right\}\\
\left(\tilde{\lambda}^P_M\right)^{H^{\pm}}\simeq-\frac{\lambda_4\lambda_P^.}{32\pi^2}\frac{A_{ud}\sin{2\beta}}{(A_{ud}+\lambda_P^Ms)}\ln\left<\frac{m_{h}^2}{m_{H}^2}\right>\\
(\kappa^2)^{H^{\pm}}\simeq\frac{1}{32\pi^2}\left\{\left(\lambda_P^{M\,2}+\lambda_P^{.\,2}\right)\ln\left<\frac{m_{h}^2m_{H}^2}{\Lambda^{4}}\right>-\lambda_P^{M\,2}\ln\left<\frac{m_{h}^2}{m_{H}^2}\right>\right\}\\
\left(\kappa^2_S\right)^{H^{\pm}}\simeq-\frac{\lambda_P^M\lambda_P^.}{8\pi^2}\frac{A_{ud}\sin{2\beta}}{(A_{ud}+\lambda_P^Ms)}\ln\left<\frac{m_{h}^2}{m_{H}^2}\right>\\
\left(\tilde{\kappa}^2_S\right)^{H^{\pm}}\simeq-\frac{\lambda_P^M\lambda_P^.}{8\pi^2}\frac{\sin{2\beta}}{(A_{ud}+\lambda_P^Ms)}(4A_{ud}+3\lambda_P^Ms)\ln\left<\frac{m_{h}^2}{m_{H}^2}\right>
\end{cases}\end{displaymath}}

{\bf Charged-Higgs contributions -- nMSSM:}\newline
The Charged-Higgs boson squared mass-matrix now reads :
{\small\begin{align}
 &({\cal M}_{H^{\pm}}^2)_{11}(S,H_{u,d}^0)=M_{EW}^2+(A_{ud}s-m_{12}^2)\tan^{-1}\beta+\lambda_.|H_u^0|^2+\lambda_3|H_d^0|^2+\lambda_P^.(|S|^2-s^2)\nonumber\\
 &({\cal M}_{H^{\pm}}^2)_{22}(S,H_{u,d}^0)=M_{EW}^2+(A_{ud}s-m_{12}^2)\tan\beta+\lambda_.|H_d^0|^2+\lambda_3|H_u^0|^2+\lambda_P^.(|S|^2-s^2)\\
 &({\cal M}_{H^{\pm}}^2)_{12}(S,H_{u,d}^0)=A_{ud}S-m_{12}^2-\lambda_4(H_u^0H_d^0)^*\nonumber
\end{align}}\noi
Applying the same recipe as in the previous paragraph, we obtain the corrections:
{\small\begin{equation}\begin{cases}
\left(\lambda_T\right)^{H^{\pm}}\simeq-\frac{A_{ud}m_{12}^2}{32\pi^2}\left\{\ln\left<\frac{m_{h}^2m_{H}^2}{\Lambda^{4}}\right>-\ln\left<\frac{m_{h}^2}{m_{H}^2}\right>\right\}\\
\left(m_{H_{u}}^2\right)^{H^{\pm}}\simeq\frac{1}{32\pi^2}\left\{\left[\frac{2(A_{ud}s-m_{12}^2)}{\sin{2\beta}}\left(\lambda_.\cos^2\beta+\lambda_3\sin^2\beta\right)+(\lambda_.+\lambda_3)(M_{EW}^2-\lambda_P^.s^2)\right]\ln\left<\frac{m_{h}^2m_{H}^2}{\Lambda^{4}}\right>\right.\\
\null\hspace{6cm}\left.+\frac{2\lambda_3(A_{ud}s-m_{12}^2)}{\sin{2\beta}}\ln\left<\frac{m_{h}^2}{m_{H}^2}\right>\right\}\\
\left(m_{H_{d}}^2\right)^{H^{\pm}}\simeq\frac{1}{32\pi^2}\left\{\left[\frac{2(A_{ud}s-m_{12}^2)}{\sin{2\beta}}\left(\lambda_.\sin^2\beta+\lambda_3\cos^2\beta\right)+(\lambda_.+\lambda_3)(M_{EW}^2-\lambda_P^.s^2)\right]\ln\left<\frac{m_{h}^2m_{H}^2}{\Lambda^{4}}\right>\right.\\
\null\hspace{6cm}\left.+\frac{2\lambda_.(A_{ud}s-m_{12}^2)}{\sin{2\beta}}\ln\left<\frac{m_{h}^2}{m_{H}^2}\right>\right\}\\
\left(m_{S}^2\right)^{H^{\pm}}\simeq\frac{1}{32\pi^2}\left\{\left[A_{ud}^2+2\lambda_P^.\left(M_{EW}^2-\lambda_P^.s^2+\frac{A_{ud}s-m_{12}^2}{\sin{2\beta}}\right)\right]\ln\left<\frac{m_{h}^2m_{H}^2}{\Lambda^{4}}\right>+\frac{2\lambda_P^.(A_{ud}s-m_{12}^2)}{\sin{2\beta}}\ln\left<\frac{m_{h}^2}{m_{H}^2}\right>\right\}\\
\left(m_{12}^2\right)^{H^{\pm}}\simeq-\frac{\lambda_4m_{12}^2}{16\pi^2}\left\{\ln\left<\frac{m_{h}^2m_{H}^2}{\Lambda^{4}}\right>-\ln\left<\frac{m_{h}^2}{m_{H}^2}\right>\right\}\\
\left(\mu_S^2\right)^{H^{\pm}}\simeq\frac{A_{ud}^2m_{12}^2\sin^2{2\beta}}{64\pi^2\left(A_{ud}s-m_{12}^2\right)}\ln\left<\frac{m_{h}^2}{m_{H}^2}\right>\\
\left(A_{ud}\right)^{H^{\pm}}\simeq\frac{\lambda_4A_{ud}}{32\pi^2}\left\{\ln\left<\frac{m_{h}^2m_{H}^2}{\Lambda^{4}}\right>-\ln\left<\frac{m_{h}^2}{m_{H}^2}\right>\right\}\\
\left(\tilde{A}_{ud}\right)^{H^{\pm}}\simeq\frac{\lambda_4\lambda_P^.A_{ud}s^2\sin{2\beta}}{32\pi^2\left(A_{ud}s-m_{12}^2\right)}\ln\left<\frac{m_{h}^2}{m_{H}^2}\right>\\
\left(A_{us}\right)^{H^{\pm}}\simeq\frac{\lambda_.A_{ud}m_{12}^2\sin{2\beta}}{32\pi^2\left(A_{ud}s-m_{12}^2\right)}\ln\left<\frac{m_{h}^2}{m_{H}^2}\right>\\
\left(A_{ds}\right)^{H^{\pm}}\simeq\frac{\lambda_3A_{ud}m_{12}^2\sin{2\beta}}{32\pi^2\left(A_{ud}s-m_{12}^2\right)}\ln\left<\frac{m_{h}^2}{m_{H}^2}\right>\\
\left(A_{S}\right)^{H^{\pm}}\simeq-\frac{3\lambda_P^.A^3_{ud}s^2\sin^3{2\beta}}{64\pi^2(A_{ud}s-m_{12}^2)^2}\ln\left<\frac{m_{h}^2}{m_{H}^2}\right>\\
\left(\tilde{A}_S\right)^{H^{\pm}}\simeq-\frac{3\lambda_P^.A_{ud}m_{12}^2\sin{2\beta}}{32\pi^2\left(A_{ud}s-m_{12}^2\right)}\ln\left<\frac{m_{h}^2}{m_{H}^2}\right>
\end{cases}\end{equation}
\begin{displaymath}\begin{cases}
\left(\lambda_1\right)^{H^{\pm}}\simeq\frac{1}{32\pi^2}\left\{(\lambda_.^2+\lambda_3^2)\ln\left<\frac{m_{h}^2m_{H}^2}{\Lambda^{4}}\right>+(\lambda_3^2-\lambda_.^2)\ln\left<\frac{m_{h}^2}{m_{H}^2}\right>\right\}\\
\left(\lambda_2\right)^{H^{\pm}}\simeq\frac{1}{32\pi^2}\left\{(\lambda_.^2+\lambda_3^2)\ln\left<\frac{m_{h}^2m_{H}^2}{\Lambda^{4}}\right>+(\lambda_.^2-\lambda_3^2)\ln\left<\frac{m_{h}^2}{m_{H}^2}\right>\right\}\\
(\lambda_3+\lambda_4)^{H^{\pm}}\simeq\frac{1}{32\pi^2}\left\{(2\lambda_.\lambda_3+\lambda_4^2)\ln\left<\frac{m_{h}^2m_{H}^2}{\Lambda^{4}}\right>-\lambda_4^2\ln\left<\frac{m_{h}^2}{m_{H}^2}\right>\right\}\\
\left(\lambda_5\right)^{H^{\pm}}\simeq\frac{\lambda^2_4}{32\pi^2}\sin{2\beta}^2\\
\left(\lambda_6\right)^{H^{\pm}}\simeq\frac{\lambda_4\lambda_3}{32\pi^2}\sin{2\beta}\ln\left<\frac{m_{h}^2}{m_{H}^2}\right>\\
\left(\lambda_7\right)^{H^{\pm}}\simeq\frac{\lambda_4\lambda_.}{32\pi^2}\sin{2\beta}\ln\left<\frac{m_{h}^2}{m_{H}^2}\right>\\
(\lambda_P^u)^{H^{\pm}}\simeq\frac{1}{32\pi^2}\left\{\lambda_P^.(\lambda_.+\lambda_3)\ln\left<\frac{m_{h}^2m_{H}^2}{\Lambda^{4}}\right>+\lambda_P^.(\lambda_--\lambda_3)\ln\left<\frac{m_{h}^2}{m_{H}^2}\right>\right\}\\
(\lambda_P^d)^{H^{\pm}}\simeq\frac{1}{32\pi^2}\left\{\lambda_P^.(\lambda_.+\lambda_3)\ln\left<\frac{m_{h}^2m_{H}^2}{\Lambda^{4}}\right>+\lambda_P^.(\lambda_3-\lambda_.)\ln\left<\frac{m_{h}^2}{m_{H}^2}\right>\right\}\\
(\tilde{\lambda}_P^u)^{H^{\pm}}\simeq\frac{(\lambda_3-\lambda_.)A_{ud}\lambda_P^.s\sin^2{2\beta}}{64\pi^2\left(A_{ud}s-m_{12}^2\right)}\ln\left<\frac{m_{h}^2}{m_{H}^2}\right>\\
(\tilde{\lambda}_P^d)^{H^{\pm}}\simeq\frac{(\lambda_.-\lambda_3)A_{ud}\lambda_P^.s\sin^2{2\beta}}{64\pi^2\left(A_{ud}s-m_{12}^2\right)}\ln\left<\frac{m_{h}^2}{m_{H}^2}\right>\\
\left(\lambda_M\right)^{H^{\pm}}\simeq-\frac{\lambda_4\lambda_P^.\sin{2\beta}}{32\pi^2(A_{ud}s-m_{12}^2)}(2A_{ud}s-m_{12}^2)\ln\left<\frac{m_{h}^2}{m_{H}^2}\right>\\
(\lambda_P^M)^{H^{\pm}}\simeq-\frac{\lambda_4\lambda_P^.sA_{ud}\sin{2\beta}}{32\pi^2(A_{ud}s-m_{12}^2)}\ln\left<\frac{m_{h}^2}{m_{H}^2}\right>\\
\left(\tilde{\lambda}^P_M\right)^{H^{\pm}}\simeq\frac{\lambda_4\lambda_P^.sA_{ud}\sin^3{2\beta}}{64\pi^2(A_{ud}s-m_{12}^2)}\ln\left<\frac{m_{h}^2}{m_{H}^2}\right>\\
(\kappa^2)^{H^{\pm}}\simeq\frac{1}{32\pi^2}\left\{\lambda_P^{.\,2}\ln\left<\frac{m_{h}^2m_{H}^2}{\Lambda^{4}}\right>-\frac{\lambda_P^.A_{ud}^2\sin{2\beta}}{A_{ud}s-m_{12}^2}\ln\left<\frac{m_{h}^2}{m_{H}^2}\right>\right\}\\
\left(\kappa^2_S\right)^{H^{\pm}}\simeq-\frac{A_{ud}^3\lambda_P^.s\sin^5{2\beta}}{64\pi^2(A_{ud}s-m_{12}^2)^2}\ln\left<\frac{m_{h}^2}{m_{H}^2}\right>\\
\left(\tilde{\kappa}^2_S\right)^{H^{\pm}}\simeq\frac{\lambda_P^.A_{ud}^2\sin^3{2\beta}}{32\pi^2(A_{ud}s-m_{12}^2)^2}(3A_{ud}s-m_{12}^2)\ln\left<\frac{m_{h}^2}{m_{H}^2}\right>
\end{cases}\end{displaymath}}

{\bf Neutral-Higgs contributions - NMSSM:}\newline
The $6\times6$ symmetric squared mass-matrix is given by its entries, in the base {\small$(h_u^0,h_d^0,h_S^0,a_u^0,a_d^0,a_S^0)$}:
{\small\begin{align}
 ({\cal M}_{H^0}^2)_{11}&=M_{EW}^2+(A_{ud}+\lambda_P^Ms)\frac{s}{\tan\beta}+\lambda_P^.(|S|^2-s^2)+\lambda_.(2|H_u^0|^2+\mbox{Re}(H_u^{0\,2}))+(\lambda_3+\lambda_4)|H_d^0|^2\nonumber\\
 ({\cal M}_{H^0}^2)_{22}&=M_{EW}^2+(A_{ud}+\lambda_P^Ms)s\tan\beta+\lambda_P^.(|S|^2-s^2)+\lambda_.(2|H_d^0|^2+\mbox{Re}(H_d^{0\,2}))+(\lambda_3+\lambda_4)|H_u^0|^2\nonumber\\
 ({\cal M}_{H^0}^2)_{12}&=-A_{ud}\mbox{Re}(S)-\lambda_P^M\mbox{Re}(S^2)+2(\lambda_3+\lambda_4)\mbox{Re}(H_u^0)\mbox{Re}(H_d^0)\nonumber\\
 ({\cal M}_{H^0}^2)_{33}&=(A_S+4\kappa^2s)s+2A_S(\mbox{Re}(S)-s)+2\kappa^2(2|S|^2+\mbox{Re}(S^2)-3s^2)+\lambda_P^.(|H_u^0|^2+|H_d^0|^2)-2\lambda_P^M\mbox{Re}(H_u^0H_d^0)\nonumber\\
 ({\cal M}_{H^0}^2)_{13}&=-A_{ud}\mbox{Re}(H_d^0)-2\lambda_P^M\mbox{Re}(S^*H_d^0)+2\lambda_P^.\mbox{Re}(S)\mbox{Re}(H_u^0)\nonumber\\
 ({\cal M}_{H^0}^2)_{23}&=-A_{ud}\mbox{Re}(H_u^0)-2\lambda_P^M\mbox{Re}(S^*H_u^0)+2\lambda_P^.\mbox{Re}(S)\mbox{Re}(H_d^0)\nonumber\\
 ({\cal M}_{H^0}^2)_{44}&=M_{EW}^2+(A_{ud}+\lambda_P^Ms)\frac{s}{\tan\beta}+\lambda_P^.(|S|^2-s^2)+\lambda_.(2|H_u^0|^2-\mbox{Re}(H_u^{0\,2}))+(\lambda_3+\lambda_4)|H_d^0|^2\nonumber\\
 ({\cal M}_{H^0}^2)_{55}&=M_{EW}^2+(A_{ud}+\lambda_P^Ms)s\tan\beta+\lambda_P^.(|S|^2-s^2)+\lambda_.(2|H_d^0|^2-\mbox{Re}(H_d^{0\,2}))+(\lambda_3+\lambda_4)|H_u^0|^2\nonumber\\
 ({\cal M}_{H^0}^2)_{45}&=A_{ud}\mbox{Re}(S)+\lambda_P^M\mbox{Re}(S^2)+2(\lambda_3+\lambda_4)\mbox{Im}(H_u^0)\mbox{Im}(H_d^0)\nonumber\\
 ({\cal M}_{H^0}^2)_{66}&=-3A_Ss-2A_S(\mbox{Re}(S)-s)+2\kappa^2(2|S|^2-\mbox{Re}(S^2)-s^2)+\lambda_P^.(|H_u^0|^2+|H_d^0|^2)+2\lambda_P^M\mbox{Re}(H_u^0H_d^0)\nonumber\\
 ({\cal M}_{H^0}^2)_{46}&=A_{ud}\mbox{Re}(H_d^0)-2\lambda_P^M\mbox{Re}(S^*H_d^0)+2\lambda_P^.\mbox{Im}(S)\mbox{Im}(H_u^0)\nonumber\\
 ({\cal M}_{H^0}^2)_{56}&=A_{ud}\mbox{Re}(H_u^0)-2\lambda_P^M\mbox{Re}(S^*H_u^0)+2\lambda_P^.\mbox{Im}(S)\mbox{Im}(H_d^0)\nonumber\\
 ({\cal M}_{H^0}^2)_{14}&=\lambda_.\mbox{Im}(H_u^{0\,2})\\
 ({\cal M}_{H^0}^2)_{15}&=A_{ud}\mbox{Im}(S)-\lambda_P^M\mbox{Im}(S^2)+2(\lambda_3+\lambda_4)\mbox{Re}(H_u^0)\mbox{Im}(H_d^0)\nonumber\\
 ({\cal M}_{H^0}^2)_{16}&=A_{ud}\mbox{Im}(H_d^0)+2\lambda_P^M\mbox{Im}(SH_d^{0\,*})+2\lambda_P^.\mbox{Im}(S)\mbox{Re}(H_u^0)\nonumber\\
 ({\cal M}_{H^0}^2)_{24}&=A_{ud}\mbox{Im}(S)-\lambda_P^M\mbox{Im}(S^2)+2(\lambda_3+\lambda_4)\mbox{Im}(H_u^0)\mbox{Re}(H_d^0)\nonumber\\
 ({\cal M}_{H^0}^2)_{25}&=\lambda_.\mbox{Im}(H_d^{0\,2})\nonumber\\
 ({\cal M}_{H^0}^2)_{26}&=A_{ud}\mbox{Im}(H_u^0)+2\lambda_P^M\mbox{Im}(SH_u^{0\,*})+2\lambda_P^.\mbox{Im}(S)\mbox{Re}(H_d^0)\nonumber\\
 ({\cal M}_{H^0}^2)_{34}&=A_{ud}\mbox{Im}(H_d^0)+2\lambda_P^M\mbox{Im}(S^*H_d^{0})+2\lambda_P^.\mbox{Re}(S)\mbox{Im}(H_u^0)\nonumber\\
 ({\cal M}_{H^0}^2)_{35}&=A_{ud}\mbox{Im}(H_u^0)+2\lambda_P^M\mbox{Im}(S^*H_u^{0})+2\lambda_P^.\mbox{Re}(S)\mbox{Im}(H_d^0)\nonumber\\
 ({\cal M}_{H^0}^2)_{36}&=-2A_S\mbox{Im}(S)+2\kappa^2\mbox{Im}(S^2)-2\lambda_P^M\mbox{Im}(H_u^0H_d^0)\nonumber
\end{align}}\noi
One can expand its eigenvalues in terms of doublet fields:
{\small\begin{equation}
 m_i^2(S,H_u^0,H_d^0)=m_i^{2\,(0)}(S)+m_i^{2\,(1)}(S,H_u^0,H_d^0)+m_i^{2\,(2)}(S,H_u^0,H_d^0)+O\left((H^0_{u,d})^5\right)
\end{equation}\noi}
The associated potential is then given by:
{\small\begin{multline}\label{RHpot}
 \delta{\cal V}_{\mbox{\small eff}}^{\Lambda,H^{0}}(H_u^0,H_d^0,S)=\frac{1}{64\pi^2}\sum_im_i^4(S,H_u^0,H_d^0)\left[\ln\left(\frac{m_i^2(S,H_u^0,H_d^0)}{\Lambda^2}\right)-\frac{3}{2}\right]\\
=\frac{1}{64\pi^2}\sum_i\left\{m_i^{4\,(0)}(S)\left[\ln\left(\frac{m_i^{2\,(0)}(S)}{\Lambda^2}\right)-\frac{3}{2}\right]+2m_i^{2\,(0)}(S)m_i^{2\,(1)}(S,H_u^0,H_d^0)\left[\ln\left(\frac{m_i^{2\,(0)}(S)}{\Lambda^2}\right)-1\right]\right.\\\left.+\left(2m_i^{2\,(0)}(S)m_i^{2\,(2)}(S,H_u^0,H_d^0)\left[\ln\left(\frac{m_i^{2\,(0)}(S)}{\Lambda^2}\right)-1\right]+m_i^{4\,(1)}(S,H_u^0,H_d^0)\ln\left(\frac{m_i^{2\,(0)}(S)}{\Lambda^2}\right)\right)+O(H^5_{u,d})\right\}
\end{multline}}\noi
The large logarithms are then those terms multiplying $\ln\left(\frac{m_i^{2\,(0)}(S)}{\Lambda^2}\right)\simeq\ln\left(\frac{m_i^{2\,(0)}(s)}{\Lambda^2}\right)+\ldots$
Thence consists our primary task in diagonalizing ${\cal M}_{H^0}^2$ perturbatively with respect to the doublet fields.

\noi We first consider ${\cal M}_{H^0}^2(S,H_{u,d}^0=0)$ in order to obtain $m_i^{2\,(0)}(S)$. We denote as $\left|E_i\right>\equiv(\delta_{ij})_{j=1,\ldots,6}$
the elements of the canonical base of $\mathbb{R}^6$. The subspaces $\mbox{Vec}\{\left|E_1\right>,\left|E_2\right>,\left|E_4\right>,\left|E_5\right>\}$
and $\mbox{Vec}\{\left|E_3\right>,\left|E_6\right>\}$ obviously decouple in ${\cal M}_{H^0}^2(S,H_{u,d}^0=0)$. In the doublet sector, one notices that the eigenstate
equation $\left({\cal M}_{H^0}^2(S,H_{u,d}^0=0)-m^2\right)\sum_{i=1,2,4,5}x_i\left|E_i\right>=0$ is equivalent to $\left(\tilde{\cal M}^2-m^2\right)(z_1,z_2)^T=0$,
where $z_1\equiv x_1+\imath x_3$, $z_2=x_2-\imath x_4$ and $\tilde{\cal M}^2$ is the $2\times2$ (complex) hermitian matrix
determined by the following entries:
\begin{align}
 \tilde{\cal M}^2_{11}&=M_{EW}^2+(A_{ud}+\lambda_P^Ms)\frac{s}{\tan\beta}+\lambda_P^.(|S|^2-s^2)\nonumber\\
 \tilde{\cal M}^2_{22}&=M_{EW}^2+(A_{ud}+\lambda_P^Ms)s\tan\beta+\lambda_P^.(|S|^2-s^2)\\
 \tilde{\cal M}^2_{12}&=-A_{ud}S^*-\lambda_P^MS^{2}\nonumber
\end{align}\noi
One recognises ${\cal M}_{H^{\pm}}^2(S,H_{u,d}^0=0)$, up to the sign of the off-diagonal terms. $\tilde{\cal M}^2$ is diagonalized by the eigenstates:
{\small\begin{multline}
 m_{h^0/H^0}^2=M_{EW}^2+\frac{(A_{ud}+\lambda_P^Ms)s}{\sin{2\beta}}+\lambda_P^{.}(|S|^2-s^2)-/+\left[\left(\frac{(A_{ud}+\lambda_P^Ms)s}{\sin{2\beta}}\right)^2+\left(\left|A_{ud}S^*+\lambda_P^MS^2\right|^2-(A_{ud}+\lambda_P^Ms)^2s^2\right)\right]^{1/2}\\
(z_1)_{h^0}=\frac{m_{H^0}^2-\tilde{\cal M}^2_{11}}{\sqrt{(m_{H^0}^2-\tilde{\cal M}^2_{11})^2+|\tilde{\cal M}^2_{12}|^2}}\equiv x_D\ \ \ ;\ \ \ (z_2)_{h^0}=\frac{-\tilde{\cal M}^{2\,*}_{12}}{\sqrt{(m_{H^0}^2-\tilde{\cal M}^2_{11})^2+|\tilde{\cal M}^2_{12}|^2}}\equiv -y^*_D\\
(z_1)_{H^0}=\frac{\tilde{\cal M}^{2}_{12}}{\sqrt{(m_{H^0}^2-\tilde{\cal M}^2_{11})^2+|\tilde{\cal M}^2_{12}|^2}}\equiv y_D\ \ \ ;\ \ \ (z_2)_{H^0}=\frac{m_{H^0}^2-\tilde{\cal M}^2_{11}}{\sqrt{(m_{H^0}^2-\tilde{\cal M}^2_{11})^2+|\tilde{\cal M}^2_{12}|^2}}\equiv x_D\hspace{0.4cm}\null
\end{multline}}\noi
The following relations will proove useful later:
\begin{equation}\label{relD}
 x_D^2+|y_D|^2=1\ \ \ ;\ \ \ x_D^2-|y_D|^2=\frac{2}{\tan{2\beta}}\frac{(A_{ud}+\lambda_P^Ms)s}{m_{h^0}^2(S)-m_{H^0}^2(S)}\ \ \ ;\ \ \ x_Dy_D=\frac{A_{ud}S^*+\lambda_P^MS^2}{m_{h^0}^2(S)-m_{H^0}^2(S)}
\end{equation}\noi
$Z_1\equiv\left((z_1)_{h^0},(z_2)_{h^0}\right)^T$ and $Z_2\equiv\left((z_1)_{H^0},(z_2)_{H^0}\right)^T$ are eigenvectors of $\tilde{\cal M}^2$ in
the complex sense. In the real sense, $\imath Z_1$ and $\imath Z_2$ form two other linearly-independant (and degenerate to $Z_1$, $Z_2$)
eigenstates. We thus obtain the doublet eigenstates of ${\cal M}_{H^0}^2(S,H_{u,d}^0=0)$:
{\small\begin{align}
 m_{h^0}^2\ \ \ ;&\ \ \begin{cases}\left|h^0_1\right>=x_D\left|E_1\right>-\mbox{Re}(y_D)\left|E_2\right>-\mbox{Im}(y_D)\left|E_4\right>\\ \left|h^0_2\right>=-\mbox{Im}(y_D)\left|E_2\right>+x_D\left|E_3\right>+\mbox{Re}(y_D)\left|E_4\right> \end{cases}\\
 \nonumber m_{H^0}^2\ \ \ ;&\ \ \begin{cases}\left|H^0_1\right>=\mbox{Re}(y_D)\left|E_1\right>+x_D\left|E_2\right>+\mbox{Im}(y_D)\left|E_3\right>\\ \left|H^0_2\right>=-\mbox{Im}(y_D)\left|E_1\right>+\mbox{Re}(y_D)\left|E_3\right>-x_D\left|E_4\right> \end{cases}
\end{align}}

\noi For the remaining singlet-states, ${\cal M}_{H^0}^2(S,H_{u,d}^0=0)$ is diagonalized by:
{\small\begin{align}
& m_{h_S^0/a_S^0}=(-A_S+2\kappa^2s)s+4\kappa^2(|S|^2-s^2)+/-2\left[|A_SS+\kappa^2S^{*\,2}|^2\right]^{1/2}\ \ \ ;\ \ \ \begin{cases}
\left|h_S^0\right>=x_S\left|E_3\right>-y_S\left|E_6\right>\\\left|a_S^0\right>=y_S\left|E_3\right>+x_S\left|E_6\right>\end{cases}\\
\nonumber&x_S\equiv\frac{{\cal M}_{H^0}^2(S,0)_{66}-m^2_{h_S^0}}{\sqrt{\left({\cal M}_{H^0}^2(S,0)_{66}-m^2_{h_S^0}\right)^2+|{\cal M}_{H^0}^2(S,0)_{36}|^2}}\ \ \ ;\ \ \ y_S\equiv\frac{{\cal M}_{H^0}^2(S,0)_{36}}{\sqrt{\left({\cal M}_{H^0}^2(S,0)_{66}-m^2_{h_S^0}\right)^2+|{\cal M}_{H^0}^2(S,0)_{36}|^2}}
\end{align}}\noi
It is convenient to introduce the notation $z_S=x_S+\imath y_S$ and note the following relations:
\begin{equation}\label{relS}
 |z_S|^2=1\ \ \ \ \ \ ;\ \ \ \ \ \ z_S^2=4\frac{A_SS+\kappa^2S^{*\,2}}{m^2_{h_S^0}(S)-m^2_{a_S^0}(S)}
\end{equation}

\noi At this stage, one can already determine the pure-singlet parameters of the potential. Similarly to the charged case (eq. \ref{CHpot}),
one can formulate the first term in eq. \ref{RHpot} in terms of $T_D=\frac{1}{2}\left(m_{H^0}^2+m_{h^0}^2\right)$, $R_D=\frac{1}{2}\left(m_{H^0}^2-m_{h^0}^2\right)$,
$T_S=m_{h^0_S}^2+m_{a^0_S}^2$ and $R_S=m_{h^0_S}^2-m_{a^0_S}^2$. Moreover, the contributions from the doublet are trivially 
identical to those in the charged case and here, as well, we give only the leading term in $\sin{2\beta}\to0$ for the coefficient multiplying 
$\ln\left<\frac{m_{h}^2}{m_{H}^2}\right>$. The logarithmicaly-enhanced parameters are then:
{\small\begin{equation}\begin{cases}
\left(\lambda_T\right)^{H^{0}}\simeq\frac{\lambda_P^.\lambda_P^Ms^3\sin{2\beta}}{32\pi^2\left(A_{ud}+\lambda_P^Ms\right)}(5A_{ud}+12\lambda_P^Ms)\ln\left<\frac{m_{h}^2}{m_{H}^2}\right>-\frac{A_Ss(A_S+2\kappa^2s)}{128\pi^2}\ln\left<\frac{m_{h_S^0}^2}{m_{a_S^0}^2}\right>\\
\left(m_{S}^2\right)^{H^{0}}\simeq\frac{1}{32\pi^2}\left\{\left[A_{ud}^2+2\lambda_P^.\left(M_{EW}^2-\lambda_P^.s^2+\frac{s(A_{ud}+\lambda_P^Ms)}{\sin{2\beta}}\right)\right]\ln\left<\frac{m_{h}^2m_{H}^2}{\Lambda^{4}}\right>-\frac{2\lambda_P^.s(A_{ud}+\lambda_P^Ms)}{\sin{2\beta}}\ln\left<\frac{m_{h}^2}{m_{H}^2}\right>\right.\\
\null\hspace{1cm}+\left.\frac{1}{2}\left[A_S^2-2 A_S\kappa^2s-4 \kappa^4 s^2\right]\ln\left<\frac{m_{h_S^0}^2m_{a_S^0}^2}{\Lambda^4}\right>-\frac{3A_S^5+10A_S^4\kappa^2s+2A_S^3\kappa^4s^2-60A_S^2\kappa^6s^3-20A_S\kappa^8s^4+8\kappa^{10}s^5}{8\left(A_S+\kappa^2s\right)^3}\ln\left<\frac{m_{h_S^0}^2}{m_{a_S^0}^2}\right>\right\}\\
\left(\mu_S^2\right)^{H^{0}}\simeq-\frac{3\lambda_P^M\lambda_P^.s^2\sin{2\beta}}{16\pi^2\left(A_{ud}+\lambda_P^Ms\right)}(2A_{ud}+3\lambda_P^Ms)\ln\left<\frac{m_{h}^2}{m_{H}^2}\right>+\frac{3A_S}{256\pi^2}\frac{A_S^4+2A_S^3\kappa^2s-6A_S^2\kappa^2s^2-32A_S\kappa^3s^3-16\kappa^4s^4}{(A_S+\kappa^2s)^3}\ln\left<\frac{m_{h_S^0}^2}{m_{a_S^0}^2}\right>\\
\left(A_{S}\right)^{H^{0}}\simeq\frac{3\lambda_P^MA_{ud}}{32\pi^2}\left\{\ln\left<\frac{m_{h}^2m_{H}^2}{\Lambda^{4}}\right>-\ln\left<\frac{m_{h}^2}{m_{H}^2}\right>\right\}\\
\null\hspace{1cm}+\frac{3\kappa^2A_S}{64\pi^2}\left\{\ln\left<\frac{m_{h_S^0}^2m_{a_S^0}^2}{\Lambda^4}\right>-\frac{A_S}{16}\frac{3A_S^4+19A_S^3\kappa^2s+30A_S^2\kappa^4s^2-120A_S\kappa^6s^3-80\kappa^8s^4}{\kappa^2s(A_S+\kappa^2s)^3}\ln\left<\frac{m_{h_S^0}^2}{m_{a_S^0}^2}\right>\right\}\\
\left(\tilde{A}_S\right)^{H^{0}}\simeq\frac{9\lambda_P^.\lambda_P^Ms\sin{2\beta}}{16\pi^2\left(A_{ud}+\lambda_P^Ms\right)}(A_{ud}+3\lambda_P^Ms)\ln\left<\frac{m_{h}^2}{m_{H}^2}\right>+\frac{9\kappa^2A_S}{1024\pi^2}\frac{A_S^4+17A_S^3\kappa^2s+42A_S^2\kappa^4s^2-8A_S\kappa^6s^3-16\kappa^8s^4}{\kappa^2s(A_S+\kappa^2s)^3}\ln\left<\frac{m_{h_S^0}^2}{m_{a_S^0}^2}\right>\\
(\kappa^2)^{H^{0}}\simeq\frac{1}{32\pi^2}\left\{\left(\lambda_P^{M\,2}+\lambda_P^{.\,2}\right)\ln\left<\frac{m_{h}^2m_{H}^2}{\Lambda^{4}}\right>-\lambda_P^{M\,2}\ln\left<\frac{m_{h}^2}{m_{H}^2}\right>\right\}\\
\null\hspace{1cm}+\frac{\kappa^4}{64\pi^2}\left\{5\ln\left<\frac{m_{h_S^0}^2m_{a_S^0}^2}{\Lambda^4}\right>-\frac{1}{64}\frac{A_S^5+14A_S^4\kappa^2s+32A_S^3\kappa^4s^2+256A_S^2\kappa^6s^3-640A_S\kappa^8s^4-256\kappa^{10}s^5}{\kappa^4s^2(A_S+\kappa^2s)^3}\ln\left<\frac{m_{h_S^0}^2}{m_{a_S^0}^2}\right>\right\}\\
\left(\kappa^2_S\right)^{H^{0}}\simeq-\frac{\lambda_P^M\lambda_P^.}{8\pi^2}\frac{A_{ud}\sin{2\beta}}{(A_{ud}+\lambda_P^Ms)}\ln\left<\frac{m_{h}^2}{m_{H}^2}\right>+\frac{\kappa^4A_S}{2048\pi^2}\frac{5A_S^4+22A_S^3\kappa^2s+144A_S^2\kappa^4s^2-352A_S\kappa^6s^3-128\kappa^8s^4}{\kappa^4s^2(A_S+\kappa^2s)^3}\ln\left<\frac{m_{h_S^0}^2}{m_{a_S^0}^2}\right>\\
\left(\tilde{\kappa}^2_S\right)^{H^{0}}\simeq-\frac{\lambda_P^M\lambda_P^.}{8\pi^2}\frac{\sin{2\beta}}{(A_{ud}+\lambda_P^Ms)}(4A_{ud}+3\lambda_P^Ms)\ln\left<\frac{m_{h}^2}{m_{H}^2}\right>-\frac{\kappa^4A_S}{512\pi^2}\frac{A_S^4+2A_S^3\kappa^2s-36A_S^2\kappa^4s^2-344A_S\kappa^6s^3-64\kappa^8s^4}{\kappa^4s^2(A_S+\kappa^2s)^3}\ln\left<\frac{m_{h_S^0}^2}{m_{a_S^0}^2}\right>
\end{cases}\end{equation}}

\noi The next step consists in considering $O(H^2)$ corrections to the neutral eigenvalues. Note that, for the eigenvalue $m_i^{2\,(0)}(S)$
of ${\cal M}_{H^0}^2(S,H_{u,d}^0=0)$, with $\mbox{Tr}_i$ the trace on the corresponding eigenspace,
\begin{equation}
 \mbox{Tr}_i[m_i^{2\,(2)}(S,H_u^0,H_d^0)]=\mbox{Tr}_i\left\{{\cal M}^{2\,(2)}_{H^0}+\sum_{j\neq i}\frac{{\cal M}^{2\,(1)}_{H^0}P_j{\cal M}^{2\,(1)}_{H^0}}{m_i^{2\,(0)}(S)-m_j^{2\,(0)}(S)}\right\}
\end{equation}\noi
where ${\cal M}^{2\,(1,2)}_{H^0}$ stand for the matrices with terms of $O(H^{1,2})$ in ${\cal M}^{2}_{H^0}$, $P_j$ corresponds to 
the projector on the eigenspace of the eigenvalue $m_j^{2\,(0)}$. Defining $\eta=1;-1$ for $D=H^0;h^0$ and $\epsilon=1;-1$ for 
$S=h_S^0;a_S^0$ and using the relations \ref{relD}, \ref{relS}, we obtain the following matrix elements:
{\small\begin{align}
&\mbox{Tr}_D\left<D\right|{\cal M}^{2\,(2)}_{H^0}\left|D\right>\equiv\left<D_S\right>+\frac{\eta}{m_{H^0}^2-m_{h^0}^2}\left<D_A\right>\nonumber\\
&\null\hspace{0.5cm}\left<D_S\right>=\left[2\lambda_.+\lambda_3+\lambda_4\right](|H_u^0|^2+|H_d^0|^2)\nonumber\\
&\null\hspace{0.5cm}\left<D_A\right>=\frac{2(A_{ud}+\lambda_P^Ms)s}{\tan{2\beta}}\left[2\lambda_.-\lambda_3-\lambda_4\right](|H_u^0|^2-|H_d^0|^2)-4(\lambda_3+\lambda_4)\left[A_{ud}\mbox{Re}(SH_u^0H_d^0)+\lambda_P^M\mbox{Re}(S^{*\,2}H_u^0H_d^0)\right]\nonumber\\
&\left<S\right|{\cal M}^{2\,(2)}_{H^0}\left|S\right>\equiv\left<S_S\right>+\frac{\epsilon}{m_{h_S^0}^2-m_{a_S^0}^2}\left<S_A\right>\nonumber\\
&\null\hspace{0.5cm}\left<S_S\right>=\lambda_P^.(|H_u^0|^2+|H_d^0|^2)\nonumber\\
&\null\hspace{0.5cm}\left<S_S\right>=-8\lambda_P^M\left[A_S\mbox{Re}(SH_u^0H_d^0)+\kappa^2\mbox{Re}(S^{*\,2}H_u^0H_d^0)\right]\nonumber\\
&\left<S\right|{\cal M}^{2\,(1)}_{H^0}P_D{\cal M}^{2\,(1)}_{H^0}\left|S\right>\equiv\left<1;1\right>+\frac{\eta}{m_{H^0}^2-m_{h^0}^2}\left<-1;1\right>+\frac{\epsilon}{m_{h_S^0}^2-m_{a_S^0}^2}\left<1;-1\right>+\frac{\eta}{m_{H^0}^2-m_{h^0}^2}\frac{\epsilon}{m_{h_S^0}^2-m_{a_S^0}^2}\left<-1;-1\right>\nonumber\\
&\null\hspace{0.5cm}\left<1;1\right>=\frac{1}{2}\left\{A_{ud}^2+4\lambda_P^{M\,2}|S|^2+2\lambda_P^{.\,2}|S|^2\right\}(|H_u^0|^2+|H_d^0|^2)-2\lambda_P^.\left[A_{ud}\mbox{Re}(SH_u^0H_d^0)+2\lambda_P^M\mbox{Re}(S^{*\,2}H_u^0H_d^0)\right]\nonumber\\
&\null\hspace{0.5cm}\left<-1;1\right>=2\lambda_P^.\left[A_{ud}^2|S|^2+2\lambda_P^{M\,2}|S|^4+3\lambda_P^MA_{ud}\mbox{Re}(S^3)\right](|H_u^0|^2+|H_d^0|^2)\\
&\null\hspace{1cm}-\frac{(A_{ud}+\lambda_P^Ms)s}{\tan{2\beta}}\left[A_{ud}^2+4\lambda_P^{M\,2}|S|^2-2\lambda_P^{.\,2}|S|^2\right](|H_u^0|^2-|H_d^0|^2)\nonumber\\
&\null\hspace{1cm}-4\lambda_P^{.\,2}|S|^2\left[A_{ud}\mbox{Re}(SH_u^0H_d^0)+\lambda_P^M\mbox{Re}(S^{*\,2}H_u^0H_d^0)\right]-8\lambda_P^MA_{ud}\left[\lambda_P^M|S|^2\mbox{Re}(SH_u^0H_d^0)+A_{ud}\mbox{Re}(S^{*\,2}H_u^0H_d^0)\right]\nonumber\\
&\null\hspace{0.5cm}\left<1;-1\right>=\left[8\lambda_P^MA_{ud}(A_S|S|^2+\kappa^2\mbox{Re}(S^3))+4\lambda_P^{.\,2}\left(A_S\mbox{Re}(S^3)+\kappa^2|S|^4\right)\right](|H_u^0|^2+|H_d^0|^2)\nonumber\\
&\null\hspace{2cm}-8\lambda_P^.\left[(\kappa^2A_{ud}+2\lambda_P^MA_S)|S|^2\mbox{Re}(SH_u^0H_d^0)+(A_SA_{ud}+2\lambda_P^M\kappa^2|S|^2)\mbox{Re}(S^{*\,2}H_u^0H_d^0)\right]\nonumber\\
&\null\hspace{0.5cm}\left<-1;-1\right>=8\lambda_P^.\left[A_{ud}(\kappa^2A_{ud}+3\lambda_P^MA_S)|S|^4+2\lambda_P^{M\,2}\kappa^2|S|^6\right.\nonumber\\
&\null\hspace{5cm}\left.+(A_SA_{ud}^2+3\lambda_P^MA_{ud}\kappa^2|S|^2+2\lambda_P^{M\,2}A_S|S|^2)\mbox{Re}(S^3)\right](|H_u^0|^2+|H_d^0|^2)\nonumber\\
&\null\hspace{1cm}-\frac{8(A_{ud}+\lambda_P^Ms)s}{\tan{2\beta}}\left[2\lambda_P^MA_{ud}\left(A_S|S|^2+\kappa^2\mbox{Re}(S^3)\right)-\lambda_P^{.\,2}\left(A_S\mbox{Re}(S^3)+\kappa^2|S|^4\right)\right](|H_u^0|^2-|H_d^0|^2)\nonumber\\
&\null\hspace{1cm}-16\lambda_P^{.\,2}\left(A_S\mbox{Re}(S^3)+\kappa^2|S|^4\right)\left[A_{ud}\mbox{Re}(SH_u^0H_d^0)+\lambda_P^M\mbox{Re}(S^{*\,2}H_u^0H_d^0)\right]\nonumber\\
&\null\hspace{1cm}-8\left[\left(\lambda_P^MA_S(A_{ud}^2+4\lambda_P^{M\,2}|S|^2)+\kappa^2A_{ud}^3\right)|S|^2\mbox{Re}(SH_u^0H_d^0)\right.\nonumber\\
&\null\hspace{3cm}\left.+A_{ud}\left(A_S(A_{ud}^2+4\lambda_P^{M\,2}|S|^2)+4\lambda_P^{M\,3}\kappa^2|S|^4\right)\mbox{Re}(S^{*\,2}H_u^0H_d^0)\right.\nonumber\\
&\null\hspace{5cm}\left.+\lambda_P^MA_{ud}^2\kappa^2\mbox{Re}(S^{4}H_u^0H_d^0)+4\lambda_P^{M\,2}A_{ud}\kappa^2\mbox{Re}(S^{*\,5}H_u^0H_d^0)\right]\nonumber
\end{align}}

\noi Getting back to \ref{RHpot}, one obtains the $O(H^2)$ coefficients:
{\small\begin{equation}
 \null\hspace{0cm}\begin{cases}
\left(m_{H_{u}}^2\right)^{H^{0}}\simeq\frac{1}{64\pi^2}\left\{\left[A_{ud}^2+4\lambda_.\left(M_{EW}^2+\lambda_P^.s^2+2\cos^2\beta\frac{(A_{ud}+\lambda_P^Ms)s}{\sin{2\beta}}\right)\right.\right.\\
\null\hspace{5cm}\left.\left.+2(\lambda_3+\lambda_4)\left(M_{EW}^2-\lambda_P^.s^2+2\sin^2\beta\frac{(A_{ud}+\lambda_P^Ms)s}{\sin{2\beta}}\right)\right]\ln\left<\frac{m_{h}^2m_{H}^2}{\Lambda^{4}}\right>\right.\\
\null\hspace{2cm}\left.+\left[A_{ud}^2-2\lambda_P^.s(A_S+2\kappa^2s)\right]\ln\left<\frac{m_{h_S^0}^2m_{a_S^0}^2}{\Lambda^4}\right>+\ldots\right\}\\
\left(m_{H_{d}}^2\right)^{H^{0}}\simeq\frac{1}{64\pi^2}\left\{\left[A_{ud}^2+4\lambda_.\left(M_{EW}^2+\lambda_P^.s^2+2\sin^2\beta\frac{(A_{ud}+\lambda_P^Ms)s}{\sin{2\beta}}\right)\right.\right.\\
\null\hspace{5cm}\left.\left.+2(\lambda_3+\lambda_4)\left(M_{EW}^2-\lambda_P^.s^2+2\cos^2\beta\frac{(A_{ud}+\lambda_P^Ms)s}{\sin{2\beta}}\right)\right]\ln\left<\frac{m_{h}^2m_{H}^2}{\Lambda^{4}}\right>\right.\\
\null\hspace{2cm}\left.+\left[A_{ud}^2-2\lambda_P^.s(A_S+2\kappa^2s)\right]\ln\left<\frac{m_{h_S^0}^2m_{a_S^0}^2}{\Lambda^4}\right>+\ldots\right\}\\
\left(A_{ud}\right)^{H^{0}}\simeq\frac{1}{32\pi^2}\left\{(\lambda_3+\lambda_4+\lambda_P^.)A_{ud}\ln\left<\frac{m_{h}^2m_{H}^2}{\Lambda^{4}}\right>+(\lambda_P^.A_{ud}+2\lambda_P^MA_S)\ln\left<\frac{m_{h_S^0}^2m_{a_S^0}^2}{\Lambda^4}\right>+\ldots\right\}\\
(\lambda_P^u)^{H^{0}}\simeq\frac{1}{32\pi^2}\left\{\left(\lambda_P^.(2\lambda_.\lambda_3+\lambda_4+\lambda_P^.)+\lambda_P^{M\,2}\right)\ln\left<\frac{m_{h}^2m_{H}^2}{\Lambda^{4}}\right>+2\left(\lambda_P^.(4\kappa^2+\lambda_P^.)+2\lambda_P^{M\,2}\right)\ln\left<\frac{m_{h_S^0}^2m_{a_S^0}^2}{\Lambda^4}\right>+\ldots\right\}\\
(\lambda_P^d)^{H^{0}}\simeq\frac{1}{32\pi^2}\left\{\left(\lambda_P^.(2\lambda_.\lambda_3+\lambda_4+\lambda_P^.)+\lambda_P^{M\,2}\right)\ln\left<\frac{m_{h}^2m_{H}^2}{\Lambda^{4}}\right>+2\left(\lambda_P^.(4\kappa^2+\lambda_P^.)+2\lambda_P^{M\,2}\right)\ln\left<\frac{m_{h_S^0}^2m_{a_S^0}^2}{\Lambda^4}\right>+\ldots\right\}\\
(\lambda_P^M)^{H^{0}}\simeq\frac{1}{32\pi^2}\left\{\lambda_P^M(\lambda_3+\lambda_4+2\lambda_P^.)\ln\left<\frac{m_{h}^2m_{H}^2}{\Lambda^{4}}\right>+2\lambda_P^{M}(\kappa^2+\lambda_P^.)\ln\left<\frac{m_{h_S^0}^2m_{a_S^0}^2}{\Lambda^4}\right>+\ldots\right\}\\
 \end{cases}
\end{equation}}\noi
The logarithms involving ratios of Higgs masses (symbolized by \ldots) are too complicated to write down explicitely. They also appear 
within contributions to the $\mathbb{Z}_3$-violating parameters. We could check however that such contributions to 
$\mathbb{Z}_3$-violating parameters vanished in relevant limits ($m_{h_S^0,a_S^0}^2\ll m_{H^0}^2\ \bigoplus\ \sin{2\beta}\to0$,
$m_{H^0}^2\ll m_{h_S^0,a_S^0}^2$, $m_{a_S^0}^2\to0$).

\noi We skip the computation of corrections to the quartic doublet parameters, here as well as for the neutralino 
contributions: such a task, although straightforward (perturbative calculation of the eigenvalues of a matrix up 
to the fourth order) promises to be technically tedious.

{\bf Neutral-Higgs contributions - nMSSM:}\newline
We can draw some conclusions from the study in the NMSSM case. For simplicity, we will confine here to contributions to pure 
singlet parameters, so that we need only consider ${\cal M}_{H^0}^2(S,H_{u,d}^0=0)$. Obviously, doublet and singlet sectors 
will decouple again, and the doublet sector will generate the same corrections to the $\l_i$'s as the charged Higgs sector. As
for the singlet sector, it is particularly simple in the nMSSM since no dependence in $S$ appears. We thus obtain the 
leading-logarithms to the pure-singlet coefficients:
{\small\begin{equation}\begin{cases}
\left(\lambda_T\right)^{H^{0}}\simeq-\frac{A_{ud}m_{12}^2}{32\pi^2}\left\{\ln\left<\frac{m_{h}^2m_{H}^2}{\Lambda^{4}}\right>-\ln\left<\frac{m_{h}^2}{m_{H}^2}\right>\right\}\\
\left(m_{S}^2\right)^{H^{0}}\simeq\frac{1}{32\pi^2}\left\{\left[A_{ud}^2+2\lambda_P^.\left(M_{EW}^2-\lambda_P^.s^2+\frac{A_{ud}s-m_{12}^2}{\sin{2\beta}}\right)\right]\ln\left<\frac{m_{h}^2m_{H}^2}{\Lambda^{4}}\right>+\frac{2\lambda_P^.(A_{ud}s-m_{12}^2)}{\sin{2\beta}}\ln\left<\frac{m_{h}^2}{m_{H}^2}\right>\right\}\\
\left(\mu_S^2\right)^{H^{0}}\simeq\frac{A_{ud}^2m_{12}^2\sin^2{2\beta}}{64\pi^2\left(A_{ud}s-m_{12}^2\right)}\ln\left<\frac{m_{h}^2}{m_{H}^2}\right>\\
\left(A_{S}\right)^{H^{0}}\simeq-\frac{3\lambda_P^.A^3_{ud}s^2\sin^3{2\beta}}{64\pi^2(A_{ud}s-m_{12}^2)^2}\ln\left<\frac{m_{h}^2}{m_{H}^2}\right>\\
\left(\tilde{A}_S\right)^{H^{0}}\simeq-\frac{3\lambda_P^.A_{ud}m_{12}^2\sin{2\beta}}{32\pi^2\left(A_{ud}s-m_{12}^2\right)}\ln\left<\frac{m_{h}^2}{m_{H}^2}\right>\\
(\kappa^2)^{H^{0}}\simeq\frac{1}{32\pi^2}\left\{\lambda_P^{.\,2}\ln\left<\frac{m_{h}^2m_{H}^2}{\Lambda^{4}}\right>-\frac{\lambda_P^.A_{ud}^2\sin{2\beta}}{A_{ud}s-m_{12}^2}\ln\left<\frac{m_{h}^2}{m_{H}^2}\right>\right\}\\
\left(\kappa^2_S\right)^{H^{0}}\simeq-\frac{A_{ud}^3\lambda_P^.s\sin^5{2\beta}}{64\pi^2(A_{ud}s-m_{12}^2)^2}\ln\left<\frac{m_{h}^2}{m_{H}^2}\right>\\
\left(\tilde{\kappa}^2_S\right)^{H^{0}}\simeq\frac{\lambda_P^.A_{ud}^2\sin^3{2\beta}}{32\pi^2(A_{ud}s-m_{12}^2)^2}(3A_{ud}s-m_{12}^2)\ln\left<\frac{m_{h}^2}{m_{H}^2}\right>
\end{cases}\end{equation}}

{\bf Summary of the analysis:}\newline
Among the potentially large logarithms, we may distinguish among those of the form $\ln\frac{m^2}{\Lambda^2}$, which compare
a given sector to the scale $\Lambda$, typically chosen as the mass of the third-generation squarks, or to another sector, and
those sensitive to hierarchies within a sector $\ln\frac{m_i^2}{m_j^2}$.

In the case of the NMSSM, the logarithms $\ln\frac{m^2}{\Lambda^2}$ obviously appear only in the corrections to the 
$\mathbb{Z}_3$-conserving parameters. Moreover, when logarithms of the type $\ln\frac{m_i^2}{m_j^2}$ appear within 
$\mathbb{Z}_3$-violating parameters, they tend to be balanced by prefactors vanishing in the hierarchical limit 
(typically $\frac{m_i^2m_j^2}{(m_i^2-m_j^2)}$), so that they cannot be regarded as an enhancement factor (contrarily
to when they appear in $\mathbb{Z}_3$-conserving parameters, where the prefactor does not necessarily vanish in this 
limit). One can thus conclude that leading-logarithms preserve the $\mathbb{Z}_3$-induced structure of the potential.

For the nMSSM, the $\mathbb{Z}_3$-symmetry is actually still present at tree-level in all the sectors of the spectrum,
with the exception of the Higgs sector, where it is explicitly violated. Consequently, large logarithms still favour 
the $\mathbb{Z}_3$-conserving terms (even though they are not all present at the classical level), while 
$\mathbb{Z}_3$-violating effects perdure in the Higgs sector. In that case, large logarithms seem likely to destroy 
the classical structure.